\documentclass[aps,pra,amsmath,amssymb,floatfix,twocolumn,amsmath,superscriptaddress,twocolumn,nofootinbib,tighten,letterpaper]{revtex4-2}
\usepackage[colorlinks,linkcolor=blue,citecolor=blue,urlcolor=blue]{hyperref}
\usepackage{multirow}
\usepackage{subfigure}
\usepackage{color}
\usepackage{mathrsfs}
\usepackage{hyperref}
\usepackage[normalem]{ulem}
\usepackage{bm}

\usepackage{amssymb}
\usepackage{amsmath}
\renewcommand\vec[1]{\ensuremath\boldsymbol{#1}}

\usepackage{tabularray}

\usepackage{array} 
\newcolumntype{P}[1]{>{\centering\arraybackslash}p{#1}}

\definecolor{RowColor}{rgb}{0.88,1,0.9}

\usepackage{amsfonts, relsize, color, mathtools, physics}
\usepackage{graphics}
\usepackage{graphicx}
\usepackage{subfigure}
\usepackage{color}
\usepackage{comment}

\begin{document}
\title{Non-Hermitian catalysis of spontaneous symmetry breaking on Euclidean and hyperbolic lattices}

\author{Christopher A. Leong}
\affiliation{Department of Physics, Lehigh University, Bethlehem, Pennsylvania 18015, USA}

\author{Bitan Roy}
\affiliation{Department of Physics, Lehigh University, Bethlehem, Pennsylvania 18015, USA}

\date{\today}

\begin{abstract}
Depending on the lattice geometry, the prototypical nearest-neighbor (NN) tight-binding model for free fermions gives rise to particle-hole symmetric emergent Dirac liquids, Fermi liquids, and (quasi)-flat bands near half filling or zero energy on (bipartite) Euclidean and hyperbolic lattices, respectively, embedded on flat and negatively curved spaces. Such noninteracting electronic fluids are characterized by a vanishing, a finite, and a diverging density of states near the half-filling, respectively. Here, we outline a general principle of realizing non-Hermitian (NH) generalizations of these scenarios in which the resulting NH operator features an all-real eigenvalue spectrum over an extended NH parameter regime where all the single-particle states remain sharp and filling factor is well defined. Most importantly, such a construction reduces the bandwidth of noninteracting systems without altering the characteristic scaling of the density of states close to the zero energy, thereby triggering the ordering propensity toward the nucleation of a family of spontaneous symmetry breaking quantum phases, named commuting-class masses, at weaker interactions. We name this phenomenon ``non-Hermitian catalysis of spontaneous symmetry breaking'' that hinges on a robust and universal algebraic criterion, ultimately liberating this mechanism from the burden of the underlying lattice specification as well as the dimensionality of the system. We establish the NH catalysis of commuting-class masses from the numerical self-consistent solutions of the charge-density-wave and spin-density-wave orders at weaker (in comparison to the counterparts in conventional or Hermitian systems) NN Coulomb and on-site Hubbard repulsions, respectively, obtained by decomposing them in the Hartree channel when combined with the biorthogonal quantum mechanics on NH Euclidean and hyperbolic lattices in which the non-Hermiticity results from an imbalance of the hopping amplitudes between the sites of two sublattices in the opposite directions. These two ordered states correspond to staggered patterns of average electronic density and spin between the NN sites, respectively, and both cause insulation in half-filled systems. We discuss the scaling of the associated mass gaps near the zero energy with the non-Hermitian parameter, showing excellent agreement with analytical predictions, and also address the finite-size scaling of the order parameters specifically on hyperbolic lattices with open boundary conditions. We conclude with an outline of possible experimental platforms to test our concrete theoretical predictions.    
\end{abstract}

\maketitle

\section{Introduction}

The relative scale between the bandwidth and electron-electron interactions determines the quantum behavior of clean electronic crystals at low temperatures. When interparticle interactions are completely ignored, the free fermion motion on crystals is succinctly captured by a tight-binding hopping model and the resulting Bloch states, giving rise to a metallic or ballistic behavior of electrons in solids. Together the hopping amplitude and lattice geometry set the corresponding bandwidth. By contrast, Coulomb or on-site repulsion localizes electronic wavefunctions, yielding insulating behavior at low temperatures. Therefore, a metal-insulator quantum phase transition can be triggered by tuning the interaction-to-bandwidth ratio in quantum crystals, although the exact path for which is not always obvious in real materials~\cite{MIT:1, MIT:2, MIT:3}. 

\begin{figure}[b!]
\includegraphics[width=1.00\linewidth]{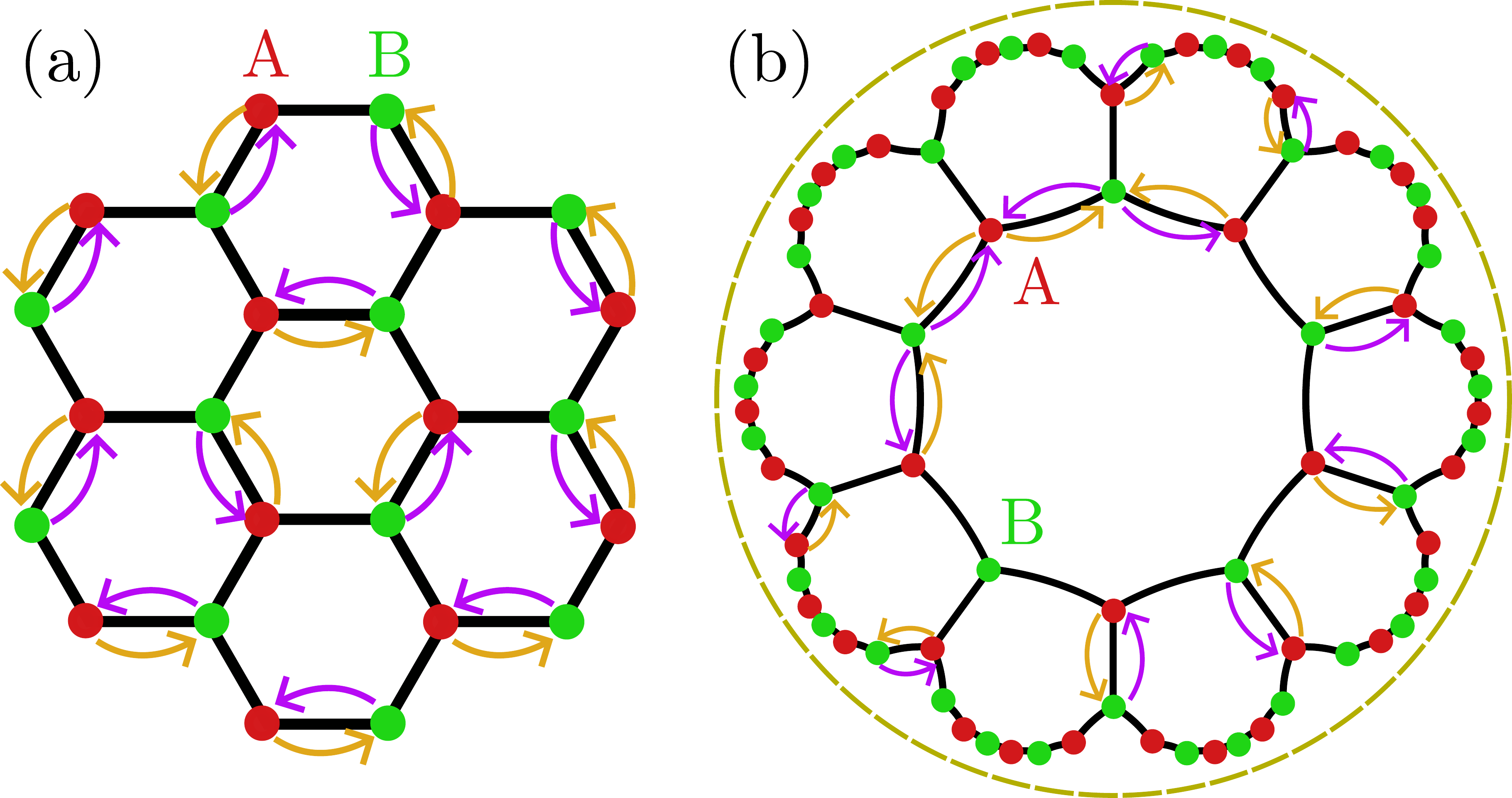}
\caption{Schematic representations of non-Hermitian (a) Euclidean honeycomb or $\{ 6, 3\}$ lattice and (b) $\{ 10, 3\}$ hyperbolic lattice, resulting from an imbalance in the hopping amplitudes between all the nearest-neighbor (NN) sites, belonging to the $A$ (red) and $B$ (green) sublattices. Unequal and opposite directional hopping amplitudes are explicitly shown by arrows with different colors for only a few pairs of the NN sites. The dashed circle represents the Poincar\'e disk in (b).
}~\label{fig:NHLattice}
\end{figure}

\begin{figure*}[t!]
\includegraphics[width=1.00\linewidth]{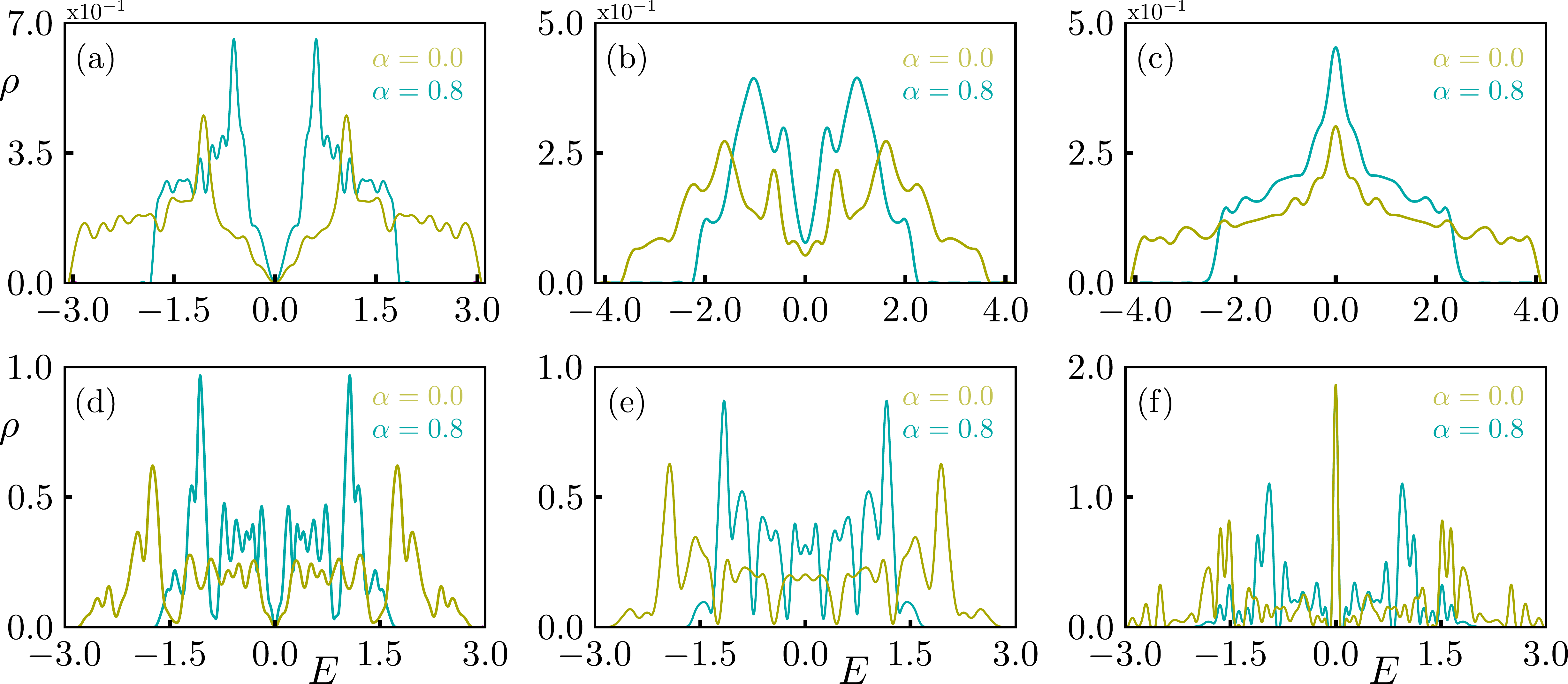}
\caption{Density of states ($\rho$) as a function of energy ($E$) in Hermitian ($\alpha=0.0$) and non-Hermitian ($\alpha=0.8$) systems for (a) Dirac liquid on an Euclidean honeycomb or $\{ 6, 3\}$ lattice, (b) Fermi liquid on an Euclidean Bernal-stacked bilayer honeycomb lattice, (c) (quasi)flat-band system on a Euclidean square or $\{ 4,4 \}$ lattice, (d) Dirac liquid on $\{ 10, 3 \}$ hyperbolic lattice, (e) Fermi liquid on $\{ 16, 3 \}$ hyperbolic lattice, and (f) flat-band system on $\{ 8, 4\}$ hyperbolic lattice. The results show that while nontrivial $\alpha$ does not change the characteristic scaling of the density of states near half-filling or zero energy, it squeezes the entire spectrum toward zero energy, thereby increasing the density of states near it. See Sec.~\ref{sec:freefermions} for a detailed discussion. Here, $E$ is measured in units of $t$ (nearest-neighbor hopping amplitude).      
}~\label{fig:DoSFreeFermion}
\end{figure*}

Here, we show that a specific class of non-Hermitian (NH) quantum crystals offer a universal route to catalyze a particular family of spontaneous symmetry-breaking quantum phases, which we name ``commuting-class masses'' (explained shortly) at weaker interactions in comparison to their counterparts in conventional or Hermitian systems, a phenomenon we name ``non-Hermitian catalysis''. Within this framework, the free-fermion NH operator hosts guaranteed all-real eigenvalues within an extended NH parameter regime, where all the single-particle states are sharp and filling factor is well defined. In such a parameter domain, the NH catalysis mechanism, predicted from a concrete and robust algebraic criterion, is anchored in lattice-based numerical calculations within the mean-field approximation. In our explicit construction, the non-Hermiticity results from the imbalance in the hopping amplitudes of free fermions in opposite directions between any pair of nearest-neighbor (NN) sites, as schematically shown in Fig.~\ref{fig:NHLattice}. In such NH systems, as explicit examples, we showcase the NH catalysis mechanism within the framework of a minimal extended Hubbard model, containing the NN Coulomb ($V$) and on-site Hubbard ($U$) repulsions, favoring charge- and spin-density-wave orders, respectively, both belonging to the family of ``commuting-class masses''. Therefore, by tuning the degree of non-Hermiticity in quantum crystals one can trigger metal-insulator transitions at desired weak coupling. These outcomes hold on both Euclidean and hyperbolic bipartite lattices. Below we present a brief summary of our central results, starting with a brief discussion on the requisite background information.

\subsection{Background}

Crystals are formed by periodically arranging regular polygons with $p$ arms (also named $p$-gons), each vertex of which is connected to $q$ equidistant NN sites. Such an arrangement gives rise to periodic crystals on a flat Euclidean plane when 
\begin{equation}~\label{eq:euclidean}
(p-2) (q-2)=4.
\end{equation}  
The above equation yields only three solutions, namely the triangular lattice with $p=3$ and $q=6$, the square lattice with $p=q=4$, and the hexagonal or honeycomb lattice with $p=6$ and $q=3$. They can also be characterized by the Schl\"afli symbol $\{ p, q\}$. On the other hand, periodic lattices on a hyperbolic plane with a constant negative curvature are realized when  
\begin{equation}~\label{eq:hyperbolic}
(p-2) (q-2)>4.
\end{equation}  
Naturally, the number of available hyperbolic lattices is infinite, and can be identified from their Schl\"afli symbols. The diverse landscape of hyperbolic lattices has attracted much attention recently, exploring a number of unconventional phenomena therein that often solely arise from the underlying negative spatial curvature~\cite{HL:1, HL:2, HL:3, HL:4, HL:5, HL:6, HL:7, HL:8, HL:9, HL:10, HL:11, HL:12, HL:13, HL:14, HL:15, HL:16, HL:17, HL:18, HL:19}. When a tight-binding model with only NN hopping processes is considered on any $\{ p,q \}$ lattice, a bipartite lattice structure emerges in the real space hopping Hamiltonian when $p$ is an even integer, which we discuss in detail in the next section. Here, we restrict ourselves to such lattice systems embedded on Euclidean and hyperbolic spaces.    

\begin{figure*}[t!]
\includegraphics[width=1.00\linewidth]{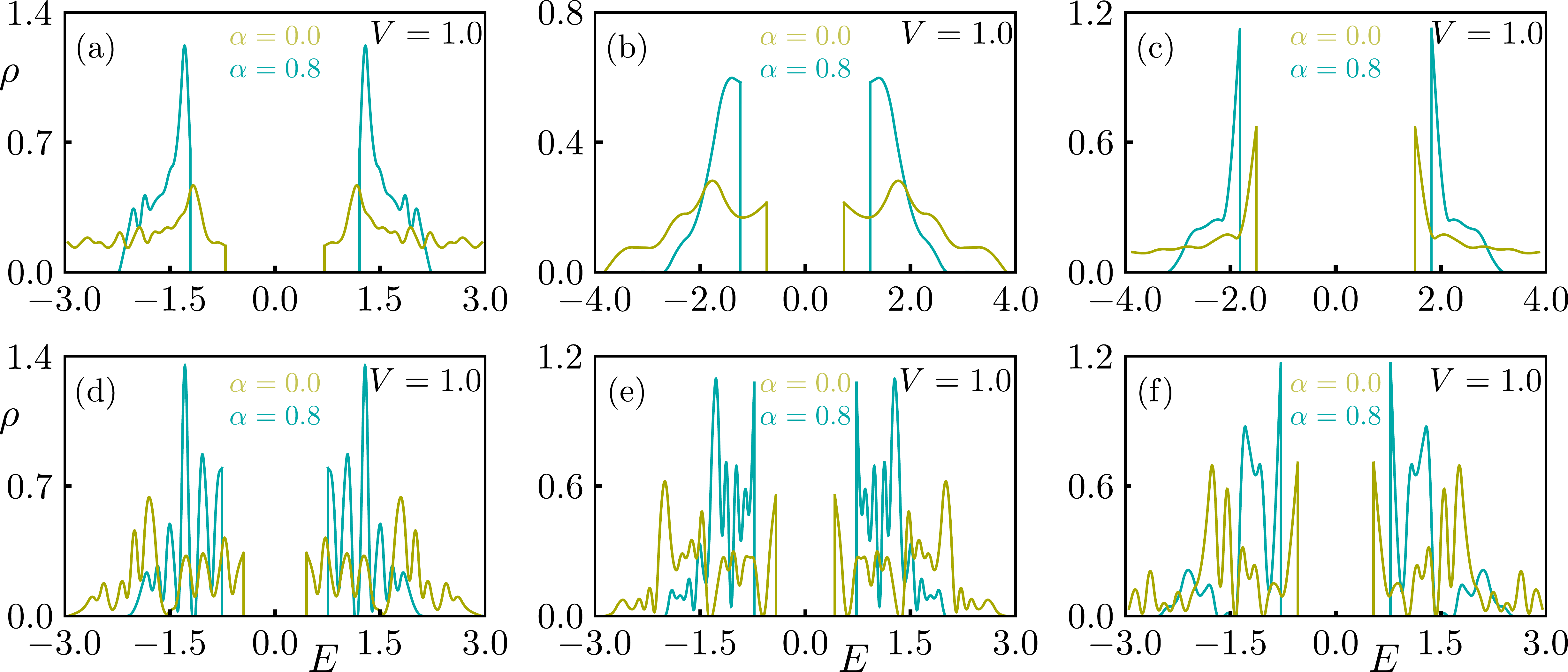}
\caption{Density of states ($\rho$) as a function of energy ($E$) in Hermitian ($\alpha=0.0$) and non-Hermitian ($\alpha=0.8$) systems for (a) Dirac liquid on an Euclidean honeycomb or $\{ 6, 3\}$ lattice, (b) Fermi liquid on an Euclidean Bernal-stacked bilayer honeycomb lattice, (c) quasiflat-band system on an Euclidean square or $\{ 4,4 \}$ lattice, (d) Dirac liquid on $\{ 10, 3 \}$ hyperbolic lattice, (e) Fermi liquid on $\{ 16, 3 \}$ hyperbolic lattice, and (f) flat-band system on $\{ 8, 4\}$ hyperbolic lattice with self-consistent solutions of the charge-density-wave order for nearest-neighbor Coulomb repulsion $V=1.0$. The chosen strength of $V$ always exceeds the critical ones in Euclidean and hyperbolic Dirac systems (Table~\ref{tab:NNCoulcric}). Results show that nontrivial $\alpha$ amplifies the gap near the zero energy, thereby endorsing the proposed non-Hermitian catalysis mechanism for spontaneous symmetry breaking in terms of charge-density-wave orders. See Secs~\ref{sec:system} and~\ref{sec:CDW} for details. Here, $E$ and $V$ are measured in units of $t$ (nearest-neighbor hopping amplitude).     
}~\label{fig:NHCatalysisCDW}
\end{figure*}

The NN tight-binding model, depending on $p$ and $q$, can facilitate emergent Dirac liquids, Fermi liquids, and flat bands near the half-filling, respectively identified by the hallmark vanishing, constant, and diverging density of states therein. Because of the bipartite nature of the underlying lattice with even $p$, all these systems can foster commensurate charge-density-wave (CDW) and spin-density-wave (SDW) orders, respectively featuring staggered patterns of average electronic density and spin between the NN sites without encountering any geometric frustration. Such orderings are favored by NN Coulomb repulsion ($V$) and on-site Hubbard repulsion ($U$), respectively, and both give rise to insulation in half-filled systems. In Dirac liquids, such orderings take place beyond critical strengths of the NN and on-site repulsions, respectively denoted by $V_c$ and $U_c$, due to the vanishing density of states near half-filling or zero energy therein. By contrast, in Fermi liquids and flat-band systems these orders appear even for infinitesimal $V$ and $U$ due to constant and diverging density of states, respectively.

\subsection{Summary of main results}

In this work, we consider a NH generalization of these scenarios in the following way. Let us denote the NN free-fermion tight-binding Hamiltonian in the real space as $\hat{h}_0$. If we can identify another operator $\hat{h}_{\rm mass}$ such that $\{ \hat{h}_0, \hat{h}_{\rm mass}\}=0$, named the mass operator or matrix, then the spectrum of $\hat{h}_0$ displays pairs of positive and negative energy eigenvalues with equal magnitudes, thereby possessing a particle-hole symmetry. Existence of such a mass matrix allows us to define an anti-Hermitian operator $\hat{h}_{\rm mass} \hat{h}_0$ that fully anticommutes with $\hat{h}_0$. And finally, in terms of such an anti-Hermitian operator, we introduce the desired NH operator~\cite{NH:1, NH:2, NH:3, NH:4, NH:5} 
\begin{equation}~\label{eq:NHgeneral}
\hat{h}_{\rm NH}= \hat{h}_0 + \alpha \; \hat{h}_{\rm mass} \; \hat{h}_0 , 
\end{equation}
where $\alpha$ is a real parameter, manifesting three key features. (1) The NH operator $\hat{h}_{\rm NH}$ continues to support an all-real eigenvalue spectrum over an extended parameter regime ($|\alpha|<1$), allowing us to define the filling accurately therein, where all the single-particle states are sharp. (2) The non-Hermiticity ($\alpha$) reduces the bandwidth of the noninteracting system by bringing all the states at positive and negative energies closer to the zero energy. Namely, the set of the eigenvalues of $\hat{h}_0$, given by $\{ E_i \} \to \sqrt{1-\alpha^2}\{ E_i \}$ in NH systems. (3) The eigenspectrum of $\hat{h}_{\rm NH}$ does not alter the characteristic scaling of the density of states near zero energy, leaving the Dirac liquid, Fermi liquid, and flat-band characterizations of noninteracting electronic fluids in the Hermitian limit unaffected even in such NH or open systems.  

\begin{figure*}[t!]
\includegraphics[width=1.00\linewidth]{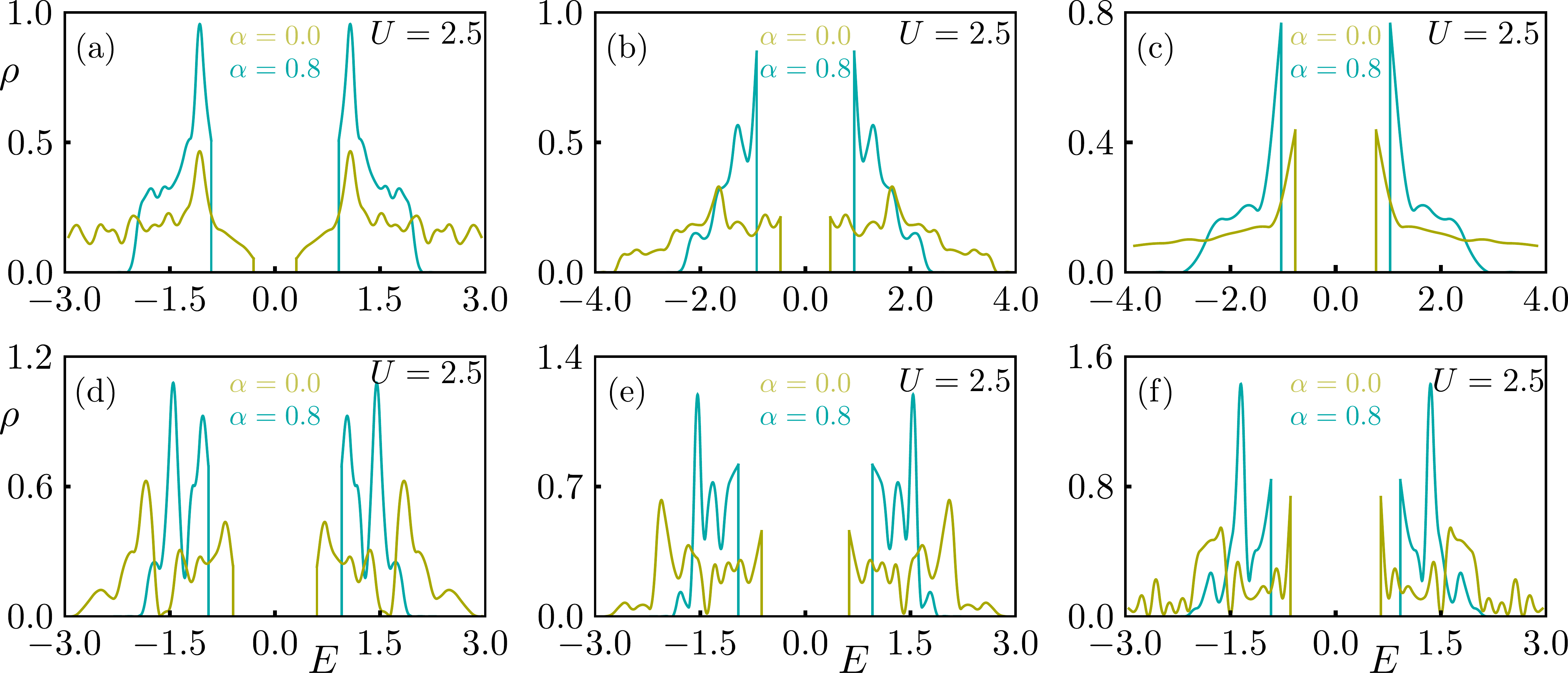}
\caption{Density of states ($\rho$) as a function of energy ($E$) in Hermitian ($\alpha=0.0$) and non-Hermitian ($\alpha=0.8$) systems for (a) Dirac liquid on an Euclidean honeycomb or $\{ 6, 3\}$ lattice, (b) Fermi liquid on a Euclidean Bernal-stacked bilayer honeycomb lattice, (c) quasiflat-band system on an Euclidean square or $\{ 4,4 \}$ lattice, (d) Dirac liquid on $\{ 10, 3 \}$ hyperbolic lattice, (e) Fermi liquid on $\{ 16, 3 \}$ hyperbolic lattice, and (f) flat-band system on $\{ 8, 4\}$ hyperbolic lattice with self-consistent solutions of the spin-density-wave order for on-site Hubbard repulsion $U=2.5$. The chosen strength of $U$ always exceeds the critical ones in Euclidean and hyperbolic Dirac systems (Table~\ref{tab:Hubbardcric}). Results show that nontrivial $\alpha$ amplifies the gap near the zero energy, thereby endorsing the proposed non-Hermitian catalysis mechanism of spontaneous symmetry breaking in terms of spin-density-wave orders. See Secs.~\ref{sec:system} and~\ref{sec:SDW} for details. Here, $E$ and $U$ are measured in units of $t$ (nearest-neighbor hopping amplitude).      
}~\label{fig:NHCatalysisSDW}
\end{figure*}

With a specific choice of $\hat{h}_{\rm mass}$, such that only its diagonal entries are nontrivial and assume uniform values with opposite signs on complementary sublattices, these outcomes are summarized in Fig.~\ref{fig:DoSFreeFermion}. In this construction, the non-Hermiticity results from an imbalance in the hopping amplitudes in the opposite directions between any pair of NN sites (characterized by $\alpha$). This construction on a few specific lattices is schematically shown in Fig.~\ref{fig:NHLattice}.

In this work, we present a concrete mathematical criterion responsible for the NH catalysis mechanism that should be operative beyond the realm of the specific systems, physical dimension, and orders considered here (discussed below). For simplicity, we consider only the orders that fully anticommute with $\hat{h}_0$, thus representing mass orders as they maximally lower the energy of the system by developing a uniform and isotropic gap near the zero energy. Such mass orders can be catalyzed in a NH system if the corresponding effective single-particle matrix operators \emph{commute} with $\hat{h}_{\rm mass}$, appearing in Eq.~\eqref{eq:NHgeneral}, as then they fully anti-commute with $\hat{h}_{\rm NH}$ as well. Together, they constitute the commuting-class mass family (because of the commutation relation with $\hat{h}_{\rm mass}$). Notice that $\hat{h}_{\rm mass}$ is a natural member of this family.

With the aforementioned specific choice of $\hat{h}_{\rm mass}$, we show that the NH parameter ($\alpha$) catalyzes the formation of the CDW and SDW orders in the following way. In Dirac systems, increasing $\alpha$ monotonically decreases $V_c$ and $U_c$, and increases the magnitude of the CDW (SDW) order for $V>V_c$ ($U>U_c$). In Fermi liquids and flat-band systems, where these two orderings develop for any finite strength of $V$ and $U$, nontrivial $\alpha$ always amplifies their magnitudes. These conclusions hold equally on Euclidean and hyperbolic bipartite lattices, which can be verified by comparing the spectral gap near zero energy in Hermitian ($\alpha=0.0$) and NH ($\alpha=0.8$) systems for fixed values of $V$ (see Fig.~\ref{fig:NHCatalysisCDW}) and $U$ (see Fig.~\ref{fig:NHCatalysisSDW}), yielding CDW and SDW orders, respectively. We arrive at these conclusions from numerical self-consistent solutions for these two orders in the real space, obtained by combining biorthogonal NH quantum mechanics and mean-field theory in the Hartree limit. It is worth mentioning that the operators describing the CDW and SDW orders commute with such a specific choice of $\hat{h}_{\rm mass}$. Therefore, altogether these findings promote the notion of \emph{non-Hermitian catalysis} of spontaneous symmetry breaking of commuting-class masses in terms of two density-wave orders on bipartite lattices in flat and curved spaces.

\begin{figure}[t!]
\includegraphics[width=1.00\linewidth]{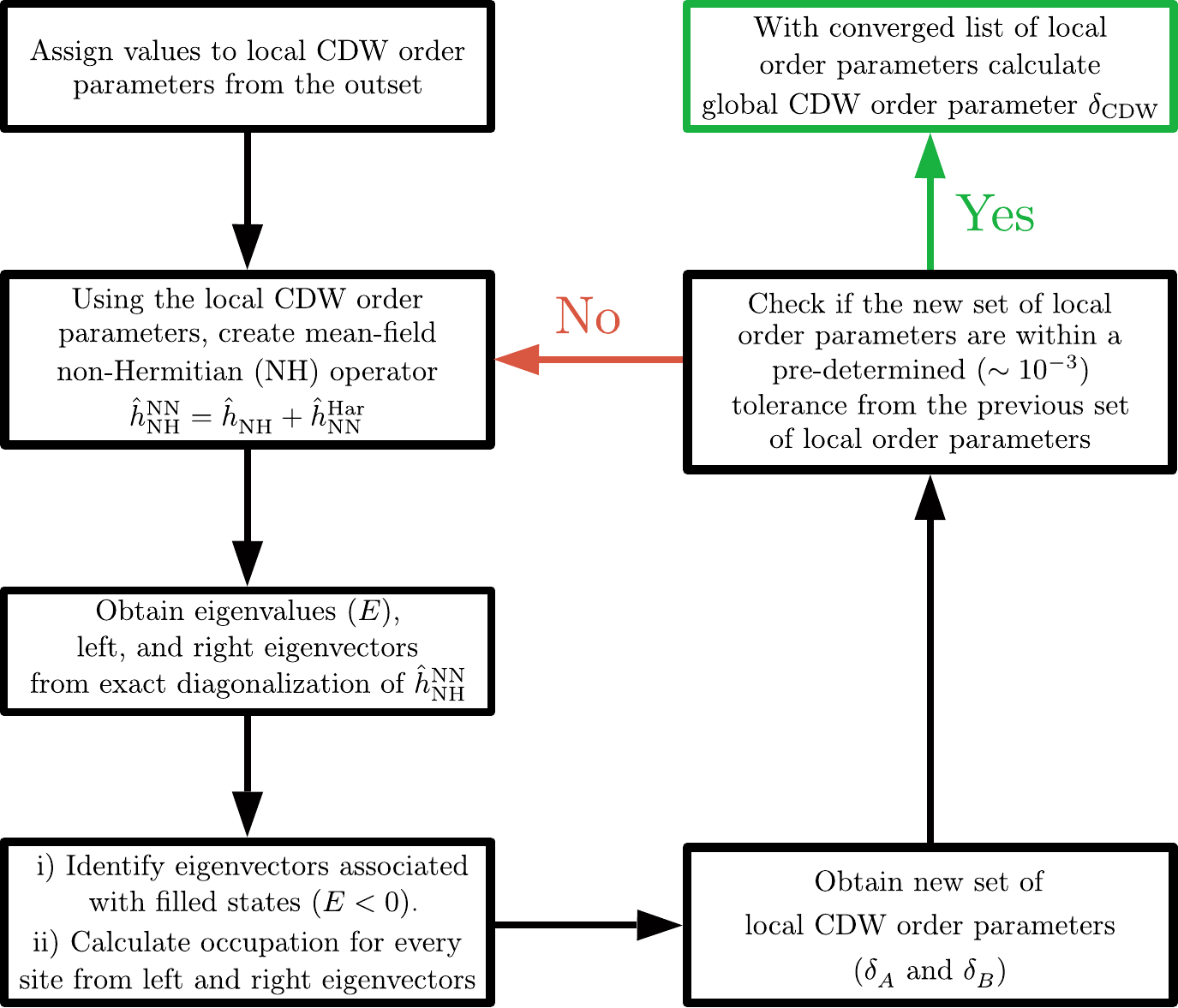}
\caption{A schematic flow chart to arrive at the self-consistent solution for the charge-density-wave order parameter ($\delta_{\rm CDW}$) in the mean-field approximation after decomposing the nearest-neighbor Coulomb repulsion ($V$) in the Hartree channel; see Sec.~\ref{sec:CDW}. A similar approach is pursued to find analogous solutions for the spin-density-wave order parameter ($\delta_{\rm SDW}$) after decomposing the on-site Hubbard repulsion ($U$) in the Hartree channel; see Sec.~\ref{sec:SDW}.           
}~\label{fig:Flowchart}
\end{figure}

\subsection{Organization}          

The rest of the paper is organized as follows. In Sec.~\ref{sec:system}, we specify the lattice configurations yielding Dirac liquids, Fermi liquids, and flat bands in noninteracting systems, their dimensions (number of lattice sites), and boundary conditions. In Sec.~\ref{sec:freefermions}, we introduce the NH tight-binding model featuring a real eigenvalue spectrum, and discuss the resulting scaling of the density of states and reduction of the bandwidth. Section~\ref{sec:NHcatalysis} is devoted to establishing a general criterion for the NH catalysis of ordered states, which we exemplify with the CDW and SDW orders by considering a specific construction of the NH operator in the noninteracting system, following the general principle outlined in Sec.~\ref{sec:freefermions}. In Sec.~\ref{sec:NHQM}, we briefly review the quintessential components of biorthogonal quantum mechanics tailored for NH quantum systems and required for the current venture. Section~\ref{sec:CDW} establishes the NH catalysis mechanism for the CDW order within the framework of the NN Coulomb repulsion, where we also discuss various scaling functions. In Sec.~\ref{sec:SDW}, we demonstrate similar outcomes for the SDW order by considering the on-site Hubbard repulsion. Finite size effects on the local gap and order parameters in hyperbolic lattices with open boundary conditions are discussed in Sec.~\ref{sec:finitesizehyperbolic}. In Sec.~\ref{sec:summary} we summarize the key findings and discuss related issues, including possible experimental platforms to test our theoretical predictions. A detailed discussion on the real-space Hamiltonian (both tight-binding and effective single-particle) on Bernal-stacked bilayer honeycomb lattice is relegated to Appendix~\ref{append:BBLG}. In Appendix~\ref{append:HypFB}, we show additional numerical results for the CDW and SDW orders in flat-band systems, realized on the $\{ 8,4 \}$ hyperbolic lattice. A competition between commuting- and anticommuting-class masses is briefly discussed in Appendix~\ref{append:competition}.


\section{System specifications}~\label{sec:system}

All our numerical results are obtained from exact numerical diagonalization of the real-space tight-binding Hamiltonian in noninteracting systems and the effective single-particle Hamiltonian in interacting systems. The latter ones are obtained after decomposing the four-fermion interaction terms, capturing the on-site Hubbard and NN Coulomb repulsions, in the Hartree channel. Therefore, it is worth mentioning the lattice systems yielding Dirac liquids, Fermi liquids, and (quasi)-flat band on Euclidean and hyperbolic planes, the system sizes (number of lattice points), and the boundary conditions we impose therein up front. 

\begin{table}[t!]
\centering
{\renewcommand{\arraystretch}{1.2}
\begin{tabular}{|P{0.5cm}|P{3.25cm}|P{3.75cm}|}
        \hline
        \multicolumn{1}{|c|}{} & \multicolumn{2}{c|}{Critical NN Coulomb repulsion $V_{c}$ for CDW order}\\
        \cline{2-3}
        \vspace{-0.22in}$\alpha$ & Honeycomb lattice & $\{10,3\}$ hyperbolic lattice\\
        \hline
        0.0 & 0.695 & 0.670\\
        0.2 & 0.680 & 0.658\\
        0.4 & 0.637 & 0.617\\
        0.6 & 0.562 & 0.541\\
        0.8 & 0.420 & 0.403\\
        \hline
\end{tabular}
}
\caption{The critical strength of the nearest-neighbor (NN) Coulomb repulsion $V_c$ for the charge-density-wave (CDW) ordering for various non-Hermitian parameter $\alpha$ in a honeycomb lattice and $\{ 10,3\}$ hyperbolic lattice, representing Dirac systems on flat and negatively curved spaces, respectively; see Fig.~\ref{fig:DiracfitCDW}. Also notice that $V_c$ is smaller in the hyperbolic Dirac system than in the Euclidean Dirac system for each $\alpha$, suggesting that the curvature-induced quantum phase transition of a mass ordering (in this case CDW) at weaker coupling remains operative even in non-Hermitian systems~\cite{HL:19}. Numerically obtained values of $V_c(\alpha)$ satisfy the scaling law $V_c(\alpha)=\sqrt{1-\alpha^2} V_c(0)$, predicted analytically in Eqs.~\eqref{eq:criticaldiscrete} and~\eqref{eq:Vccontinuum}.     
}~\label{tab:NNCoulcric}
\end{table}
 
A Euclidean Dirac (Fermi) liquid is realized on a half-filled monolayer~\cite{grpahene:Wallace} (Bernal-stacked bilayer~\cite{graphene:RMP}) honeycomb lattice, and a quasi flat band is found on a half-filled square lattice, resulting from the van Hove singularity displaying logarithmically diverging density of states close to the zero energy~\cite{GF:DOS}. On all the Euclidean lattices we always impose periodic boundary conditions in both directions. All the numerical calculations are performed on a monolayer (Bernal-stacked bilayer) honeycomb lattice containing 2400 (4800) lattice sites, and a square lattice that is composed of 3600 sites. Therefore, in the Bernal-stacked bilayer honeycomb lattice each layer contains 2400 sites and periodic boundary conditions are imposed on each layer individually.

\begin{table}
\centering
{\renewcommand{\arraystretch}{1.2}
\begin{tabular}{|P{0.5cm}|P{3.25cm}|P{3.75cm}|}
        \hline
        \multicolumn{1}{|c|}{} & \multicolumn{2}{c|}{Fitting parameter $\kappa_{_{\rm CDW}}$ for CDW order}\\
        \cline{2-3}
        \vspace{-0.14in}$\alpha$ & Bernal-stacked bilayer honeycomb lattice & \vspace{-0.14cm}$\{16,3\}$ hyperbolic lattice\\
        \hline
        0.0 & 5.00 & 1.67\\
        0.2 & 4.95 & 1.65\\
        0.4 & 4.60 & 1.57\\
        0.6 & 4.00 & 1.36\\
        0.8 & 3.10 & 1.02\\
        \hline
\end{tabular}
}
\caption{The fitting parameter $\kappa_{_{\rm CDW}}$ appearing in the BCS-like scaling form for the charge-density-wave (CDW) order for various non-Hermitian parameter $\alpha$ in Bernal-stacked bilayer honeycomb lattice and $\{16,3\}$ hyperbolic lattice, hosting Fermi liquids in the noninteracting limit on flat Euclidean and negatively curved spaces, respectively. See also Fig.~\ref{fig:FLfitCDW} for the scaling of $\kappa_{_{\rm CDW}}(\alpha)/\kappa_{_{\rm CDW}}(0)$ in these two systems. Numerical values of $\kappa_{\rm CDW}(\alpha)$ satisfy the scaling law $\kappa_{\rm CDW}(\alpha)=\sqrt{1-\alpha^2} \kappa_{\rm CDW}(0)$, predicted analytically in Eq.~\eqref{eq:BCSCDWfittinganalytical}.
}~\label{tab:CDWFLfitting}
\end{table}

Dirac liquid, Fermi liquid, and flat-band systems are also found on $\{ 10, 3\}$, $\{ 16, 3\}$, and $\{ 8, 4\}$ hyperbolic lattices, respectively~\cite{HL:19}. All the numerical calculations are performed on the third-generation $\{ 10, 3\}$, second-generation $\{ 16, 3\}$, and third-generation $\{ 8, 4\}$ hyperbolic lattices, containing 2880, 2704, and 1968 lattice sites, respectively, and with open boundary conditions. On any hyperbolic lattice, the center plaquette corresponds to the zeroth generation, and each successive layer of plaquettes constitutes its progressively next generation; see Fig.~\ref{fig:NHLattice}(b). Since the self-consistent solutions for the CDW and SDW orders are obtained on hyperbolic lattices with open boundary conditions, only for these systems we analyze their spatial variation and finite size effects.

\section{NH tight-binding model}~\label{sec:freefermions}

A tight-binding model for noninteracting fermions, allowed to hop only between the NN sites of the underlying lattice, takes the form
\begin{equation}~\label{eq:HamilTB}
H_0 = - \sum_{\langle i,j \rangle} \: \sum_{\sigma=\uparrow, \downarrow} \; t_{ij} \; c^\dagger_{i\sigma} c_{j \sigma}.
\end{equation}
Here, $c^\dagger_{i \sigma}$ ($c_{i \sigma}$) is the fermionic creation (annihilation) operator on the $i$th site with spin projection $\sigma=\uparrow, \downarrow$ and $\langle \cdots \rangle$ restricts the summation to NN sites. Throughout, we consider the NN hopping amplitude $t_{ij}$ to be spin independent and constant $t$, and set $t=1$ for convenience. Therefore, the spin degrees of freedom lead to only a mere doubling of the Hamiltonian in the noninteracting system. In any lattice, denoted by the Schl\"afli symbol $\{ p,q\}$ with an even integer $p$, the NN tight-binding model gives rise to an emergent bipartite structure in the hopping Hamiltonian, which can be appreciated by labeling the NN sites with the $A$ and $B$ sublattices; see Fig.~\ref{fig:NHLattice}. Such a labeling is of course arbitrary in the sense that under $A \leftrightarrow B$ the hopping Hamiltonian retains all its symmetry properties, thereby manifesting its Ising-like sublattice exchange symmetry. 

\begin{table}[t!]
\centering
{\renewcommand{\arraystretch}{1.2}
\begin{tabular}{|P{0.5cm}|P{3.5cm}|P{3.5cm}|}
        \hline
        \multicolumn{1}{|c|}{} & \multicolumn{2}{c|}{Fitting parameter on \{4, 4\} or square lattice}\\
        \cline{2-3}
        \vspace{-0.2in}$\alpha$ & \vspace{-0.27cm} For CDW order ($\eta_{\rm CDW}^{}$) & \vspace{-0.27cm} For SDW order ($\eta_{\rm SDW}^{}$)\\
        \hline
        0.0 & 1.50 & 3.20\\
        0.2 & 1.47 & 3.10\\
        0.4 & 1.45 & 3.10\\
        0.6 & 1.35 & 2.80\\
        0.8 & 1.15 & 2.45\\
        \hline
\end{tabular}
}
\caption{The fitting parameters for the scaling of the charge-density-wave ($\eta_{\rm CDW}$) and spin-density-wave ($\eta_{\rm SDW}$) orders for various values of the non-Hermitian parameter $\alpha$ in a \{4, 4\} or square lattice with periodic boundary conditions. This system describes a quasiflat-band system, with a logarithmically diverging density of states near zero energy. These parameters are obtained by the fit curves displayed in Figs.~\ref{fig:QFBfitCDW} and~\ref{fig:QFBfitSDW} for charge- and spin-density-wave, respectively. Numerically obtained values of $\eta_{\rm CDW}(\alpha)$ and $\eta_{\rm SDW}(\alpha)$ satisfy the scaling laws $\eta_{\rm CDW}(\alpha)=\left( 1-\alpha^2 \right)^{1/4} \eta_{\rm CDW}(0)$ and $\eta_{\rm SDW}(\alpha)=\left( 1-\alpha^2 \right)^{1/4} \eta_{\rm SDW}(0)$, respectively, predicted analytically in Eqs.~\eqref{eq:futtingSLCDW} and~\eqref{eq:futtingSLSDW}.
}~\label{tab:QFBfitting}
\end{table}

Next, we define a $2N$-component spinor $\Psi^\top=(\Psi_\uparrow, \Psi_\downarrow)$, where $N$ is the total number of sites in the system and $\top$ denotes the transposition. The $N$-dimensional spinor for each spin projection $\sigma$ takes the form $\Psi^\top_\sigma= (c_{A\sigma}, c_{B\sigma})$, where $c_{A\sigma}$ and $c_{B\sigma}$ are $N/2$-dimensional spinors constituted by the annihilation operators on the sites belonging to the $A$ and $B$ sublattices, respectively, with spin projection $\sigma$. In this basis, the Hamiltonian operator associated with $H_0$ from Eq.~\eqref{eq:HamilTB} takes the form $\hat{h}_0=\sigma_0 \otimes \hat{h}^{\rm spinless}_0$. Here the set of Pauli matrices $\{ \sigma_\mu \}$ for $\mu=0,1,2,3$ operates on the spin index and 
\begin{equation}
\hat{h}^{\rm spinless}_0 = \left(\begin{array}{cc}
{\boldsymbol 0} & {\bf t} \\
{\bf t}^\top & {\boldsymbol 0}
\end{array}
\right)   
\end{equation}
captures the hopping Hamiltonian for spinless fermions, which we consider in Sec.~\ref{sec:CDW}. Here ${\boldsymbol 0}$ is a $N/2$-dimensional null matrix and ${\bf t}$ is a real $N/2$-dimensional intersublattice hopping matrix.

\begin{figure*}[t!]
\includegraphics[width=1.00\linewidth]{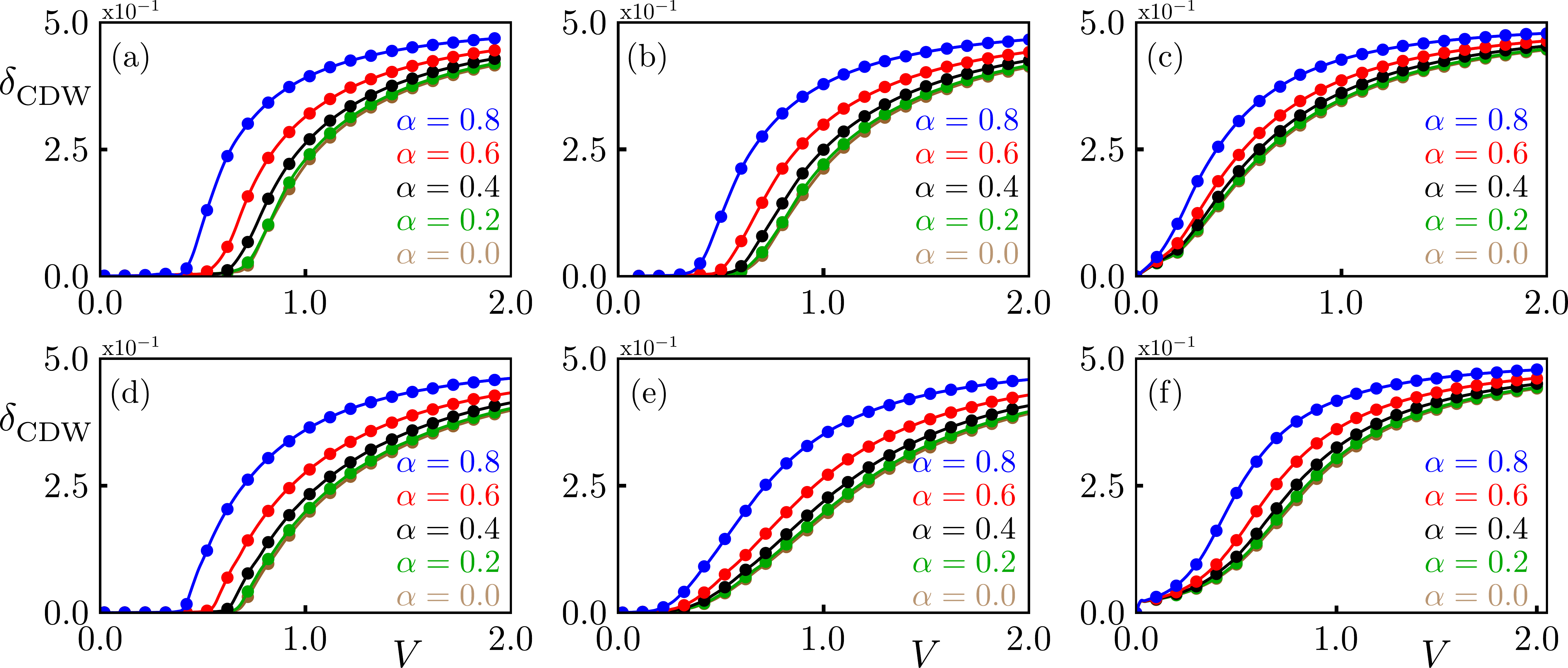}
\caption{Scaling of the self-consistent solutions of the charge-density-wave (CDW) order ($\delta_{\rm CDW}$) on (a) Euclidean Dirac system (honeycomb or $\{ 6, 3\}$ lattice), (b) Euclidean Fermi liquid system (Bernal-stacked bilayer honeycomb lattice), (c) Euclidean flat-band system (square or $\{ 4,4 \}$ lattice), (d) hyperbolic Dirac system ($\{ 10, 3 \}$ lattice), (e) hyperbolic Fermi liquid system ($\{ 16, 3 \}$ lattice), and (f) hyperbolic flat-band system ($\{ 8, 4\}$ lattice) as functions of the strength of the nearest-neighbor Coulomb repulsion ($V$) with varying non-Hermiticity $\alpha$, including the Hermitian system ($\alpha=0.0$). In Dirac systems, increasing $\alpha$ reduces the requisite critical strength of the nearest-neighbor Coulomb repulsion for the CDW ordering; see also Fig.~\ref{fig:DiracfitCDW}. In Fermi liquid and flat-band systems, where such an ordering develops for even infinitesimal $V$, increasing $\alpha$ increases the magnitude of $\delta_{\rm CDW}$ for any fixed $V$. Altogether, these results show that nontrivial $\alpha$ catalyzes the formation of the CDW order (a member of commuting-class mass family) in the proposed non-Hermitian bipartite lattices (Sec.~\ref{sec:freefermions}). See Sec.~\ref{sec:CDW} for details. Here, $V$ is measured in units of $t$ (nearest-neighbor hopping amplitude).      
}~\label{fig:CDWScalingNH}
\end{figure*}

Our proposed NH generalization of the hopping Hamiltonian, resulting from the imbalance in the hopping amplitudes in the opposite directions between any pair of NN sites, rests on the identification of an operator $\hat{h}_{\rm mass} \equiv \sigma_0 \otimes \hat{h}_{\rm CDW}$ such that $\hat{h}_{\rm CDW}$ anticommutes with $\hat{h}^{\rm spinless}_0$ and it is given by 
\begin{equation}~\label{eq:CDWspinlessmatrix}
\hat{h}_{\rm CDW}= \left(\begin{array}{cc}
{\bf I}_{N/2} & {\boldsymbol 0} \\
{\boldsymbol 0} & -{\bf I}_{N/2}
\end{array}
\right).   
\end{equation}
Here, ${\bf I}_{N/2}$ is a $N/2$-dimensional identity matrix. We discuss the physical meaning of this operator shortly. Then, in terms of an anti-Hermitian operator $\hat{h}_{\rm CDW} \hat{h}^{\rm spinless}_0$, we introduce the desired NH hopping operator following the general principle of construction from Eq.~\eqref{eq:NHgeneral}
\allowdisplaybreaks[4] 
\begin{eqnarray}~\label{eq:NHHamil}
\hat{h}_{\rm NH} &=& \sigma_0 \otimes \; \left[ \hat{h}^{\rm spinless}_0 + \alpha  \hat{h}_{\rm CDW} \hat{h}^{\rm spinless}_0  \right] \nonumber \\
& = & \sigma_0 \otimes 
\left(\begin{array}{cc}
{\boldsymbol 0} & (1+ \alpha) \; {\bf t} \\
(1- \alpha) \; {\bf t}^\top & {\boldsymbol 0}
\end{array}
\right),
\end{eqnarray}
where $\alpha$ is a real parameter. Therefore, the above NH hopping operator corresponds to hopping amplitudes $1+\alpha$ and $1-\alpha$ between any pair of NN sites in opposite directions, as schematically shown in Fig.~\ref{fig:NHLattice}. By virtue of the anticommutation relation $\{ \hat{h}^{\rm spinless}_0 , \hat{h}_{\rm CDW} \}=0$, it can be shown that if the eigenvalues of $\hat{h}^{\rm spinless}_0$ constitute a set $\{ E_i \}$, then the eigenspectrum for the NH operator $\hat{h}_{\rm NH}$ is given by $\sqrt{1-\alpha^2} \{ E_i \}$ for any $i=1, \cdots, N$. Therefore, the entire spectrum gets squeezed toward the zero energy owing to the nontrivial non-Hermiticity in the system, parameterized by $\alpha$. Concomitantly any finite $\alpha$ reduces the bandwidth of the system. But, nontrivial $\alpha$ does not alter the scaling of the density of states near the zero energy. We also note that all the eigenvalues of $\hat{h}_{\rm NH}$ are purely real when $|\alpha|<1$, but are purely imaginary when $|\alpha|>1$. Throughout, we restrict ourselves within the former parameter regime. For $\alpha=\pm 1$, all the eigenvalues of $\hat{h}_{\rm NH}$ are zero, yielding an exceptional point flat band. This is a singular point, which is not considered in this work. Next, we numerically verify these outcomes for the NH free-fermion tight-binding model. Notice that each eigenvalue is two fold degenerate due to the spin degrees of freedom. It should be noted that even though the bandwidth of all the systems can be reduced by decreasing the strength of the nearest-neighbor-hopping amplitude ($t$) in a Hermitian setup, it always keeps the amplitude of two-point correlation functions between the NN sites in the opposite directions equal to ensure the Hermiticity of the systems. By contrast, the bandwidth reduction in our NH setup is accomplished by making the magnitude of the two-point correlation functions between the NN sites in the opposite directions different.

\subsection{Numerical verification}~\label{subsec:TBnumerics}

 As representative Dirac systems, here we consider the honeycomb or $\{6,3 \}$ lattice in the Euclidean space and the $\{ 10, 3\}$ lattice in the hyperbolic space. Fermi liquids in these two spaces are represented by the Bernal-stacked bilayer honeycomb lattice and the $\{ 16, 3 \}$ hyperbolic lattice, respectively. Although there exists no bipartite Euclidean lattice that features perfect flat bands at zero energy resulting from the NN hopping Hamiltonian, on a square or $\{ 4,4 \}$ lattice it produces a quasi-flat band showing logarithmically diverging density of states as energy $|E| \to 0$ (van Hove singularity), which we consider here a representative of the flat-band system on the flat Euclidean plane. On a negatively curved hyperbolic plane, the $\{ 8, 4 \}$ lattice yields a flat band at half-filling.

\begin{figure}[t!]
\includegraphics[width=1.00\linewidth]{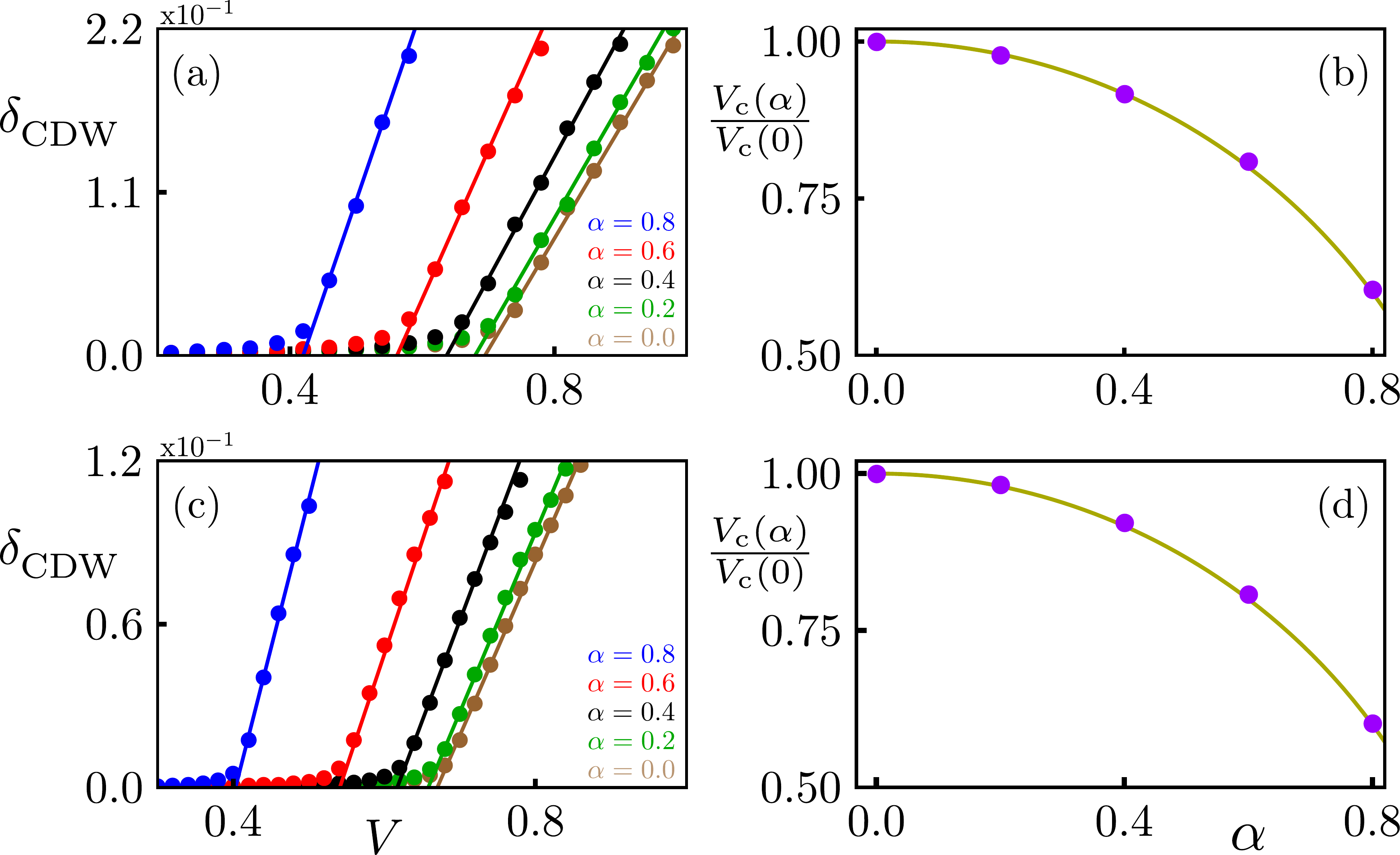}
\caption{(a) Computation of the critical strength of the nearest-neighbor Coulomb repulsion $V_c (\alpha)$ as a function of the non-Hermitian parameter $\alpha$ on a Euclidean Dirac system of honeycomb or $\{ 6,3 \}$ lattice. We obtain $V_c (\alpha)$ from the linear fit of the self-consistent solution of the charge-density-wave order parameter $\delta_{\rm CDW}$ as a function of $V$ when it appreciably rises from the trivial value. At $V=V_c(\alpha)$, such a linear fit crosses the $\delta_{\rm CDW}=0$ axis. (b) The scaling of the ratio $V_c(\alpha)/V_c(0)$ as a function of $\alpha$, where the isolated dots correspond to the numerically computed values and the solid line corresponds to the analytical prediction of $V_c(\alpha)/V_c(0) = \sqrt{1-\alpha^2}$. Subfigures (c) and (d) are analogues of (a) and (b), respectively, but for the hyperbolic Dirac system on the $\{ 10, 3\}$ lattice. Explicit values of $V_c (\alpha)$ for these two systems are shown in Table~\ref{tab:NNCoulcric}. Here, $V$ is measured in units of $t$ (nearest-neighbor hopping amplitude).             
}~\label{fig:DiracfitCDW}
\end{figure}

A brief comment on the Bernal-stacked bilayer honeycomb lattice is due at this stage. It is constituted by two layers of the honeycomb lattice, stacked in such a way that the sites belonging to the $B$ sublattice on the layer 1 reside right beneath  the sites belonging to the sublattice $A$ living on the layer 2. There exists a direct or dimer hopping ($t_\perp$) between the sites from these two sublattices on complementary layers; the amplitude of which is also set to be unity ($t_\perp=1$). Note that the eigenmodes localized on these sites are at energy $t_\perp$ or higher and at energy $-t_\perp$ or lower. Therefore, the modes near the zero energy predominantly reside on the sites belonging to sublattice $A$ on layer 1 and sublattice $B$ on layer 2. Hence, the low-energy theory at half-filling is effectively constituted by a bipartite lattice~\cite{graphene:RMP}. However, the general formalism we are discussing here remains equally operative when all the sites belonging to the $A$ and $B$ sublattices, living on layer 1 and layer 2, are taken into account. Additional discussion on this system is relegated to Appendix~\ref{append:BBLG}.

Upon numerically diagonalizing the tight-binding hopping Hamiltonian for zero and finite $\alpha$, we extract the resulting density of states ($\rho$) by counting the number of states within various energy windows, yielding $\rho$ as a function of $E$. These results are shown in Fig.~\ref{fig:DoSFreeFermion}. Indeed, we find that nontrivial $\alpha$ does not change the Dirac liquid, Fermi liquid, and flat-band nature of the systems, but squeezes the entire spectrum closer to the zero energy. A comment on the scaling of the density of states near the zero energy in flat-band systems is due at this state. Notice that the NN tight-binding model on a square lattice only supports logarithmically diverging density of states as $E \to 0$, which strictly does not correspond to a perfect flat band. Thus the large density of states near the zero energy increases with increasing $\alpha$ in this system; see Fig.~\ref{fig:DoSFreeFermion}(c). By contrast, the $\{ 8,4 \}$ hyperbolic lattice features a perfect flat band at $E=0$. Therefore, the height of the corresponding spike in the density of states near $E=0$ is almost insensitive to $\alpha$; see Fig.~\ref{fig:DoSFreeFermion}(f). Next, we proceed to discuss the impact of $\alpha$ on various ordered states, leading to the notion of NH catalysis of spontaneous symmetry breaking.

\section{Non-Hermitian catalysis}~\label{sec:NHcatalysis}

Any ordered state can be represented by a Hermitian matrix operator $\sigma_\mu \otimes \hat{h}_{\rm order}$. In principle, lattice-based quantum systems can foster various ordered states, and each of them breaks distinct discrete and/or continuous symmetries of the noninteracting system. Here, we focus on such orders that are energetically most favored. For simplicity, we neglect any competition among various ordered states. At and near zero temperature, the ordered states that open uniform and isotropic mass gaps near zero energy in the half-filled system are thus most favored as there is no competition with the entropy.

\begin{figure}[t!]
\includegraphics[width=1.00\linewidth]{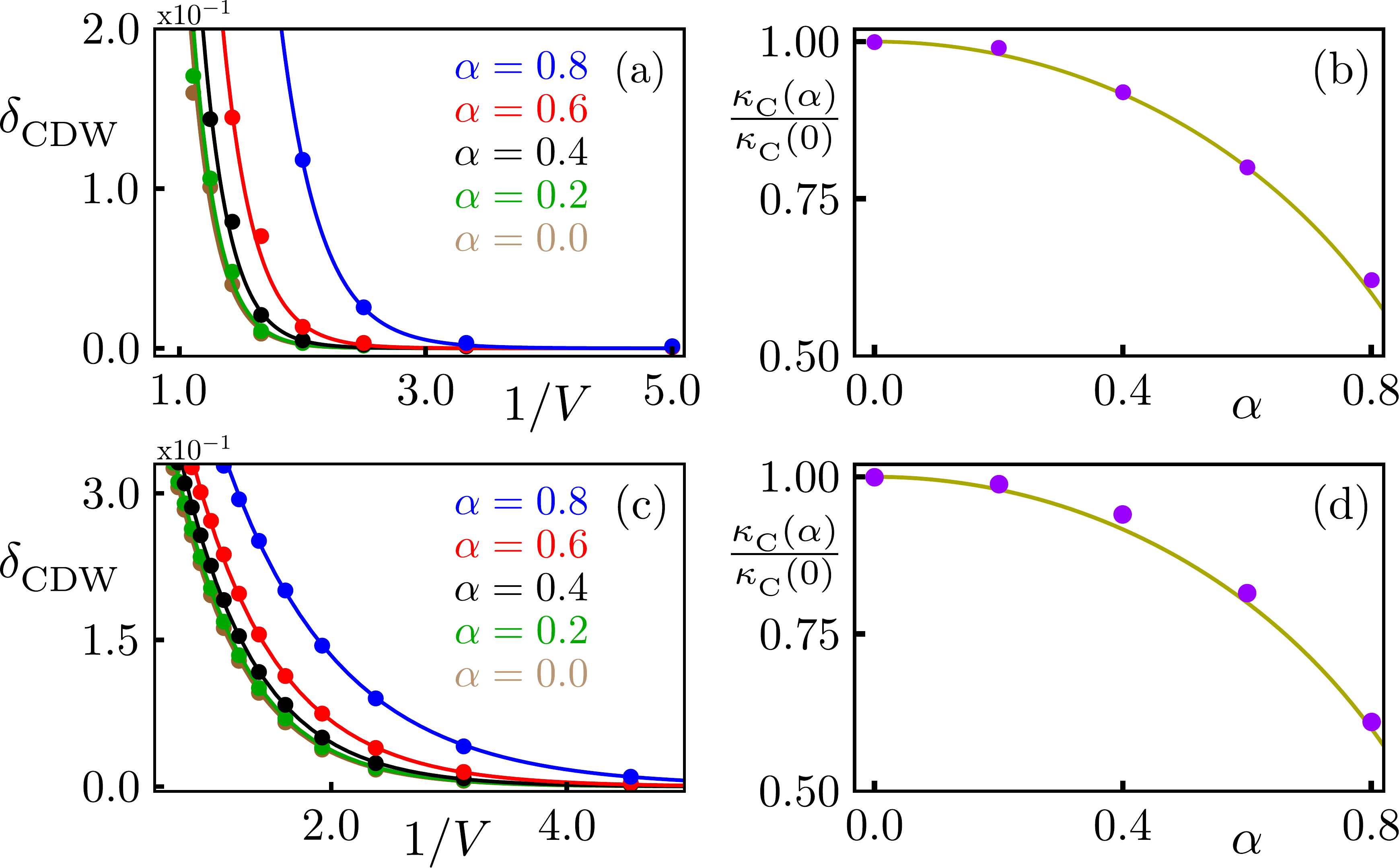}
\caption{BCS-like scaling for the charge-density-wave order parameter $\delta_{\rm CDW}$ with $1/V$ in (a) a Euclidean Fermi liquid system (Bernal-stacked bilayer honeycomb lattice) and (c) a hyperbolic Fermi liquid system ($\{ 16, 3\}$ lattice). Here, we use the fitting function $\delta_{\rm CDW} = A \exp (-\kappa_{_{\rm CDW}}(\alpha)/V)$, where $A$ and $\kappa_{_{\rm CDW}}(\alpha)$ are $\alpha$- and system-dependent fitting parameters, and $V$ denotes the strength of the nearest-neighbor Coulomb repulsion. The scaling of the ratio $\kappa_{_{\rm CDW}}(\alpha)/\kappa_{_{\rm CDW}}(0)$ as a function of $\alpha$ in these two systems are shown in (b) and (d), respectively, where the isolated dots correspond to the numerically computed values and the solid lines correspond to the analytical prediction of $\kappa_{_{\rm CDW}}(\alpha)/\kappa_{_{\rm CDW}}(0)=\sqrt{1-\alpha^2}$. Values of $\kappa_{_{\rm CDW}}(\alpha)$ for these two systems are listed in Table~\ref{tab:CDWFLfitting}. Here, $V$ is measured in units of $t$ (nearest-neighbor hopping amplitude).
}~\label{fig:FLfitCDW}
\end{figure}

We first consider Hermitian systems. Then, the matrix operators representing such mass orders, fully anticommute with $\hat{h}_0$. In the announced $2N$-component spinor basis, we immediately identify \emph{two} such matrix operators, representing mass orders, that are given by 
\allowdisplaybreaks[4]
\begin{eqnarray}~\label{eq:massoperators}
{\mathcal O}_{\rm CDW}=\sigma_0 \otimes \left( 
\begin{array}{cc}
{\boldsymbol \Delta} & {\boldsymbol 0} \\
{\boldsymbol 0} & -{\boldsymbol \Delta}
\end{array}
\right) \nonumber \\
\text{and} \: {\mathcal O}_{\rm SDW}= \sigma_j \otimes \left( 
\begin{array}{cc}
{\boldsymbol \Delta} & {\boldsymbol 0} \\
{\boldsymbol 0} & -{\boldsymbol \Delta}
\end{array}
\right),
\end{eqnarray} 
where $j=1,2,3$, and ${\boldsymbol \Delta}$ is a $N/2$-dimensional diagonal matrix, whose entries can be identical or different, as discussed in Secs.~\ref{sec:CDW} and~\ref{sec:SDW}. Irrespective of these details, we find that $\{ \hat{h}_0, {\mathcal O}_{\rm CDW} \} = \{ \hat{h}_0, {\mathcal O}_{\rm SDW} \} =0$. Physically, ${\mathcal O}_{\rm CDW}$ represents a CDW order on a bipartite lattice, resulting from the staggered pattern of the average electronic density between the neighboring sites belonging to sublattices $A$ and $B$. By contrast, ${\mathcal O}_{\rm SDW}$ corresponds to a staggered pattern of electronic spin between any pair of NN sites. This phase is also known as antiferromagnet. Both orders break the Ising-like sublattice exchange symmetry, while the SDW order in addition breaks the SU(2) spin rotational symmetry, generated by ${\boldsymbol \sigma} \otimes {\bf I}_N$. 

By virtue of the above anticommutation relations, whenever one of these two operators acquires a vacuum expectation value, the spectrum becomes fully gapped near the zero energy, yielding insulating behavior at half-filling. It can be seen by considering a simpler situation when all the entries of ${\boldsymbol \Delta}$ are identical and equal to $\Delta$. Then, ${\boldsymbol \Delta}$ becomes proportional to a $N/2$-dimensional identity matrix, with ${\mathcal O}_{\rm CDW}=\Delta \sigma_0 \otimes \hat{h}_{\rm CDW}$ and ${\mathcal O}_{\rm SDW}=\Delta \sigma_j \otimes \hat{h}_{\rm CDW}$. Then any eigenvalue of $\hat{h}_0$, given by ${\rm sgn}(E_i) |E_i|$ in the absence of any ordering, goes to ${\rm sgn}(E_i) \sqrt{E^2_i +\Delta^2}$ for any $i$. This outcome can be verified by diagonalizing the effective single-particle Hamiltonian in the presence of these orderings, given by $\hat{h}_0 + {\mathcal O}_{j}$ with $j={\rm CDW}$ or ${\rm SDW}$. Therefore, the spectrum becomes fully gapped near the zero energy with the onset of CDW or SDW order, and the system becomes an insulator at half-filling.

Notice that at half-filling all the states at negative (positive) energies are completely filled (empty). Therefore, the ground state energy in the noninteracting system is given by $-2\sum^{\prime}_{i} |E_i|$, while in the presence of any mass ordering it is $-2\sum^{\prime}_{i} [E^2_i+\Delta^2]^{1/2}$. The summation is performed over all the filled states, denoted by the prime symbol, and the factor of 2 accounts for the twofold degeneracy of each state. Therefore, formation of mass orders yields maximal gain in condensation energy, and their nucleation is energetically most favored at and close to the zero temperature.         

\begin{figure}[t!]
\includegraphics[width=1.00\linewidth]{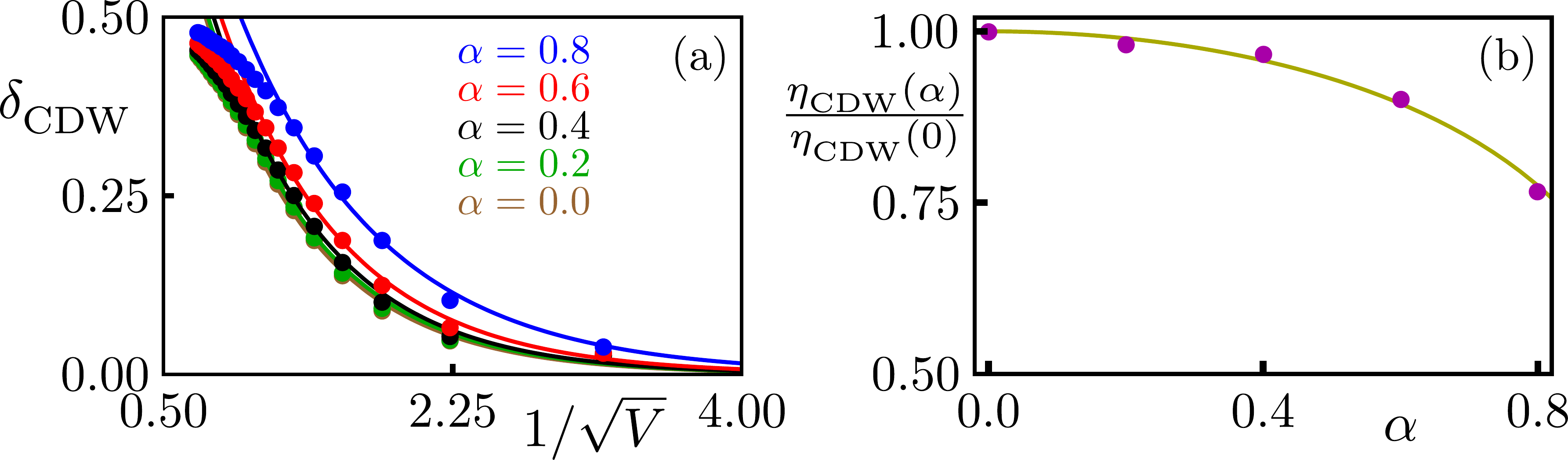}
\caption{(a) Fitting of the charge-density-wave order parameter $\delta_{\rm CDW}$ as a function of $1/\sqrt{V}$ [Eq.~\eqref{eq:CDWFitsquareLat}] for the Euclidean \{4, 4\} or square lattice with periodic boundary conditions, and with various values of the non-Hermitian parameter $\alpha$. The fitting function reads as $\delta_{\rm CDW} = A \exp(-\eta_{\rm CDW} / \sqrt{V})$ where $A$ and $\eta_{\rm CDW}$ are system-specific ($\alpha$-dependent) fitting parameters and $V$ is the coupling strength of the nearest-neighbor Coulomb repulsion interaction. (b) The scaling of the ratio $\eta_{\rm CDW}(\alpha) / \eta_{\rm CDW}(0)$ as a function of the non-Hermitian parameter $\alpha$. The numerically obtained results are represented by the magenta points, while the yellow curve is the expected theoretical scaling relation given by $\eta_{CDW}(\alpha) / \eta_{\rm CDW}(0) = (1 - \alpha^2)^{1/4}$ as shown in Eq.~\eqref{eq:futtingSLCDW}. See Table~\ref{tab:QFBfitting} for the values of $\eta_{\rm CDW}$ for various $\alpha$. Here, $V$ is measured in units of $t$ (nearest-neighbor hopping amplitude).   
}~\label{fig:QFBfitCDW}
\end{figure}

\begin{figure*}[t!]
\includegraphics[width=1.00\linewidth]{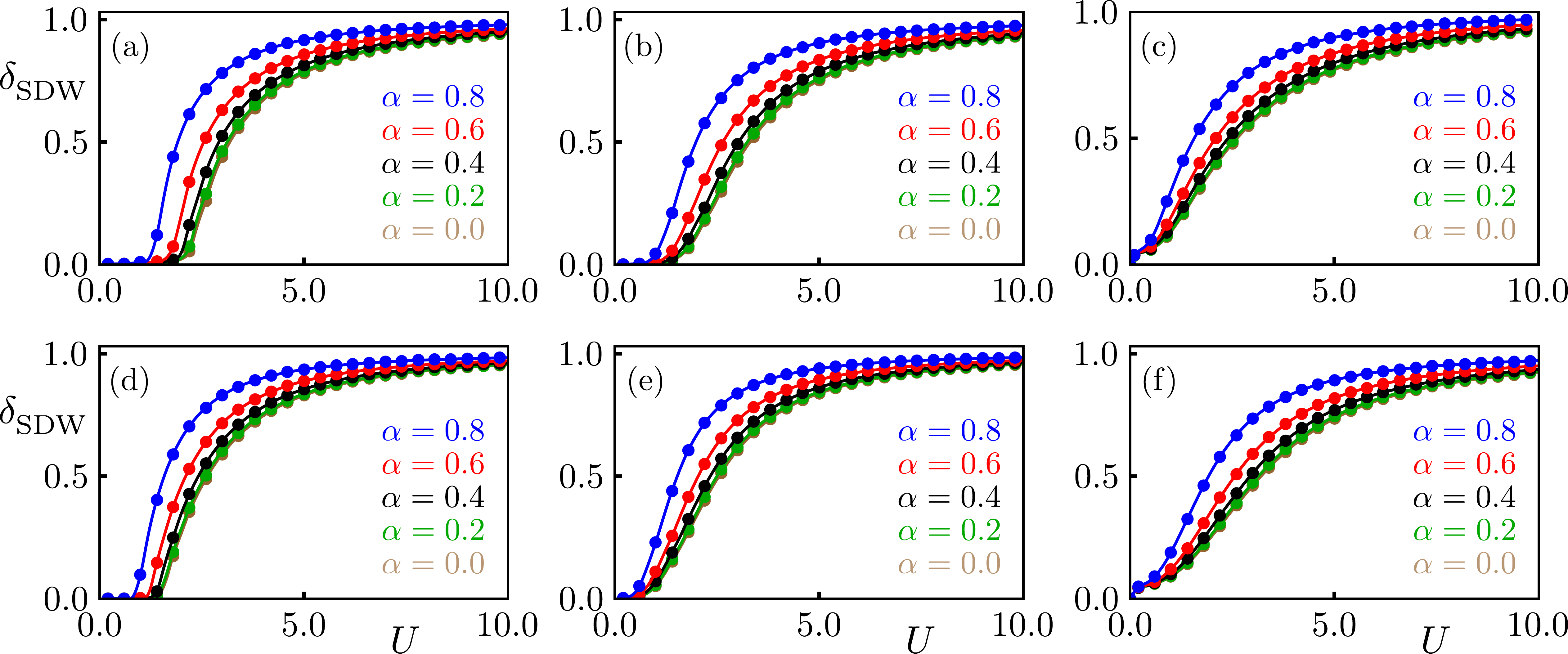}
\caption{Scaling of the self-consistent solutions of the spin-density-wave (SDW) order ($\delta_{\rm SDW}$) on (a) Euclidean Dirac system (honeycomb or $\{ 6, 3\}$ lattice), (b) Euclidean Fermi liquid system (Bernal-stacked bilayer honeycomb lattice), (c) Euclidean flat-band system (square or $\{ 4,4 \}$ lattice), (d) hyperbolic Dirac system ($\{ 10, 3 \}$ lattice), (e) hyperbolic Fermi liquid system ($\{ 16, 3 \}$ lattice), and (f) hyperbolic flat-band system ($\{ 8, 4\}$ lattice) as functions of the strength of the on-site Hubbard repulsion ($U$) with varying non-Hermiticity $\alpha$, including the Hermitian system ($\alpha=0.0$). In Dirac systems, increasing $\alpha$ reduces the requisite critical strength of the on-site Hubbard repulsion for the SDW ordering; see also Fig.~\ref{fig:DiracfitSDW}. In Fermi liquid and flat-band systems, where such an ordering develops for even infinitesimal $U$, increasing $\alpha$ increases the magnitude of $\delta_{\rm SDW}$ for any fixed $U$. Altogether, these results show that nontrivial $\alpha$ catalyzes the formation of the SDW order (a member of the commuting-class mass family) in the proposed non-Hermitian bipartite lattices (Sec.~\ref{sec:freefermions}). See Sec.~\ref{sec:SDW} for details. Here, $U$ is measured in units of $t$ (nearest-neighbor hopping amplitude).     
}~\label{fig:SDWScalingNH}
\end{figure*}

The stage is now set to promote the notion of the NH catalysis of these two density-wave orders, which we subsequently generalize to showcase a robust mathematical criterion for this mechanism to be operative. Once again, we restrict the discussion to mass orders as they are energetically most favored. Recall that the hopping Hamiltonian for the NH system is given by $\hat{h}_{\rm NH}$ in Eq.~\eqref{eq:NHHamil}. Any mass order for which the corresponding operator \emph{commutes} with $\hat{h}_{\rm mass} \equiv \sigma_0 \otimes \hat{h}_{\rm CDW}$ (entering the anti-Hermitian part of $\hat{h}_{\rm NH}$), anticommutes with $\hat{h}_{\rm NH}$, which we name commuting-class mass. In the presence of such mass ordering the eigenvalues of the total effective single-particle operator $\hat{h}_{\rm NH} + {\mathcal O}_{j}$ with $j={\rm CDW}$ or ${\rm SDW}$ constitute a set $\{ {\rm sgn}(E_i) [(1-\alpha^2) E^2_i + \Delta^2]^{1/2}\}$. Therefore, the ground-state energy of the half-filled NH system at zero temperature and in the presence of commuting-class mass is 
\begin{equation}
E_{\rm ground}= -2 \sideset{}{^\prime} \sum_i [(1-\alpha^2) E^2_i + \Delta^2]^{1/2},
\end{equation}
where the summation is restricted over the filled states at negative energies.

In order to recognize the NH catalysis mechanism, we consider the free energy per site at zero temperature
\begin{equation}
F= \frac{\Delta^2}{2 g} - \frac{2}{N} \sideset{}{^\prime} \sum_i [(1-\alpha^2) E^2_i + \Delta^2]^{1/2},
\end{equation} 
where the strength of the four-fermion interaction $g$ is proportional to $V$ and $U$ for the CDW and SDW orders, respectively, and $N$ is the number of lattice sites in the system. The above free-energy density ($F$) is obtained by first performing a Hubbard-Stratonovich decoupling of the four-fermion interaction terms $g \left( \Psi^\dagger {\mathcal O}_{\rm CDW} \Psi \right)^2$ and $g \left( \Psi^\dagger {\mathcal O}_{\rm SDW} \Psi \right)^2$ favoring CDW and SDW orders, respectively, in terms of the corresponding order parameters $\Delta=\Psi^\dagger {\mathcal O}_{\rm CDW} \Psi$ and $\Psi^\dagger {\mathcal O}_{\rm SDW} \Psi$. Subsequently, we integrate out the fermionic degrees of freedom from the effective quadratic action in the path integral formalism~\cite{neegleorland}. Minimizing $F$ with respect to $\Delta$, we arrive at the self-consistent gap equation. Besides the trivial solution $\Delta=0$, it also permits a nontrivial solution of $\Delta$ that can be obtained from \begin{equation}~\label{eq:gapgeneral}
\frac{1}{2g} = \frac{2}{N} \sideset{}{^\prime} \sum_i \frac{1}{[(1-\alpha^2) E^2_i + \Delta^2]^{1/2}}.
\end{equation}
Note that for a fixed desired magnitude of $\Delta$, the right-hand side of the above gap equation increases monotonically with increasing $\alpha$. Thus, for the same nontrivial solution of $\Delta$, the requisite strength of the four-fermion interaction $g$ decreases monotonically with increasing $\alpha$. Alternatively, if we fixed the strength of $g$ from the outset, the self-consistent solution of nontrivial $\Delta$ is amplified with increasing $\alpha$. Therefore, increasing non-Hermiticity in the system boosts the propensity toward the nucleation of CDW and SDW orders, both belonging to the commuting-class mass family. We name this phenomenon non-Hermitian catalysis of spontaneous symmetry breaking, operative on commuting-class masses.

Note that the NH catalysis phenomenon occurs for any finite strength of $V$ and $U$ in Fermi liquids and flat-band systems, where the CDW and SDW orderings develop even for infinitesimal strengths of the NN and Hubbard repulsions, respectively. In Dirac liquids, on the other hand, the NH catalysis for these two density-wave orders shows up only when they acquire nontrivial vacuum expectation values for $V>V_c$ and $U>U_c$, respectively. Nonetheless, there exists a complementary approach to appreciate the NH catalysis mechanism in Dirac systems from the scaling of the critical strength of $g$ as a function of $\alpha$, denoted by $g_{_c} (\alpha)$, for which we set $\Delta=0$ in Eq.~\eqref{eq:gapgeneral}, leading to 
\begin{equation}
\frac{1}{g_{_c}(\alpha)}= \frac{4}{N} \sideset{}{^\prime} \sum_i \frac{1}{\sqrt{1-\alpha^2} |E_i|} = \frac{1}{\sqrt{1-\alpha^2}} \frac{1}{g_{_c}(0)}.
\end{equation}   
Therefore, for commuting class masses in Dirac systems
\begin{equation}~\label{eq:criticaldiscrete}
g_{_c}(\alpha) = \sqrt{1-\alpha^2} \; g_{_c}(0) 
\: \Rightarrow \:
\frac{g_{_c}(\alpha)}{g_{_c}(0)} = \sqrt{1-\alpha^2}, 
\end{equation}
showing that $g_{_c}(\alpha) < g_{_c}(0)$, thereby promoting the NH catalysis mechanism. It should be noted that $g_{_c}(\alpha) \to 0$ for any $\alpha$ (including $\alpha=0$ corresponding to the Hermitian limit) in Fermi liquids and flat-band systems, as such systems always host a finite (possibly two) and a large number of states with $E_i \approx 0$ (within the numerical accuracy in sufficiently large systems), respectively.  

\subsection{General criterion for NH catalysis}

The above discussion lays the foundation to stage the general mathematical criterion operative behind the NH catalysis. Consider the general form of the NH operator in noninteracting systems from Eq.~\eqref{eq:NHgeneral}. In the examples we discussed so far $\hat{h}_{\rm mass} \equiv \sigma_0 \otimes \hat{h}_{\rm CDW}$ [see Eq.~\eqref{eq:CDWspinlessmatrix}]. For such an explicit choice of $\hat{h}_{\rm mass}$, we find two mass operators (${\mathcal O}_{\rm mass}$), namely ${\mathcal O}_{\rm CDW}$ and ${\mathcal O}_{\rm SDW}$ [see Eq.~\eqref{eq:massoperators}] that anticommute with $\hat{h}_0$ (by the definition of any mass order) and commute with $\hat{h}_{\rm mass}$ (thereby belonging to the commuting-class mass family), and hence anticommute with the total NH operator $\hat{h}_{\rm NH}$ [see Eqs.~\eqref{eq:NHgeneral} and~\eqref{eq:NHHamil}]. As such, the entire discussion on the NH catalysis mechanism rests on these (anti)commutation relations, irrespective of any microscopic details. Therefore, within the general principle of constructing NH operators in noninteracting bipartite lattice systems by supplementing the tight-binding Hamiltonian ($\hat{h}_0$) with a masslike (since $\{ \hat{h}_0, \hat{h}_{\rm mass} \}=0$) anti-Hermitian operator $\hat{h}_{\rm mass} \hat{h}_0$, the nucleation of any mass order (${\mathcal O}_{\rm mass}$) can be catalyzed by the non-Hermiticity in the system (parameterized by $\alpha$) when it belongs to the commuting-class mass family satisfying $\left[{\mathcal O}_{\rm mass}, \hat{h}_{\rm mass} \right]=0$ and $\{{\mathcal O}_{\rm mass}, \hat{h}_0 \}=0$.

\begin{table}[t!]
\centering
{\renewcommand{\arraystretch}{1.2}
\begin{tabular}{|P{0.5cm}|P{3.25cm}|P{3.75cm}|}
        \hline
        \multicolumn{1}{|c|}{} & \multicolumn{2}{c|}{Critical on-site interactions $U_{\rm c}^{}$ for SDW}\\
        \cline{2-3}
        \vspace{-0.22in}$\alpha$ & Honeycomb lattice & $\{10,3\}$ hyperbolic lattice\\
        \hline
        0.0 & 2.08 & 1.46\\
        0.2 & 2.06 & 1.43\\
        0.4 & 1.91 & 1.34\\
        0.6 & 1.68 & 1.16\\
        0.8 & 1.24 & 0.87\\
        \hline
\end{tabular}
}
\caption{The critical strength of the on-site Hubbard repulsion $U_c$ for the spin-density-wave (SDW) ordering for various non-Hermitian parameter $\alpha$ in honeycomb lattice and $\{ 10,3\}$ hyperbolic lattice, representing Dirac systems on the flat and negatively curved spaces, respectively; see Fig.~\ref{fig:DiracfitSDW}. Also notice that $U_c$ is smaller in the hyperbolic Dirac systems than in the Euclidean Dirac system for each $\alpha$, suggesting that the curvature-induced quantum phase transition of a mass ordering (in this case SDW) at weaker coupling remains operative even in non-Hermitian systems~\cite{HL:19}. Numerically obtained values of $U_c(\alpha)$ satisfy the scaling law $U_c(\alpha)=\sqrt{1-\alpha^2} U_c(0)$, predicted analytically in Eqs.~\eqref{eq:criticaldiscrete} and~\eqref{eq:Uccontinuum}.    
}~\label{tab:Hubbardcric}
\end{table}

Next, from lattice-based numerical self-consistent calculations, we demonstrate the NH catalysis of two density-wave orders on bipartite Euclidean and hyperbolic lattices, fostering Dirac liquids, Fermi liquids, and flat bands, within the framework of a minimal extended Hubbard model with NN Coulomb and on-site Hubbard repulsions in the Hartree limit; see Secs.~\ref{sec:CDW} and~\ref{sec:SDW}.

\section{Biorthogonal quantum mechanics}~\label{sec:NHQM}

For the lattice-based numerical self-consistent mean-field calculations of the CDW and SDW orders, we need to compute the spin-independent and spin-dependent average fermionic density at each site of the underlying lattice, respectively. These quantities are tied to the probability of finding a particle at a given point from the solution of the appropriate lattice-based Schr\"odinger equation. In Hermitian systems, these quantities can be readily found from the square of the amplitude of a wave function at a given point in space. However, in a NH system one needs to generalize the definition of the probability of finding a particle in terms of the inner product between the left and right eigenvectors (biorthogonal product) of the corresponding NH operator. Biorthogonal quantum mechanics is a vast and rich subject~\cite{biorthogonalQM}. So, here we focus only on its components that are required for our investigation.

\begin{figure}[t!]
\includegraphics[width=1.00\linewidth]{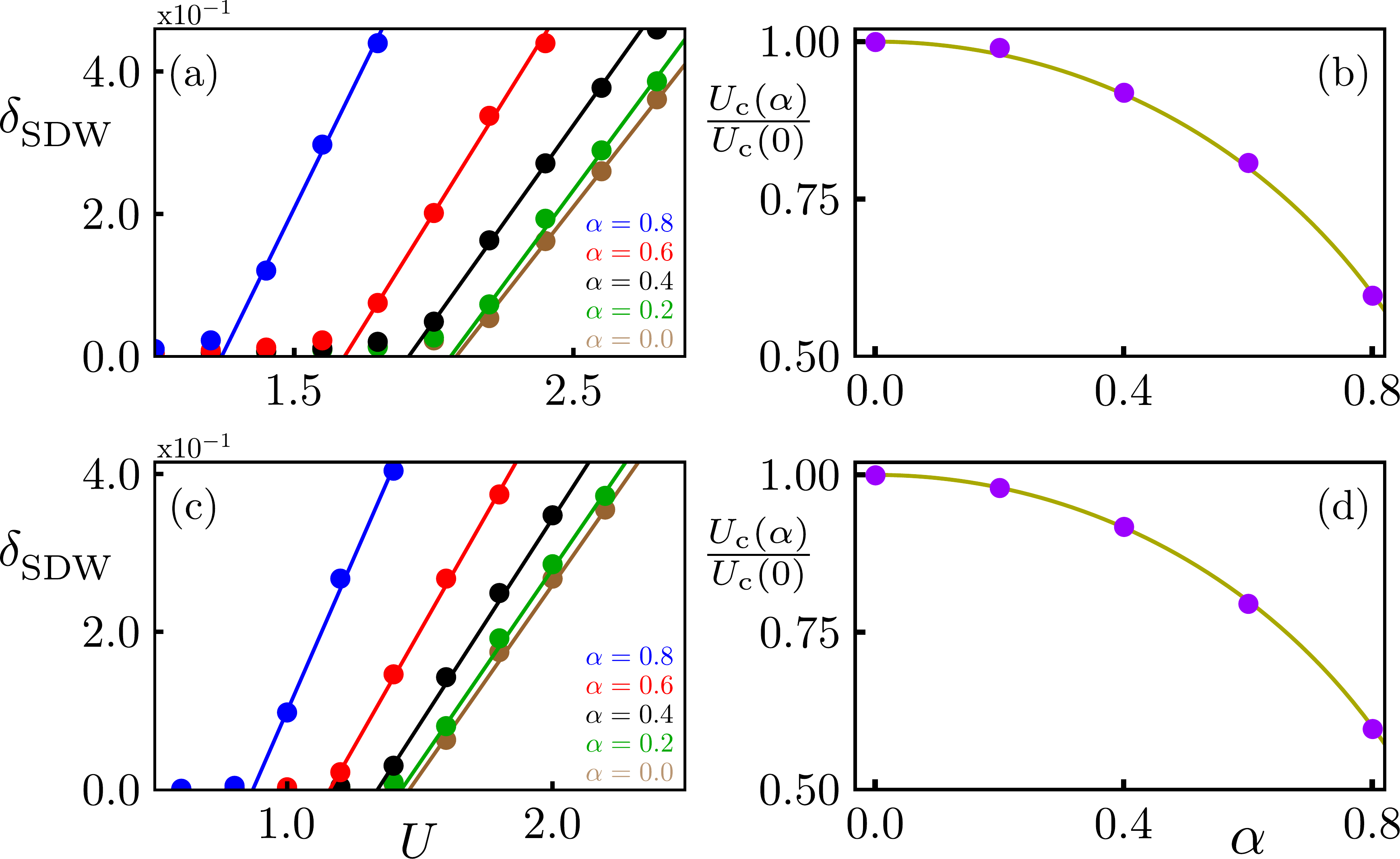}
\caption{(a) Computation of the critical strength of the on-site Hubbard repulsion $U_c (\alpha)$ as a function of the non-Hermitian parameter $\alpha$, obtained from the linear fit of the self-consistent solution of the spin-density-wave order parameter $\delta_{\rm SDW}$ as a function of $U$ when it appreciably rises from the trivial value on a Euclidean Dirac system of honeycomb or $\{ 6,3 \}$ lattice. At $U=U_c(\alpha)$, such a linear fit crosses the $\delta_{\rm SDW}=0$ axis. (b) The scaling of the ratio $U_c(\alpha)/U_c(0)$ as a function of $\alpha$, where the isolated dots correspond to the numerically computed values and the solid line corresponds to the analytical prediction of $U_c(\alpha)/U_c(0) = \sqrt{1-\alpha^2}$. Subfigures (c) and (d) are analogues of (a) and (b), respectively, but for the hyperbolic Dirac system on the $\{ 10, 3\}$ lattice. Explicit values of $U_c(\alpha)$ are shown in Table~\ref{tab:Hubbardcric}. Here, $U$ is measured in units of $t$ (nearest-neighbor hopping amplitude).         
}~\label{fig:DiracfitSDW}
\end{figure}

Let ${\mathcal H}_{\rm NH}$ be a generic NH operator, whose eigenvalues are nondegenerate (leaving aside the Kramers' degeneracy). The secular equations for the right ($R$) and left ($L$) eigenvectors of ${\mathcal H}_{\rm NH}$ with a real eigenvalue $E_\alpha$ are 
\begin{equation}
{\mathcal H}_{\rm NH} \; \ket{E_\alpha}_R = E_\alpha \; \ket{E_\alpha}_R
\:\: \text{and} \:\:
\prescript{}{L}{\bra{E_\alpha}} \; {\mathcal H}_{\rm NH} = \prescript{}{L}{\bra{E_\alpha}} \; E_\alpha,
\end{equation}
respectively. In our study, ${\mathcal H}_{\rm NH}$ corresponds to the hopping operator $\hat{h}_{\rm NH}$ [Eq.~\eqref{eq:NHHamil}] and the effective single-particle operator, discussed in the following sections. In NH quantum mechanics the orthonormality condition is defined with respect to the biorthogonal product between the left and right eigenvectors according to 
\begin{equation}~\label{eq:biorthonormal}
\prescript{}{L}{\bra{E_\alpha}\ket{E_\beta}_R}=\delta_{\alpha \beta},
\end{equation}
where $\delta_{\alpha \beta}$ is the Kronecker delta symbol. The probability of finding the particle at the $i$th site of the lattice in a state with eigenvalue $E_\alpha$ of ${\mathcal H}_{\rm NH}$ is given by 
\begin{equation}~\label{eq:probability}
P_{i}(E_\alpha) = \prescript{}{L}{\bra{i}\ket{E_\alpha}_R} \:\: \prescript{}{L}{\bra{E_\alpha}\ket{i}_R}, 
\end{equation}
where $\prescript{}{L}{\bra{i}}$ and $\ket{i}_R$ are the left and right eigenvectors for the site-localized Wannier state on the $i$th site. For the sake of simplicity, here we have omitted the spin degrees of freedom. Nonetheless, the above definition can immediately be generalized to find the probability of a particle at any site $i$ with spin projection $\sigma=\uparrow, \downarrow$. In the following two sections, we subscribe to these definitions to compute the local CDW and SDW orders self-consistently at each site of the lattice, from which we compute the corresponding global order parameters in the system.

Regarding the numerical assurance of the biorthonormality condition from Eq.~\eqref{eq:biorthonormal}, a comment is due at this stage. The numerical packages on Python are extremely efficient to deal with Hermitian operators and ensure the conventional orthonormality condition between any pair of eigenstates with different eigenvalues. However, for NH operators Python-based numerical packages do not necessarily respect the biorthonormality condition always, especially when the corresponding eigenvalues are extremely close to each other. In order to ensure the biorthonormality condition, we then add an extremely small amount of disorder at each site, drawn randomly and uniformly from a box distribution within the range $[-W/2,W/2]$ with $W \sim 10^{-3}$ denoting the strength of disorder. We pick the disorder configurations for which Eq.~\eqref{eq:biorthonormal} is satisfied. Our results are, however, insensitive to the disorder configuration and do not affect the numerically obtained self-consistent solutions of the CDW and SDW order parameters as they are typically at least an order of magnitude larger than $W$. A deterministic approach based on the presence of a small symmetry-breaking perturbation to numerically ensure the biorthonormality condition for the NH operators is presently unknown, and stands as an important open problem.

\section{NN Coulomb repulsion}~\label{sec:CDW}

In a half-filled bipartite lattice, the Coulomb repulsions between electrons living on the NN sites can be conducive to the nucleation of the CDW order as it produces a uniform and isotropic gap near the zero energy, thereby yielding insulating behavior in the system. As the corresponding order parameter is insensitive to the spin degrees of freedom, we consider a collection of spinless fermions for the sake of simplicity in this section. Then the Hamiltonian for the NN Coulomb repulsion reads as 
\begin{equation}~\label{eq:NNCoul}
H^{\rm Coul}_{\rm NN} = \frac{V}{2} \sum_{\langle i,j \rangle} n_i n_j - \mu {\mathcal N},
\end{equation}
where $n_i$ is the fermionic density on the $i$th site of the underlying lattice, ${\mathcal N}$ is the total number of spinless fermions in the half-filled system, $\mu$ is the chemical potential, and $V$ is the strength of the NN Coulomb repulsion. Here, we showcase the emergence of the CDW order from NN Coulomb repulsion within the mean-field approximation in NH bipartite lattices, embedded in the flat Euclidean and curved hyperbolic spaces, after decomposing the first term in Eq.~\eqref{eq:NNCoul} in the Hartree channel. Notice that the CDW order has been discussed from the continuum models of monolayer~\cite{MLGInt:1, MLGInt:2, MLGInt:3, MLGInt:4, MLGInt:5, MLGInt:6, MLGInt:7} and Bernal-stacked bilayer~\cite{BBLGInt:1, BBLGInt:2, BBLGInt:3, BBLGInt:4, BBLGInt:5} graphene.

Performing the Hartree decomposition of $H^{\rm Coul}_{\rm NN}$, we arrive at the following effective single-particle Hamiltonian 
\begin{equation}
H^{\rm Har}_{\rm NN} = V \sum_{\langle i,j \rangle} \bigg( \langle n_{B,i} \rangle n_{A, j} + \langle n_{A,i} \rangle n_{B, j} \bigg) 
- \mu {\mathcal N},
\end{equation}
where $\langle n_{A,i} \rangle$ and $\langle n_{B,i} \rangle$ correspond to the site ($i$) dependent self-consistent average electronic density on the $A$ and $B$ sublattices, respectively, at half-filling~\cite{HF:1, HF:2, HF:3, HF:4}. We measure these two quantities relative to the uniform density at half-filling according to 
\begin{equation}
\langle n_{A,i}\rangle = \frac{1}{2} + \delta_{A,i} 
\quad \text{and} \quad 
\langle n_{B,i}\rangle = \frac{1}{2} - \delta_{B,i}.
\end{equation} 
The half-filling condition is maintained by choosing $\mu=V/2$ and ensuring that 
\begin{equation}
\sum_{i} \delta_{A,i} - \sum_{i} \delta_{B,i} =0.
\end{equation} 
Then the positive definite quantities $\delta_A$ and $\delta_B$ yield the local CDW order parameter 
\begin{equation}~\label{eq:CDWOP}
\delta^{\rm local}_{\rm CDW}=\frac{1}{2} \left( \delta_{A} + \delta_{B} \right).   
\end{equation}
We numerically compute $\delta_{A}$ and $\delta_{B}$, and subsequently the global CDW order parameter ($\delta_{\rm CDW}$) in the entire system with periodic boundary conditions on Euclidean lattices and open boundary conditions on hyperbolic lattices (see Sec.~\ref{sec:system}), for a wide range of $V$. In Appendix~\ref{append:BBLG}, we generalize this formulation for Bernal-stacked honeycomb bilayer system upon including the layer indices.

The average fermionic density at each site of all the half-filled NH lattices is computed from the definition of the probability of finding such a particle at that site; see Eq.~\eqref{eq:probability}. Namely, we diagonalize the effective single-particle NH operator $\hat{h}^{\rm NN}_{\rm NH}=\hat{h}_{\rm NH} + \hat{h}^{\rm Har}_{\rm NN}$, and compute $\langle n_{A,i}\rangle$ and $\langle n_{B,i}\rangle$ in the half-filled system from all the states with negative eigenvalues of $\hat{h}^{\rm NN}_{\rm NH}$. Here, the operator $\hat{h}^{\rm Har}_{\rm NN}$ is obtained by casting $H^{\rm Har}_{\rm NN}$ in the $N$-dimensional spinor basis $\Psi$ without invoking the spin degrees of freedom. The numerical procedure to find self-consistent solutions for the CDW order parameters is summarized in Fig.~\ref{fig:Flowchart}. Next we discuss the results.

\subsection{NH catalysis of charge-density-wave: Results}~\label{subsec:CDW} 

To promote the NH catalysis of the CDW order, we first consider a fixed and moderate strength of the NN Coulomb repulsion $V=1.0$, such that it always exceeds the critical strength for this ordering in Euclidean and hyperbolic lattice-based Dirac systems (see Table~\ref{tab:NNCoulcric}). For such a fixed $V$, we compute and compare the density of states from the self-consistent solutions of the CDW order parameter on Euclidean and hyperbolic lattices, featuring Dirac liquids, Fermi liquids, and flat bands in the noninteracting limit (Sec.~\ref{sec:system}), in Hermitian ($\alpha=0.0$) and NH ($\alpha=0.8$) systems. The results are shown in Fig.~\ref{fig:NHCatalysisCDW}. It shows that the magnitude of the spectral gap near the zero energy gets amplified by nontrivial $\alpha$, thereby strongly endorsing the proposed NH catalysis mechanism of the CDW order, a member of the commuting-class mass family in our construction.

To gain further insights into the NH catalysis mechanism, we scrutinize the scaling of the CDW order parameter with varying strength of the NN Coulomb repulsion $V$ and for a few choices of the NH parameter $\alpha$ on Euclidean and hyperbolic lattices, mentioned in Sec.~\ref{sec:system}. The results are shown in Fig.~\ref{fig:CDWScalingNH}. In Euclidean and hyperbolic Dirac systems, the CDW order develops only beyond a critical strength of the NN Coulomb repulsion, denoted by $V_c (\alpha)$. Nevertheless, $V_c (\alpha)$ decreases monotonically with increasing $\alpha$ and the magnitude of the CDW order increases with increasing $\alpha$ for any fixed $V>V_c(\alpha)$. By contrast, in Euclidean and hyperbolic Fermi liquids and flat-band systems, the CDW orders develops for even infinitesimal NN Coulomb repulsion, which then grows monotonically with stronger $V$. In these systems, the magnitude of the CDW order increases monotonically with increasing $\alpha$ for any finite $V$. These outcomes strongly promote the proposed non-Hermitian catalysis mechanism. However, we note that the growth of the CDW order with $V$ in flat band systems is faster than that in Fermi liquid systems, irrespective of the $\alpha$ values. This feature can be attributed to the fact that the density of states near the zero energy is finite and diverging in Fermi liquids and flat-band systems, respectively, for any $|\alpha|<1$ (including $\alpha=0$).

\begin{table}[t!]
\centering
{\renewcommand{\arraystretch}{1.2}
\begin{tabular}{|P{0.5cm}|P{3.25cm}|P{3.75cm}|}
        \hline
        \multicolumn{1}{|c|}{} & \multicolumn{2}{c|}{Fitting parameter $\kappa_{_{\rm SDW}}$ for SDW order}\\
        \cline{2-3}
        \vspace{-0.14in}$\alpha$ & Bernal-stacked bilayer honeycomb lattice & \vspace{-0.14cm}$\{16,3\}$ hyperbolic lattice\\
        \hline
        0.0 & 8.40 & 3.50\\
        0.2 & 8.10 & 3.45\\
        0.4 & 7.70 & 3.20\\
        0.6 & 6.90 & 2.85\\
        0.8 & 5.10 & 2.20\\
        \hline
\end{tabular}
}
\caption{The fitting parameter $\kappa_{_{\rm SDW}}$ appearing in the BCS-like scaling form for the spin-density-wave (SDW) order for various non-Hermitian parameter $\alpha$ in Bernal-stacked bilayer honeycomb lattice and $\{16,3\}$ hyperbolic lattice, hosting Fermi liquids in the noninteracting limit on flat Euclidean and negatively curved spaces, respectively. See also Fig.~\ref{fig:FLfitSDW} for the scaling of $\kappa_{_{\rm SDW}}(\alpha)/\kappa_{_{\rm SDW}}(0)$ in these two systems. Numerical values of $\kappa_{\rm SDW}(\alpha)$ satisfy the scaling law $\kappa_{\rm SDW}(\alpha)=\sqrt{1-\alpha^2} \kappa_{\rm SDW}(0)$, predicted analytically [Eq.~\eqref{eq:BCSSDWfittinganalytical}].
}~\label{tab:SDWFLfitting}
\end{table}

Although already discussed in Sec.~\ref{sec:NHcatalysis}, next we present the scaling of the critical strength of the NN Coulomb repulsion $V_c$ for the CDW ordering with the NH parameter $\alpha$ in Dirac systems from a continuum description. This additional exercise provides a prescription to extract $V_c$ from the self-consistent solutions for the CDW order. Notice that $V_c$ can be obtained from the corresponding self-consistent mean-field gap equation, now given by 
\begin{equation}~\label{eq:gapcontinuum}
\frac{1}{V}=\int^{E_{c}(\alpha)}_0 \frac{\rho(E,\alpha)}{\sqrt{\Delta^2 + E^2}} \; dE,
\end{equation} 
where $\Delta$ is the CDW order parameter, and $E_{c}(\alpha)$ is the high-energy cut-off up to which the density of states in a $d$-dimensional Dirac system scales as $\rho(E,\alpha) = |E|^{d-1}/[v^d_{_{\rm F}}(\alpha)]$ with $v_{_{\rm F}}(\alpha)$ bearing the dimension of the Fermi velocity. We note that 
\begin{eqnarray}
E_{c}(\alpha) = E_{c}(0) \sqrt{1-\alpha^2} 
\:\:\: \text{and} \:\:\:  
v_{_{\rm F}}(\alpha) = v_{_{\rm F}}(0) \sqrt{1-\alpha^2}. \nonumber \\ 
\end{eqnarray}
The critical strength of the NN repulsion for the CDW ordering $V_c(\alpha)$ is obtained by setting $V=V_c(\alpha)$ and $\Delta=0$ in Eq.~\eqref{eq:gapcontinuum}, yielding 
\allowdisplaybreaks[4]
\begin{eqnarray}
&& \frac{1}{V_c(\alpha)} = \frac{E^{d-1}_c(\alpha)}{(d-1)v^d_{_{\rm F}}(\alpha)} \nonumber \\
&=& \frac{E^{d-1}_c(0)}{(d-1)v^d_{_{\rm F}}(0)} \; \frac{1}{\sqrt{1-\alpha^2}} 
=\frac{1}{V_c(0)} \; \frac{1}{\sqrt{1-\alpha^2}}.
\end{eqnarray}  
Therefore, in Euclidean and hyperbolic Dirac systems in any spatial dimension, we find
\begin{equation}~\label{eq:Vccontinuum}
V_c(\alpha)=\sqrt{1-\alpha^2} \: V_c(0),
\end{equation}
which is in agreement with Eq.~\eqref{eq:criticaldiscrete}.

\begin{figure}[t!]
\includegraphics[width=1.00\linewidth]{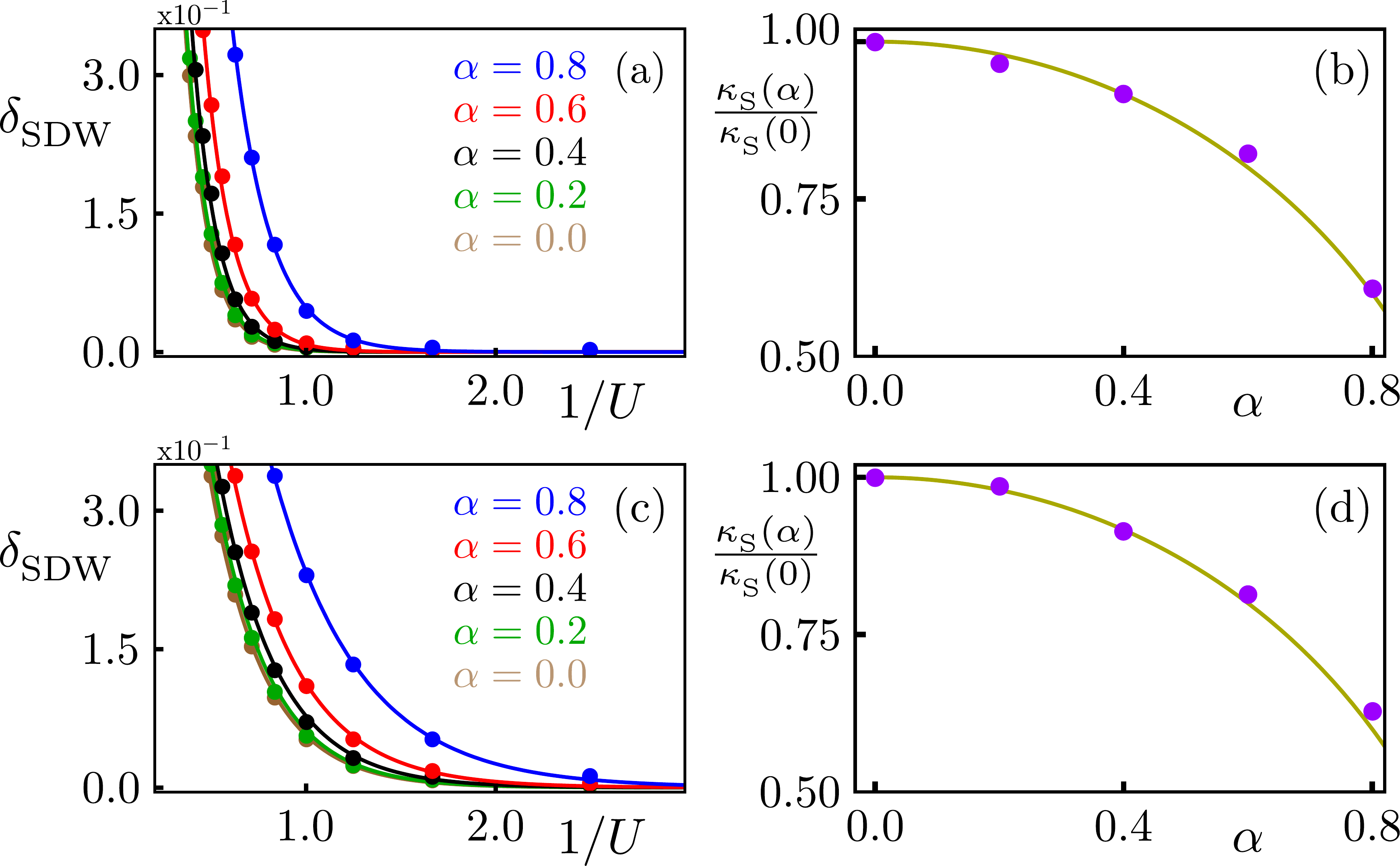}
\caption{BCS-like scaling for the spin-density-wave order parameter $\delta_{\rm SDW}$ with $1/U$ in (a) a Euclidean Fermi liquid system (Bernal-stacked bilayer honeycomb lattice) and (c) a hyperbolic Fermi liquid system ($\{ 16, 3\}$ lattice). Here, we use the fitting function $\delta_{\rm SDW} = B \exp (-\kappa_{_{\rm SDW}}/U)$, where $B$ and $\kappa_{_{\rm SDW}}$ are $\alpha$ and system dependent fitting parameters, and $U$ denotes the strength of the on-site Hubbard repulsion. The scaling of the ratio $\kappa_{_{\rm SDW}}(\alpha)/\kappa_{_{\rm SDW}}(0)$ as a function of $\alpha$ in these two systems are shown in (b) and (d), respectively, where the isolated dots correspond to the numerically computed values and the solid lines correspond to the analytical prediction of $\kappa_{_{\rm SDW}}(\alpha)/\kappa_{_{\rm SDW}}(0)=\sqrt{1-\alpha^2}$. Here, $U$ is measured in units of $t$ (nearest-neighbor hopping amplitude).
}~\label{fig:FLfitSDW}
\end{figure}

For all $\alpha$ the critical strength of the NN repulsion for the CDW ordering can be obtained by rewriting the gap equation from Eq.~\eqref{eq:gapcontinuum} in two spatial dimensions as 
\begin{equation}
\frac{1}{V_c(\alpha)}-\frac{1}{V}= \frac{1}{v^2_{_{\rm F}}(\alpha)} \int^{E_{c}(\alpha)}_0 \left( 1-\frac{E}{\sqrt{E^2+\Delta^2}}\right) \; dE.
\end{equation}  
Sufficiently close to the CDW ordering when $\Delta \ll E_c(\alpha)$, we obtain 
\begin{equation}
\Delta \approx \frac{v^2_{_{\rm F}}(\alpha)}{V V_c(\alpha)} \: \left[ V- V_c(\alpha) \right].
\end{equation} 
In terms of the dimensionless parameters, defined in systems featuring linearly vanishing density of states in two spatial dimensions according to $\bar{\delta}_{\rm CDW}(\alpha)=\Delta/E_c(\alpha)$, $V E_c(\alpha)/v^2_{_{F}} \to V$, and $V_c(\alpha) E_c(\alpha)/v^2_{_{F}} \to V_c(\alpha)$~\cite{scaling:1}, the above relationship can be cast as 
\begin{equation}~\label{eq:CDWnearcrit}
\bar{\delta}_{\rm CDW} (\alpha) = \frac{1}{V V_c(\alpha)} \: \left[ V- V_c(\alpha) \right].
\end{equation}
Therefore, when the numerically obtained self-consistent solutions of the CDW order ($\delta_{\rm CDW}$) rise from its trivial value, by fitting $\delta_{\rm CDW}$ linearly with $V$ we obtain the corresponding critical strength of the NN Coulomb repulsion $V_c(\alpha)$, as shown in Figs.~\ref{fig:DiracfitCDW}(a) and~\ref{fig:DiracfitCDW}(c) for Euclidean and hyperbolic Dirac systems, respectively. In both systems, we also find that $V_c(\alpha)/V_c(0)$ respects the scaling form displayed in Eqs.~\eqref{eq:criticaldiscrete} and~\eqref{eq:Vccontinuum}, as shown in Figs.~\ref{fig:DiracfitCDW}(b) and~\ref{fig:DiracfitCDW}(d). The explicit values for $V_c(\alpha)$ for various $\alpha$ are listed in Table~\ref{tab:NNCoulcric}.

\begin{figure}[t!]
\includegraphics[width=1.00\linewidth]{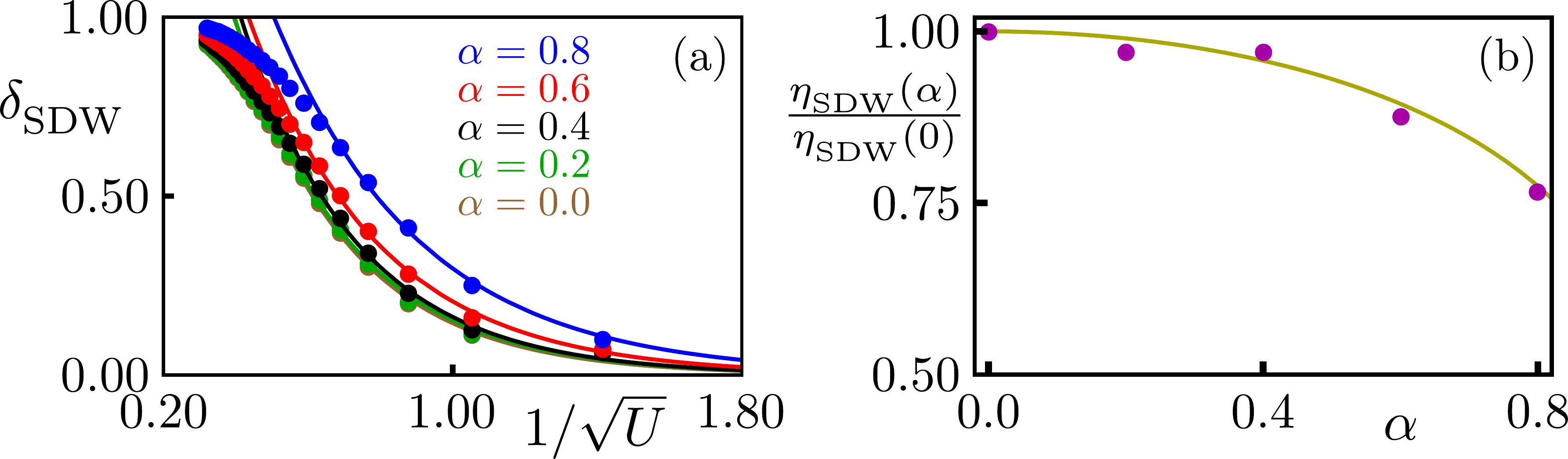}
\caption{(a) Fitting of the spin-density-wave order parameter $\delta_{\rm SDW}$ as a function of $1/\sqrt{U}$ [Eq.~\eqref{eq:SDWFitsquareLat}] for the Euclidean \{4, 4\} or square lattice with periodic boundary conditions, and with various values of the non-Hermitian parameter $\alpha$. The fitting function reads as $\delta_{\rm SDW} = B \exp(-\eta_{\rm SDW} / \sqrt{U})$ where $B$ and $\eta_{\rm SDW}$ are system-specific ($\alpha$-dependent) fitting parameters and $U$ is the coupling strength of the on-site Hubbard repulsion interaction. (b) The scaling of the ratio $\eta_{\rm SDW}(\alpha) / \eta_{\rm SDW}(0)$ as a function of the non-Hermitian parameter $\alpha$. The numerically obtained results are represented by the magenta points, while the yellow curve is the expected theoretical scaling relation given by $\eta_{SDW}(\alpha) / \eta_{\rm SDW}(0) = (1 - \alpha^2)^{1/4}$, as shown in Eq.~\eqref{eq:futtingSLSDW}. See Table~\ref{tab:QFBfitting} for explicit values of $\eta_{\rm SDW}$ for various $\alpha$. Here, $U$ is measured in units of $t$ (nearest-neighbor hopping amplitude).   
}~\label{fig:QFBfitSDW}
\end{figure}

The gap equation from Eq.~\eqref{eq:gapcontinuum} can also be employed in Fermi liquid systems, where the density of states $\rho(E,\alpha)=C(\alpha)$ is a constant as the energy $E \to 0$ that depends on the NH parameter $\alpha$ according to $C(\alpha)= C(0)/\sqrt{1-\alpha^2}$. In such systems, we find 
\allowdisplaybreaks[4]
\begin{eqnarray}
\frac{1}{V} &=& C(\alpha) \int^{E_{c}(\alpha)}_0 \frac{dE}{\sqrt{E^2+\Delta^2}} \nonumber \\
&=& C(\alpha) \; \ln \left( \frac{E_c(\alpha)+\sqrt{E^2_c(\alpha)+\Delta^2}}{\Delta}\right).
\end{eqnarray} 
In the weak interaction regime, where $\Delta \ll E_c(\alpha)$, in terms of the dimensionless parameter $\bar{\delta}_{\rm CDW}(\alpha)=\Delta/E_c(\alpha)$, we obtain the celebrated BCS-like scaling~\cite{scaling:2}
\begin{equation}
\bar{\delta}_{\rm CDW}(\alpha)= 2 \exp[-\frac{1}{V \; C(\alpha)}] \equiv 2 \exp[-\frac{\kappa_{_{\rm CDW}}(\alpha)}{V}], 
\end{equation}
where $\kappa_{_{\rm CDW}}(\alpha)=1/C(\alpha)$. Notice that in two-dimensional systems with constant density of states, $V$ is dimensionless in the above expression, following the general scaling argument. We find good agreement with this scaling form by comparing the numerically obtained self-consistent solutions for $\delta_{\rm CDW}$ with $1/V$, as shown for the Fermi liquid systems on Euclidean and hyperbolic lattices in Figs.~\ref{fig:FLfitCDW}(a) and~\ref{fig:FLfitCDW}(c), respectively, where $\kappa_{_{\rm CDW}}(\alpha)$ is treated as an $\alpha$-dependent fitting parameter. Explicit values of $\kappa_{_{\rm CDW}}(\alpha)$ for various $\alpha$ are presented in Table~\ref{tab:CDWFLfitting}. 

\begin{figure*}[t!]
\includegraphics[width=1.00\linewidth]{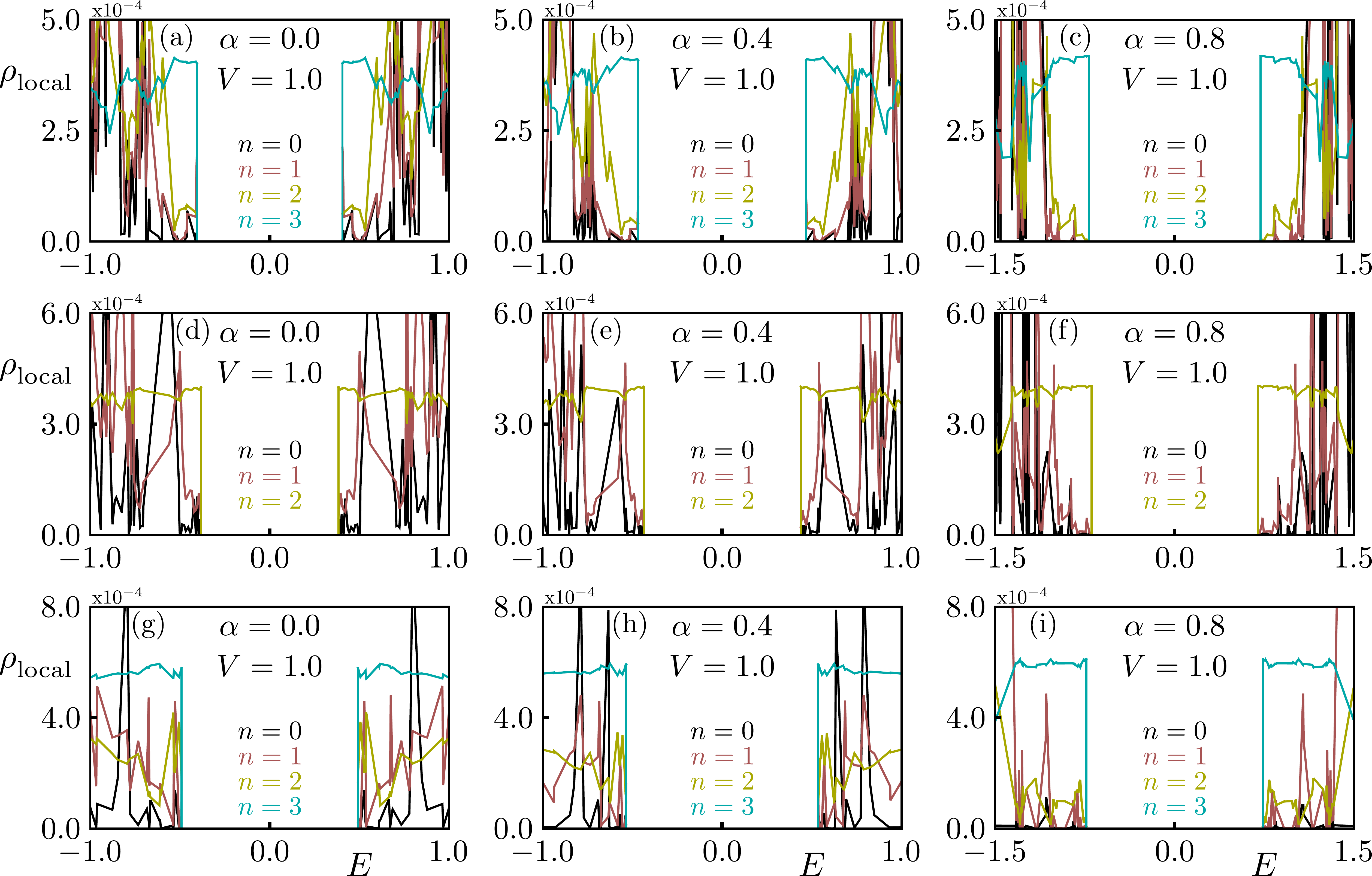}
\caption{Local density of states ($\rho_{\rm local}$) as a function of energy ($E$) at each generation $n$ (see Sec.~\ref{sec:system}) of the (a)-(c) $\{ 10, 3\}$, (d)-(f) $\{ 16, 3\}$, and (g)-(i) $\{ 8, 4\}$ hyperbolic lattices for various values of the non-Hermitian parameter $\alpha$ (quoted in each subfigure) and nearest-neighbor Coulomb repulsion $V=1.0$, which always exceeds its critical strength for the charge-density-wave ordering in the Dirac system of $\{10,3\}$ hyperbolic lattice (Table~\ref{tab:NNCoulcric}). This analysis shows that the gap at half-filling is finite in every generation of all the hyperbolic lattices with the onset of the global charge-density-wave order. For details see Sec.~\ref{sec:finitesizehyperbolic}. Here, $E$ and $V$ are measured in units of $t$ (nearest-neighbor hopping amplitude).           
}~\label{fig:LDOSHyperCDW}
\end{figure*}

In order to further confirm the BCS-like scaling form, we compute the ratio 
\begin{equation}~\label{eq:BCSCDWfittinganalytical}
\frac{\kappa_{_{\rm CDW}}(\alpha)}{\kappa_{_{\rm CDW}}(0)}=\sqrt{1-\alpha^2}.
\end{equation}   
We find good agreement with the above theoretical prediction in Euclidean and hyperbolic Fermi liquid systems, as shown in Figs.~\ref{fig:FLfitCDW}(b) and~\ref{fig:FLfitCDW}(d), respectively.

In flat-band systems, we cannot predict any analytically closed form of the scaling function for the CDW order parameter as a function of the NN Coulomb repulsion ($V$) and the NH parameter ($\alpha$) as the density of states $\rho(\alpha, E)$ diverges therein when $|E| \to 0$. Nevertheless, such a diverging density of states gives rise to stronger growth of the CDW order with increasing $V$, which gets further amplified by $\alpha$, when compared with its counterpart in Fermi liquid systems, where $\rho(\alpha, E)$ becomes a constant as $|E| \to 0$, as can be seen from Fig.~\ref{fig:CDWScalingNH}. Nonetheless, as the density of states for the quasiflat-band system on a square or $\{4, 4 \}$ lattice diverges logarithmically as $|E| \to 0$, the scaling of CDW order parameter in such system is given by~\cite{scaling:3} 
\begin{equation}~\label{eq:CDWFitsquareLat}
\delta_{\rm CDW} \sim \exp \left( - \pi \sqrt{E_\Lambda/V}\right),
\end{equation}
where $E_\lambda$ is proportional to the bandwidth. We fit the self-consistent solution of CDW order with the above form by taking $\pi \sqrt{E_\Lambda} = \eta_{\rm CDW}$ in Fig.~\ref{fig:QFBfitCDW}(a), where $\eta_{\rm CDW}$ is a $\alpha$-dependent fitting parameter, reported in Table~\ref{tab:QFBfitting}. As the scaling of the band width with the NH parameter is given by $E_\Lambda \sim \sqrt{1-\alpha^2}$, we arrive at the following scaling form of the fitting parameter 
\begin{equation}~\label{eq:futtingSLCDW}
\frac{\eta_{\rm CDW}(\alpha)}{\eta_{\rm CDW}(0)} = \left(1-\alpha^2 \right)^{1/4}.
\end{equation}
We find reasonable agreement with the above scaling form for the CDW order on a square lattice; see Fig.~\ref{fig:QFBfitCDW}(b).

In this context, we point out that while in all the systems we present the scaling of the CDW order as a function of $V$ and $\alpha$ after averaging $\delta_{\rm CDW}$ over the entire system, for hyperbolic flat-band system on the $\{ 8,4\}$ lattice we only consider the average $\delta_{\rm CDW}$ at the center plaquette of the system, constituted by the sites belonging to the zeroth generation of this lattice. When averaged over the entire system, $\delta_{\rm CDW}$ shows a putative jump for small $V$, which we discuss in more detail in Appendix~\ref{append:HypFB}. This feature, however, does not affect the proposed NH catalysis mechanism in this system. Finally, notice that the maximal value of $\delta_{\rm CDW}=0.5$ which is obtained self-consistently as $V \to \infty$. In this limit, fermions reside only on the sites of the $A$ sublattice, while those belonging to the $B$ sublattice become completely empty or vice versa. A generalization of this scenario on Bernal-stacked bilayer honeycomb lattice is discussed in Appendix~\ref{append:BBLG}. Next we establish the NH catalysis mechanism for the SDW order, yet another member of the commuting-class mass family in our construction.  

\begin{figure*}[t!]
\includegraphics[width=1.00\linewidth]{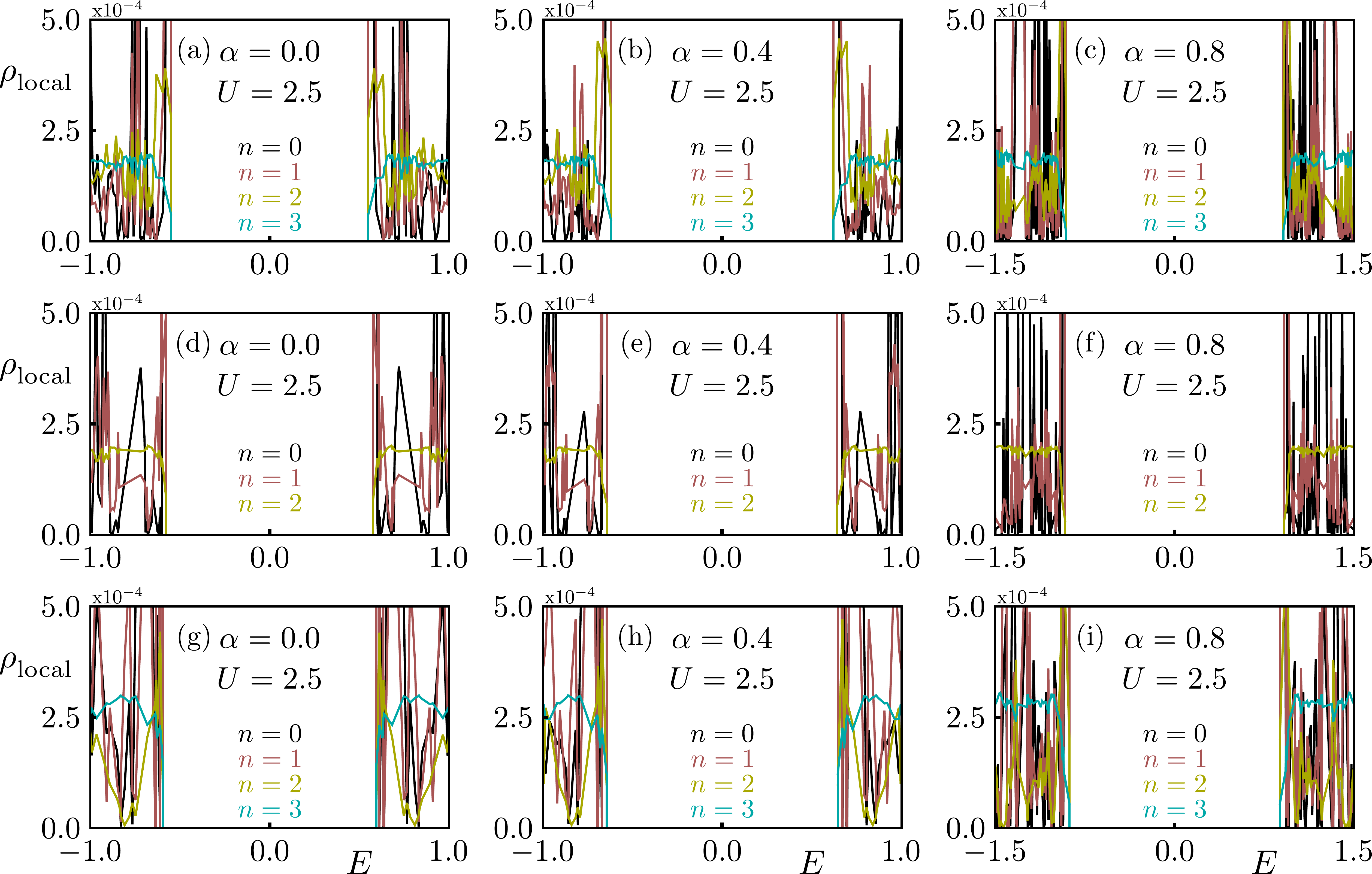}
\caption{Local density of states ($\rho_{\rm local}$) as a function of energy ($E$) at each generation $n$ (see Sec.~\ref{sec:system}) of the (a)-(c) $\{ 10, 3\}$, (d)-(f) $\{ 16, 3\}$, and (g)-(i) $\{ 8, 4\}$ hyperbolic lattices for various values of the non-Hermitian parameter $\alpha$ (quoted in each subfigure) and on-site Hubbard repulsion $U=2.5$, which always exceeds its critical strength for the spin-density-wave ordering in the Dirac system of $\{10,3\}$ hyperbolic lattice (Table~\ref{tab:Hubbardcric}). This analysis shows that the gap at half-filling is finite in every generation of all the hyperbolic lattices with the onset of the global spin-density-wave order. For details see Sec.~\ref{sec:finitesizehyperbolic}. Here, $E$ and $U$ are measured in units of $t$ (nearest-neighbor hopping amplitude).}~\label{fig:LDOSHyperSDW}
\end{figure*}

\section{On-site Hubbard repulsion}~\label{sec:SDW}

In this section, we consider a collection of fermions with spin degrees of freedom (electrons). To capture the antiferromagnetic or SDW order on NH bipartite lattices, we focus on the on-site Hubbard repulsion for which the Hamiltonian is given by 
\begin{equation}
H^{\rm Hub}_{\rm OS}= U \sum_{i} \left( n_{i, \uparrow} -\frac{1}{2} \right) \; \left( n_{i, \downarrow} -\frac{1}{2} \right) -\mu {\mathcal N},
\end{equation}
where $n_{i,\uparrow/\downarrow}$ is the electronic density at site $i$ with spin projection $\uparrow/\downarrow$ in the $z$ direction, ${\mathcal N}$ is the total number of electrons in the half-filled system, and $U$ denotes the strength of the Hubbard repulsion. The Hartree decomposition of $H^{\rm Hub}_{\rm OS}$ leads to the following effective single-particle Hamiltonian~\cite{HF:1, HF:2, HF:3, HF:4} 
\begin{eqnarray}
H^{\rm Har}_{\rm OS} &=& \sum_{x=A,B} \bigg\{ \left( \langle n_{x,\uparrow} \rangle -\frac{1}{2}\right) \left( n_{x,\downarrow} -\frac{1}{2} \right) \nonumber \\
&+& \left( \langle n_{x,\downarrow} \rangle -\frac{1}{2}\right) \left( n_{x,\uparrow} -\frac{1}{2} \right) \bigg\} -\mu {\mathcal N}.
\end{eqnarray}
Next we choose the following ansatz 
\begin{equation}
\langle n_{A,\sigma} \rangle= \frac{1}{2} +\sigma \delta_{A,\sigma}(\vec{r}) 
\:\:\: \text{and} \:\:\:
\langle n_{B,\sigma} \rangle= \frac{1}{2} -\sigma \delta_{B,\sigma}(\vec{r}). 
\end{equation}
In the last expression $\sigma=+ \; (-) \equiv \; \uparrow (\downarrow)$, and $\vec{r}$ measures the position of a site. Charge neutrality (half-filling) of the system is now maintained by choosing $\mu=0$ and
\begin{equation}
\sum_{\sigma=\pm} \; \sum_{\vec{r}} \; \sigma \; \delta_{A,\sigma} (\vec{r}) - 
\sum_{\sigma=\pm} \; \sum_{\vec{r}} \; \sigma \; \delta_{B, \sigma} (\vec{r}) =0,
\end{equation}
which also keeps the average electronic density equal to one at each site of the lattice. Then, a SDW ground state with $\delta_{A/B, \uparrow/\downarrow}>0$ is characterized by the local order parameter
\begin{equation}~\label{eq:AFMOP}
\delta^{\rm local}_{\rm SDW}=\frac{1}{2} \left( \delta_{A,\uparrow} + \delta_{A,\downarrow} + \delta_{B,\uparrow} + \delta_{B,\downarrow} \right),
\end{equation} 
where the site from the $B$ sublattice is one of the NN sites for a given site living on the $A$ sublattice. We numerically compute $\delta_{A,\uparrow/\downarrow}$ and $\delta_{B,\uparrow/\downarrow}$, and subsequently the global SDW order parameter ($\delta_{\rm SDW}$) in the entire system with periodic boundary conditions on Euclidean lattices and open boundary conditions on hyperbolic lattices (see Sec.~\ref{sec:system}), for a wide range of the on-site Hubbard repulsion $U$. In Appendix~\ref{append:BBLG}, we generalize this formulation for the Bernal-stacked honeycomb bilayer system upon including the layer indices. Notice that the SDW order has also been discussed from the continuum models of monolayer graphene~\cite{MLGInt:1, MLGInt:2, MLGInt:3, MLGInt:4, MLGInt:5, MLGInt:6, MLGInt:7} and Bernal-stacked bilayer graphene~\cite{BBLGInt:1, BBLGInt:2, BBLGInt:3, BBLGInt:4, BBLGInt:5}.

The average electronic density at each site of all the half-filled NH lattices for each projection of electronic spin is computed from the definition of the probability of finding such a particle at that site by generalizing Eq.~\eqref{eq:probability} upon invoking the spin indices therein. Namely, we diagonalize the effective single-particle NH operator $\hat{h}^{\rm OS}_{\rm NH}=\hat{h}_{\rm NH} + \hat{h}^{\rm Har}_{\rm OS}$, and compute $\langle n_{A,\uparrow/\downarrow}\rangle$ and $\langle n_{B,\uparrow/\downarrow}\rangle$ in the half-filled system from all the states with negative eigenvalues of $\hat{h}^{\rm OS}_{\rm NH}$. Here, the operator $\hat{h}^{\rm Har}_{\rm OS}$ is obtained by casting $H^{\rm Har}_{\rm OS}$ in the $2N$-dimensional spinor basis $\Psi$. The rest of the numerical procedure to find self-consistent solutions for the SDW order parameters is similar to the one for the CDW order, which is summarized in Fig.~\ref{fig:Flowchart}. Next we discuss the results.

\begin{figure*}[t!]
\includegraphics[width=1.00\linewidth]{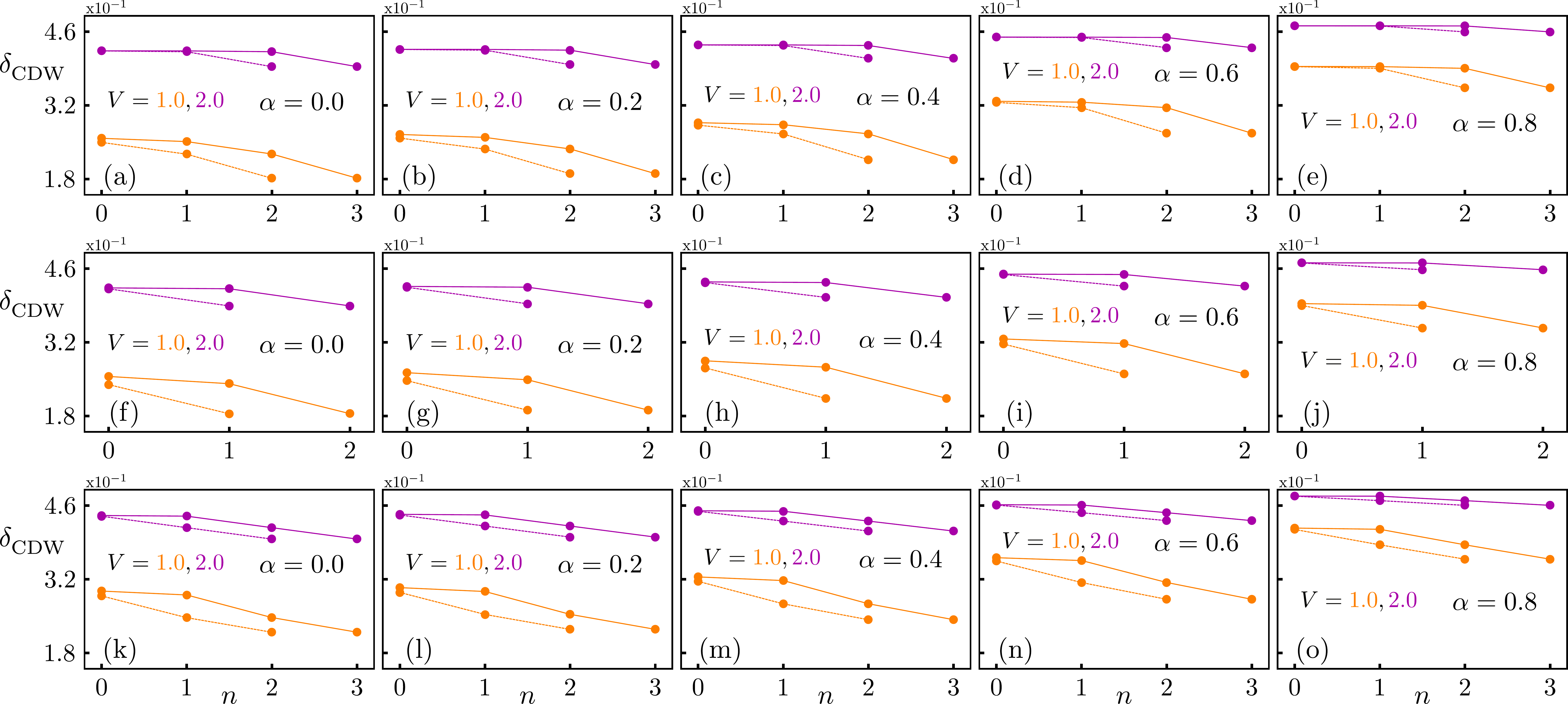}
\caption{Self-consistent solution of the charge-density-wave order parameter ($\delta_{\rm CDW}$) for $V=1.0$ (orange) and $V=2.0$ (purple) as a function of the generation number number $n$ (see Sec.~\ref{sec:system}). These quantities are computed by averaging $\delta_{\rm CDW}$ over the sites belonging to the $n$th generation of the (a)-(e) second (dashed lines) and third (solid lines) generation $\{ 10,3 \}$, (f)-(j) first (dashed lines) and second (solid lines) generation $\{ 16,3 \}$, and (k)-(o) second (dashed lines) and third (solid lines) generation $\{ 8,4 \}$ hyperbolic lattices for the non-Hermitian parameter $\alpha=0.0$ [(a), (f), (k)], $0.2$ [(b), (g), (l)], $0.4$ [(c), (h), (m)], $0.6$ [(d), (i), (n)], and $0.8$ [(e), (j), (o)]. For details see Sec.~\ref{sec:finitesizehyperbolic}. The strength of $V$ always exceeds it critical value in the Dirac system, $\{ 10, 3 \}$ hyperbolic lattice; see Table~\ref{tab:NNCoulcric}. Here, $V$ is measured in units of $t$ (nearest-neighbor hopping amplitude).             
}~\label{fig:GenerationCDW}
\end{figure*}

\subsection{NH catalysis of spin-density-wave: Results}

The results supporting the NH catalysis mechanism for the SDW order, also a member of the commuting-class mass family in our construction, are qualitatively similar to those for the CDW order, which we have discussed in detail in Sec.~\ref{subsec:CDW}. Therefore, here we only highlight the main results. The NH catalysis of the SDW order is primarily identified by computing and comparing the density of states from the self-consistent solutions of the SDW order parameter on Euclidean and hyperbolic lattices, featuring Dirac liquids, Fermi liquids, and flat bands in the noninteracting limit (Sec.~\ref{sec:system}), in Hermitian ($\alpha=0.0$) and NH ($\alpha=0.8$) systems. The results are displayed in Fig.~\ref{fig:NHCatalysisSDW}, showing that the magnitude of the spectral gap near the zero energy gets amplified by nontrivial $\alpha$. The strength of $U=2.5$ always exceeds its critical strength for the SDW ordering in any Dirac system (see Table~\ref{tab:Hubbardcric}). This observation promotes the proposed NH catalysis mechanism of the SDW order.

Furthermore, we scrutinize the scaling of the SDW order parameter ($\delta_{\rm SDW}$) with varying strength of the on-site Hubbard repulsion $U$ for a few choices of the NH parameter $\alpha$ on Euclidean and hyperbolic lattices, mentioned in Sec.~\ref{sec:system}. The results are shown in Fig.~\ref{fig:SDWScalingNH}. In Euclidean and hyperbolic Dirac systems, where the SDW order develops only beyond a critical strength of the on-site Hubbard repulsion, denoted by $U_c (\alpha)$, the magnitude of $\delta_{\rm SDW}$ increases with increasing $\alpha$ for any $U>U_c(\alpha)$. On the other hand, the SDW order develops for even infinitesimal on-site Hubbard repulsion in Euclidean and hyperbolic Fermi liquids and flat-band systems, where the magnitude of the SDW order increases with increasing $\alpha$ for any finite $U$. These outcomes strongly support the non-Hermitian catalysis mechanism for commuting-class masses, proposed in this work. In Dirac systems, this mechanism can also be appreciated by noting that the critical strength of the on-site Hubbard repulsion for the SDW ordering $U_c(\alpha)$ decreases monotonically with increasing non-Hermiticity ($\alpha$) in the system.

The solutions of the gap equations for the SDW order in Dirac and Fermi liquid systems, obtained from their continuum descriptions, are similar to the ones we previously discussed for the CDW order in detail in Sec.~\ref{subsec:CDW}. Therefore, here we only quote the requisite final expressions and relate them to the results obtained from the lattice-based numerical self-consistent calculation for the SDW order with the on-site Hubbard repulsion. In Dirac systems, the critical strength of the on-site Hubbard repulsion for the ordering follows the scaling law given by 
\begin{equation}~\label{eq:Uccontinuum}
U_c(\alpha)=\sqrt{1-\alpha^2} \: U_c(0),
\end{equation}
which is analogous to Eq.~\eqref{eq:Vccontinuum} and also is in agreement with Eq.~\eqref{eq:criticaldiscrete}. In terms of the dimensionless parameters $\bar{\delta}_{\rm SDW}(\alpha)=\Delta/E_c(\alpha)$, $U E_c(\alpha)/v^2_{_{F}} \to U$, and $U_c(\alpha) E_c(\alpha)/v^2_{_{F}} \to U_c(\alpha)$~\cite{scaling:1}, in close proximity to the onset of the SDW ordering we find 
\begin{equation}~\label{eq:SDWnearcrit}
\bar{\delta}_{\rm SDW} (\alpha) = \frac{1}{U U_c(\alpha)} \: \left[ U- U_c(\alpha) \right],
\end{equation}
taking a form similar to Eq.~\eqref{eq:CDWnearcrit}. Therefore, when the numerically obtained self-consistent solutions of the SDW order ($\delta_{\rm SDW}$) rise from its trivial value, by fitting $\delta_{\rm SDW}$ linearly with $U$ we obtain the corresponding critical strength of the on-site Hubbard repulsion $U_c(\alpha)$, as shown in Figs.~\ref{fig:DiracfitSDW}(a) and~\ref{fig:DiracfitSDW}(c) for Euclidean and hyperbolic Dirac systems, respectively. In both the systems, we also find that $U_c(\alpha)/U_c(0)$ respects the scaling form given in Eqs.~\eqref{eq:criticaldiscrete} and~\eqref{eq:Uccontinuum}, as shown in Figs.~\ref{fig:DiracfitSDW}(b) and~\ref{fig:DiracfitSDW}(d). The explicit values of $U_c(\alpha)$ for various $\alpha$ in these two systems are displayed in Table~\ref{tab:Hubbardcric}.

Similarly, in the Fermi liquid systems, we find the BCS-like scaling for the dimensionless SDW order~\cite{scaling:2}
\begin{equation}
\bar{\delta}_{\rm SDW}(\alpha)= 2 \exp[-\frac{1}{U \; C(\alpha)}] \equiv 2 \exp[-\frac{\kappa_{_{\rm SDW}}(\alpha)}{U}], 
\end{equation}
where $\kappa_{_{\rm SDW}}(\alpha)=1/C(\alpha)$. Notice that in two-dimensional systems with a constant density of states $U$ is also dimensionless in the above expression. We find good agreement with this scaling form by comparing the numerically obtained self-consistent solutions for $\delta_{\rm SDW}$ with $1/U$, as shown for the Fermi liquid systems on Euclidean and hyperbolic lattices in Figs.~\ref{fig:FLfitSDW}(a) and~\ref{fig:FLfitSDW}(c), respectively, where $\kappa_{_{\rm SDW}}(\alpha)$ is treated as an $\alpha$-dependent fitting parameter. Explicit values of $\kappa_{_{\rm SDW}}(\alpha)$ for various $\alpha$ are presented in Table~\ref{tab:SDWFLfitting}, satisfying 
\begin{equation}~\label{eq:BCSSDWfittinganalytical}
\frac{\kappa_{_{\rm SDW}}(\alpha)}{\kappa_{_{\rm SDW}}(0)}=\sqrt{1-\alpha^2}
\end{equation}
For the flat-band systems there exist no closed analytical expression for the SDW order parameter as a function of $U$ and $\alpha$. Nonetheless, as shown in Figs.~\ref{fig:SDWScalingNH}(c) and~\ref{fig:SDWScalingNH}(f), in both Euclidean and hyperbolic flat-band systems, the SDW order develops for weak enough $U$, which then increases with increasing $U$ and larger $\alpha$ amplifies $\delta_{\rm SDW}$ even further, altogether supporting the proposed mechanism of NH catalysis for the SDW order.

Since the density of states for quasiflat-band system on a square or $\{4, 4 \}$ lattice diverges logarithmically as $|E| \to 0$, the scaling of SDW order parameter in such system is given by~\cite{scaling:3} 
\begin{equation}~\label{eq:SDWFitsquareLat}
\delta_{\rm SDW} \sim \exp \left( - \pi \sqrt{E_\Lambda/U}\right),
\end{equation}
where $E_\lambda$ is proportional to the bandwidth. We fit the self-consistent solution of SDW order with the above form by taking $\pi \sqrt{E_\Lambda} = \eta_{\rm SDW}$ in Fig.~\ref{fig:QFBfitSDW}(a), where $\eta_{\rm SDW}$ is a $\alpha$-dependent fitting parameter, reported in Table~\ref{tab:QFBfitting}. As the scaling of the bandwidth with the NH parameter is given by $E_\Lambda \sim \sqrt{1-\alpha^2}$, we arrive at the following scaling form of the fitting parameter 
\begin{equation}~\label{eq:futtingSLSDW}
\frac{\eta_{\rm SDW}(\alpha)}{\eta_{\rm SDW}(0)} = \left(1-\alpha^2 \right)^{1/4}.
\end{equation}
We find good agreement with the above scaling form for the SDW order on a square lattice; see Fig.~\ref{fig:QFBfitSDW}(b).

Notice that in the hyperbolic flat-band system we compute $\delta_{\rm SDW}$ by averaging the self-consistent solutions only over the sites belonging to the zeroth generation of the $\{ 8, 4\}$ lattice. Additional discussion on this issue is relegated to Appendix~\ref{append:HypFB}. Finally, notice that in the limit $U \to \infty$, the self-consistent SDW order parameter $\delta_{\rm SDW}=1.0$. Then the sites belonging to the $A$ ($B$) sublattice are occupied by electrons with only the spin projection $\sigma=\uparrow$ ($\sigma=\downarrow$) or vice versa. A generalization of this scenario to Bernal-stacked bilayer honeycomb lattice is discussed in Appendix~\ref{append:BBLG}.

\section{Finite-size effects on hyperbolic lattices}~\label{sec:finitesizehyperbolic}

In this work, the self-consistent numerical solutions for the CDW and SDW orders are obtained on Euclidean lattices containing a large number of lattice sites with periodic boundary conditions. Thus these solutions are spatially uniform and do not suffer from the existence of edges. On the other hand, similar sets of solutions are obtained on hyperbolic lattices, each of which also contains a large number of sites, but with open boundary conditions. See Sec.~\ref{sec:system} for details. Consequently, the self-consistent solutions for the CDW and SDW orders display spatial modulations on hyperbolic lattices. Therefore, we need to scrutinize the generation dependence of the spectral gap and the order parameters, as well as the finite-size effects of $\delta_{\rm CDW}$ and $\delta_{\rm SDW}$ on hyperbolic lattices. In addition, while the ratio of the number of edge sites ($N_{\rm edge}$) to the number of bulk sites ($N_{\rm bulk}$) vanishes on Euclidean lattices in the thermodynamic limit, this ratio saturates to a finite number on hyperbolic lattices in that limit. Namely, this ratio is given by 
\begin{equation}~\label{eq:bulkedgeratio}
\frac{N_{\rm edge}}{N_{\rm bulk}} \; \approx \; 4.828, \:\: 10.916, \:\: 4.828 
\end{equation} 
on $\{ 10,3 \}$, $\{ 16,3 \}$, and $\{ 8,4 \}$ hyperbolic lattices, respectively, in the thermodynamic limit. From a realistic point of view, even though realization of hyperbolic lattices in electronic materials still remains far from reality, whenever such systems are synthesized, they will be accompanied by open boundaries. Therefore, it is of fundamental and practical importance to establish the notion of quantum orderings and their NH catalysis on hyperbolic lattices with open boundary conditions.

First, from the self-consistent solutions for the CDW and SDW order parameters, we compute the local density of states on each generation of the hyperbolic lattices. The results are shown in Fig.~\ref{fig:LDOSHyperCDW} for a fixed $V=1.0$ and in Fig.~\ref{fig:LDOSHyperSDW} for a fixed $U=2.5$. These interaction strengths always exceed the critical ones for the CDW and SDW orderings in all hyperbolic Dirac systems, respectively, for any $\alpha$ (see Tables~\ref{tab:NNCoulcric} and~\ref{tab:Hubbardcric}). We find that when $\delta_{\rm CDW}$ and $\delta_{\rm SDW}$ are nontrivial in the entire system, the spectral gap on each generation of the hyperbolic lattice is also finite and comparable. Otherwise, the magnitudes of such local gaps increase with increasing $V$ and $U$, when $\delta_{\rm CDW}$ and $\delta_{\rm SDW}$ increase in the whole system, which we do not show explicitly. These outcomes are qualitatively insensitive to the non-Hermiticity ($\alpha$) in the system. From this observation, we can conclude that when these two order parameters acquire nontrivial self-consistent solutions, the entire system becomes an insulator at half-filling.

\begin{figure*}[t!]
\includegraphics[width=1.00\linewidth]{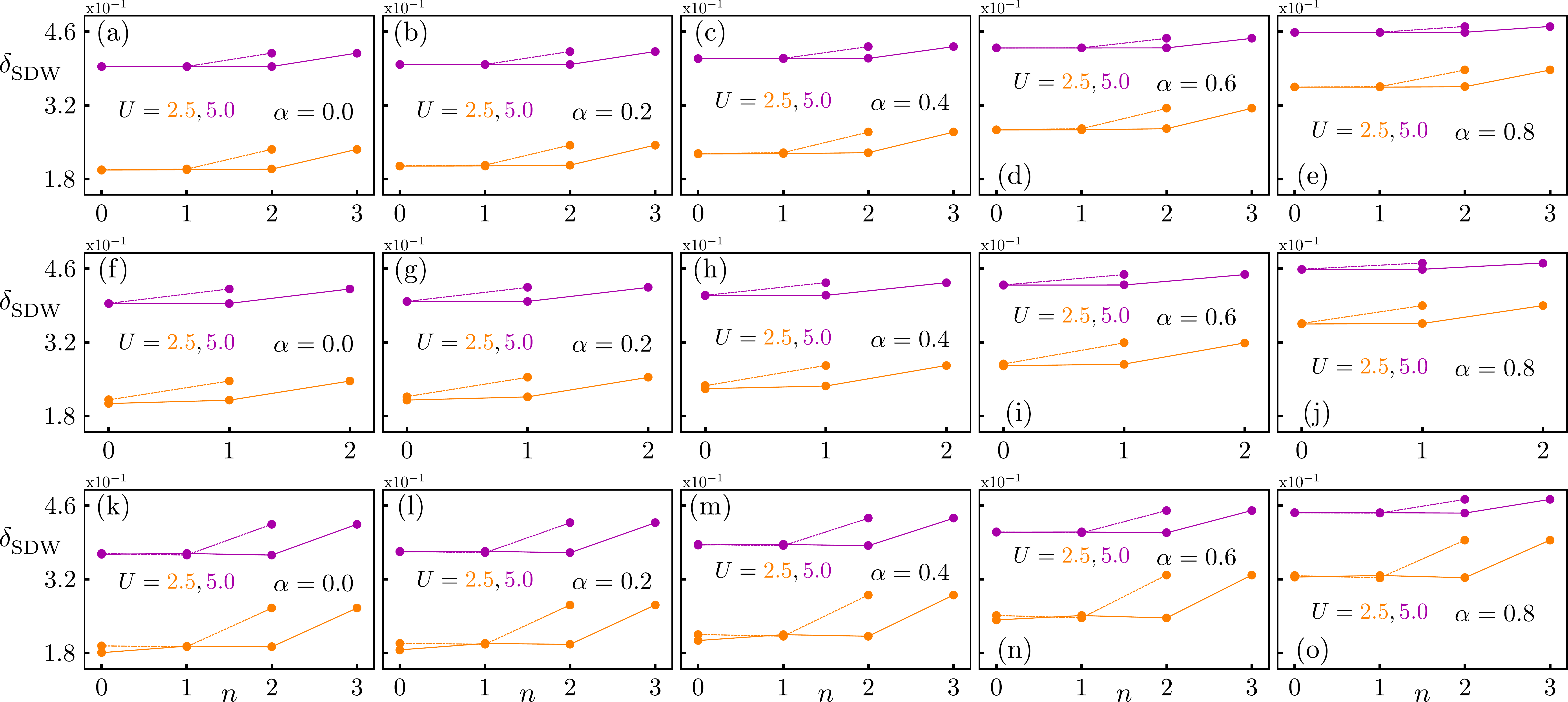}
\caption{Self-consistent solution of the spin-density-wave order parameter ($\delta_{\rm SDW}$) for $U=2.5$ (orange) and $U=5.0$ (purple) as a function of the generation number $n$ (see Sec.~\ref{sec:system}). These quantities are computed by averaging $\delta_{\rm SDW}$ over the sites belonging to the $n$th generation of the (a)-(e) second (dashed lines) and third (solid lines) generation $\{ 10,3 \}$, (f)-(j) first (dashed lines) and second (solid lines) generation $\{ 16,3 \}$, and (k)-(o) second (dashed lines) and third (solid lines) generation $\{ 8,4 \}$ hyperbolic lattices for the non-Hermitian parameter $\alpha=0.0$ [(a), (f), (k)], $0.2$ [(b), (g), (l)], $0.4$ [(c), (h), (m)], $0.6$ [(d), (i), (n)], and $0.8$ [(e), (j), (o)]. For details see Sec.~\ref{sec:finitesizehyperbolic}. The strength of $U$ always exceeds it critical value in the Dirac system, $\{ 10, 3 \}$ hyperbolic lattice; see Table~\ref{tab:Hubbardcric}. Here, $U$ is measured in units of $t$ (nearest-neighbor hopping amplitude).            
}~\label{fig:GenerationSDW}
\end{figure*}

Next, we examine the generation dependence of $\delta_{\rm CDW}$ and $\delta_{\rm SDW}$. For this purpose, we consider each hyperbolic lattice with two different total generation number ($n_{\rm total}$). Namely, we consider $\{10,3\}$ and $\{8,4\}$ hyperbolic lattices with $n_{\rm total}=2$ and $3$, and $\{16,3\}$ hyperbolic lattice with $n_{\rm total}=1$ and $2$. In addition, we consider the self-consistent solutions of $\delta_{\rm CDW}$ for $V=1.0$ and $2.0$, and $\delta_{\rm SDW}$ for $U=2.5$ and $5.0$ for various choices of $\alpha$. Therefore, the sites belonging to the $n$th generation of the hyperbolic lattice constitute the edge of the system when $n=n_{\rm total}$ and the bulk of the system when $0 \leq n < n_{\rm total}$. Once again, recall that the chosen values of $V$ and $U$ always exceed the critical strengths for the CDW and SDW orderings on $\{10,3\}$ hyperbolic lattice for any $\alpha$. Consult Tables~\ref{tab:NNCoulcric} and~\ref{tab:Hubbardcric}.

Finally, we make the following observations from the generation dependence of the CDW order parameter $\delta_{\rm CDW}(n)$, displayed in Fig.~\ref{fig:GenerationCDW}. Notice that the self-consistent solutions for $\delta_{\rm CDW}$ remain almost constant in the bulk, but dips slightly on the edges. Such a dip in $\delta_{\rm CDW}(n_{\rm total})$ decreases with increasing $V$, suggesting a uniform condensation of the CDW order throughout the entire hyperbolic lattice in the limit $V \to \infty$. Furthermore, we note that the magnitude of $\delta_{\rm CDW}(n_{\rm total})$ (at the edge of the system) is almost insensitive to $n_{\rm total}$. In addition, $\delta_{\rm CDW}(j)$ on a hyperbolic lattice with a total $n_{\rm total}$ generations is almost equal to $\delta_{\rm CDW}(j-1)$ on hyperbolic lattice with a total $n_{\rm total}-1$ generations, where $j= 1, \cdots, n_{\rm total}-1$. These outcomes are insensitive to $V$ and $\alpha$. Together these results strongly suggest that the magnitude of the CDW order parameter at each generation of the hyperbolic lattice with open boundary conditions is solely determined by $V$, and in NH systems by $\alpha$ as well. In turn, these numerical observations show that CDW ordering can occur in the entire hyperbolic lattice with the assistance of the NN Coulomb repulsion in the thermodynamic limit, despite a finite fraction of lattice sites belonging to the edge or boundary of the system. The generation dependence of the SDW order is qualitatively similar to those we discussed so far for the CDW order, as shown in Fig.~\ref{fig:GenerationSDW} with one difference. While $\delta_{\rm CDW}$ dips near the boundary of all the hyperbolic lattices, $\delta_{\rm SDW}$ shows an upturn therein. Hence, SDW order can also be realized on hyperbolic lattices with open boundary conditions in the thermodynamic limit, when the on-site Hubbard repulsion dominates in the system. 

As the final remark of this section, we note that the estimation of the critical couplings ($V_c$ and $U_c$) on hyperbolic lattices with open boundary conditions can, in principle, be affected by the system size, as a finite fraction of the total number of sites reside at the boundary of the system. However, with the system sizes considered in this work (see Sec.~\ref{sec:system}), the ratio of the number of sites living on the edge ($N_{\rm edge}$) to the bulk ($N_{\rm bulk}$) of the system is sufficiently close to the one in the thermodynamic limit, quoted in Eq.~\eqref{eq:bulkedgeratio}. Therefore, we expect finite-size effects on the estimation of $V_c$ and $U_c$ to be minimal. Although on hyperbolic Fermi liquids and flat bands, there is no critical coupling for any orderings, their self-consistent solutions are expected to be minimally affected by the system size as well for the same reason. Most importantly, the analytically predicted scaling forms of the critical couplings and magnitude of various orders match very well with the numerical findings in all hyperbolic systems. The same conclusion also holds on all the Euclidean systems, where we impose periodic boundary conditions to eliminate any boundary effects, despite a tiny fraction of sites living therein. Altogether, these findings suggest that the boundary conditions in Euclidean and hyperbolic lattices, and the fact that on hyperbolic lattices with open boundary conditions a large fraction of sites live near the boundary, play a minimal role in the validity of the proposed mechanism of NH catalysis of spontaneous symmetry breaking. Finally, we note that although PBC on hyperbolic lattices can, in principle, be imposed, presently there is no unique or unambiguous recipe for its implementation in the literature.

\section{Summary and discussion}~\label{sec:summary}

To summarize, here we propose a universal mechanism to boost the ordering tendencies of various quantum phases in interacting NH systems. The proposed non-Hermitian catalysis of a selected group of quantum orders rests on a set of simple but robust mathematical criteria that in turn make this mechanism operative on a vast number of quantum systems. (1) If the tight-binding Hamiltonian of a collection of noninteracting fermions (spinless or spinful) is described by a particle-hole symmetric Hermitian operator $\hat{h}_0$ and if there exists another Hermitian operator $\hat{h}_{\rm mass}$ such that these two operators mutually anticommute, then following the general principle of construction [see Eq.~\eqref{eq:NHgeneral}] one can always introduce a NH operator $\hat{h}_{\rm NH}$ that supports a purely real eigenvalue spectrum over an extended NH parameter regime. (2) In such NH systems, the propensity toward the nucleation of any mass ordered state (favored by an appropriate local or short-range four-fermion interaction), represented by a Hermitian operator (${\mathcal O}_{\rm mass}$) such that $\{ \hat{h}_0, {\mathcal O}_{\rm mass} \}=0$, can be amplified by the NH parameter ($\alpha$) near half-filling, if ${\mathcal O}_{\rm mass}$ is a member of the commuting-class mass family satisfying the commutation relation $[ \hat{h}_{\rm mass}, {\mathcal O}_{\rm mass} ]=0$. Condensation of such mass order yields insulation in the half-filled system by producing a spectral gap near zero energy.

The first requirement for the NH catalysis can also be cast in a slightly different language. Notice that this requirement is always fulfilled whenever the free-fermion tight-binding Hamiltonian (not necessarily NN) possesses a unitary particle-hole or sublattice or chiral symmetry, generated by a unitary operator $\hat{U}_{\rm chiral}$, such that $\{ \hat{h}_0, \hat{U}_{\rm chiral}\}=0$ and $\hat{U}_{\rm chiral}$ squares to a unity matrix. Then we can choose $\hat{h}_{\rm mass} \equiv \hat{U}_{\rm chiral}$. In the tenfold classification scheme there are five Altland-Zirnbauer symmetry classes in which the quadratic Hamiltonian for effectively noninteracting charged and neutral Majorana fermions satisfy such a chiral symmetry~\cite{tenfold:1, tenfold:2, tenfold:3}, on which our general principle for the NH catalysis mechanism can readily be applicable in any dimension. In a future investigation we will attempt to generalize this construction when $\hat{h}_0$ belongs to any other of the five symmetry classes, devoid of such chiral or sublattice symmetry.

Depending on the scaling of the density of states near zero energy, we can classify noninteracting electronic fluids into three broad categories. Namely, Dirac liquids, Fermi liquids, and flat-band systems that are respectively characterized by a vanishing, a constant, and a diverging density of states near the zero energy or half-filling. In the latter two systems, mass orderings develop even for sufficiently weak suitable local four-fermion interactions, while Dirac liquids become unstable toward any mass ordering only beyond a critical strength of such an interaction. The NH catalysis in correlated Dirac liquids manifests via the reduction of such critical strength of the interaction, beyond which the amplitude of the order parameter amplifies with increasing strength of non-Hermiticity in the system. By contrast, in the remaining two classes of electronic liquids a larger NH parameter ($\alpha$) always yields a bigger order parameter amplitude for any finite strength of interactions.

Here we present strong analytical arguments to support the NH catalysis mechanism, which we further substantiate through extensive numerical calculations by considering lattice-regularized Dirac liquids, Fermi liquids, and flat-band systems residing on a flat Euclidean plane and a hyperbolic plane with constant negative curvature. For the latter venture, we consider a specific realization of the NH operator in noninteracting systems, in which the non-Hermiticity stems from an imbalance in the hopping amplitudes in the opposite directions between any pair of NN sites. Then, within the framework of a minimal extended Hubbard model containing NN Coulomb and on-site Hubbard repulsions, we promote the NH catalysis of CDW and SDW orders, respectively, in all three classes of NH electronic systems. It should be noted that with the specific construction of the NH operator (Fig.~\ref{fig:NHLattice}), both the CDW and SDW orders belong to the family of commuting class masses. Numerical results are obtained by combining biorthogonal quantum mechanics with lattice-based self-consistent calculations, after decomposing NN Coulomb and on-site Hubbard interactions in the Hartree channels.

Although we primarily focus on lattices on which the NN tight-binding model gives rise to an emergent bipartite structure, in Euclidean space a Fermi liquid system can only be realized on Bernal-stacked honeycomb bilayer lattice. In such a system, the unit cell is constituted by four sites. Nevertheless, our proposed NH catalysis of CDW and SDW orders remain equally operative therein. This outcome liberates the NH catalysis mechanism from the burden of an underlying bipartite (emergent or microscopic) lattice as long as the aforementioned criteria are fulfilled. It should also be noted that this mechanism, resting on a few (anti)commutation relations, is not restricted to any lattice-based system of specific dimensionality. In other words, NH catalysis should be operative in any spatial dimensions.

Exact numerical diagonalization and quantum Monte Carlo simulations possibly constitute the best testbeds for the proposed NH catalysis. Both approaches take into account the effects of quantum fluctuations. While the former method is typically limited to smaller systems only, quantum Monte Carlo simulations on sufficiently large systems can nowadays be performed in all the lattice systems we discussed in this work without encountering the infamous sign problem~\cite{QMC:1, QMC:2, QMC:3, QMC:4, QMC:5, QMC:6, QMC:7, QMC:8, QMC:9, QMC:10, QMC:11, QMC:12, QMC:13, QMC:14, QMC:15, QMC:16, QMC:17, QMC:18}. Fascinatingly, this advanced numerical tool can also be employed in NH systems, at least with a similar construction we subscribed to for our mean-field calculations. Remarkably, even the results from quantum Monte Carlo simulation show reduction in the critical strength of the on-site Hubbard interaction for SDW mass ordering in a NH Euclidean Dirac system (graphene)~\cite{QMC:18}, following the scaling law we analytically predict in this work, see Eqs.~\eqref{eq:criticaldiscrete} and~\eqref{eq:Uccontinuum}, with which the results obtained from numerical self-consistent mean-field calculations agree well.

Any controlled synthesis of NH quantum crystals still remains far from the reality. Therefore, at this moment cold atomic setups constitute the ideal and most promising platform to test the validity of our proposed NH catalysis of various quantum orders in real materials~\cite{OpLat:1, OpLat:2, OpLat:3, OpLat:4, OpLat:5, OpLat:6, OpLat:7, OpLat:8}. In this arrangement, the specific NH lattice models with an imbalance in the hopping amplitudes in the opposite directions between any pair of NN sites can, in principle, be realized by generalizing a similar construction on a one-dimensional chain, proposed in Appendix~F of Ref.~\cite{NHOL:proposal}. Consider two copies of the same two-dimensional lattice,  occupied by neutral atoms living in the ground state and first excited state. These two lattices are then coupled by running waves along the NN bonds and the sites constituted by the excited state atoms undergo a rapid loss, thereby sourcing non-Hermiticity in the system. When the wavelength of the running wave is equal to the lattice spacing, we realize the desired NH operator on optical lattices. On such a platform, various conventional or Hermitian Euclidean two-dimensional lattices (including the square lattice, monolayer honeycomb lattice, and bilayer honeycomb lattice) have already been realized, although hyperbolic lattices are yet to be engineered therein. On the same platform, at least the strength of the on-site Hubbard repulsion can be tuned efficiently to test the predicted NH catalysis of the SDW order. It is also worth noting that three-dimensional Hermitian cubic lattices with NN hopping have also been built on optical lattices, where the strength of on-site Hubbard repulsion can also be tuned. Our proposed NH catalysis mechanism for the SDW order remains equally operative on NH cubic lattices, and therefore predictions in this context can be tested on interacting NH optical cubic lattices. Recent progress in realizing various NH physics in cold atomic systems, among which the NH skin effect is one of the prominent ones~\cite{OpLat:8}, keeps our theoretical proposal for NH catalysis, triggered by Hubbard-like local short range interactions, within the reach of available experimental tools and facilities. The proposed NH catalysis should therefore be of particular interest for the realization of the SDW order in NH graphene at weaker on-site Hubbard repulsion, which has already been observed in Hermitian optical honeycomb lattice~\cite{OpLat:4}, thereby opening up a realistic route to experimentally observe relativistic quantum criticality in NH systems.

Even though our analytical arguments and exact numerical results within the mean-field approximation in support of the NH catalysis mechanism are presented at zero temperature, the results are consequential in predicting the qualitative behavior of the resulting quantum orders at finite temperatures. Notice that the transition temperature for the CDW order $T^{\rm CDW}_c \propto \delta_{\rm CDW}$ as this order only breaks the discrete Ising-like sublattice exchange symmetry dynamically. On the other hand, the SDW order is accompanied by two massless Nambu-Goldstone modes [resulting from the spontaneous lifting of the SU(2) spin rotational symmetry] that destroy any true long-range order at any finite temperature in two dimensions~\cite{goldstone:1, goldstone:2, MarminWagner:1, MarminWagner:2, MarminWagner:3}. This phenomenon, however, goes beyond the scope of any mean-field theory. Still in moderately large systems there exists a putative transition temperature for the SDW order $T^{\rm SDW}_c \propto \delta_{\rm SDW}$. But, in the true thermodynamic limit $T^{\rm SDW}_c$ is only a crossover temperature. Therefore, the transition temperature (genuine or putative) for a commuting-class mass order is expected to get enhanced in NH quantum materials by the NH parameter $\alpha$. In the future, we will explicitly demonstrate this outcome by extending the present mean-field calculation to finite temperatures.

This observation opens up an interesting possibility when various superconducting orders become members of the commuting-class mass family in NH systems. Recall that pairing orders can be supported by electron-phonon interactions and attractive short-range electron-electron interactions. The former interactions, although retarded in nature, can be modeled by effective nonretarded local attractive electronic interactions. At this point, it is worth noting that on optical lattices it is conceivable to access the attractive side of the on-site Hubbard interaction, which typically favors an $s$-wave pairing. Our proposed mechanism therefore suggests that the transition temperature of such a paired state can be enhanced in a suitable NH environment, thereby paving a possibly realistic route to harness the long sought high-$T_c$ superconductors. In the future, it will also be interesting to identify the appropriate NH platforms that can boost the propensity toward the condensation of various topological superconductors, featuring topologically robust Majorana modes that can be useful for quantum computations, for example. NH catalysis assisted high-$T_c$ topological superconductors should make them useful for various practical applications, operative at higher temperatures. While the present analyses lay the foundation of these exciting research avenues, we leave their explicit demonstrations as subjects for thorough future investigations.

We close the discussion by outlining an open question in the context of NH catalysis of symmetry breaking quantum phases. Notice that so far we exclusively focused on commuting-class mass orders for which the single-particle matrix operator $\hat{{\mathcal O}} = \sigma_\mu \otimes \hat{h}_{\rm order}$ anticommutes with $\hat{h}_0$, but commutes with $\hat{h}_{\rm mass}$, such that $\hat{{\mathcal O}}$ fully anticommutes with the total NH operator $\hat{h}_{\rm NH}$. In this case, the effective single-particle Hamiltonian in the presence of ordering of amplitude $\Delta$ of commuting-class mass, given by $\hat{h}_{\rm NH} + \Delta \hat{{\mathcal O}}$ gives all-real eigenvalues for $|\alpha|<1$. Then all the single-particle states remain sharp and filling factor is well defined, such that a mean-field analysis can be carried out at any arbitrary filling. A question, therefore, should arise naturally regarding the catalysis of \emph{anticommuting} class masses, characterized by the anticommutation relations $\{ \hat{h}_0,  \hat{{\mathcal O}}\}=0$ (to qualify as a mass) and $\{ \hat{h}_{\rm mass}, \hat{{\mathcal O}} \}=0$ (to qualify as a anticommuting-class mass), in such NH systems. Unfortunately, any attempt to capture the nucleation of anticommuting-class mass order encounters a severe challenge for which there exists no resolution at this moment. Readers can convince themselves that for such mass orders, the spectrum of the effective single-particle NH operator features complex eigenvalues (as $\hat{{\mathcal O}}$ commutes with the anti-Hermitian component of $\hat{h}_{\rm NH}$). Hence, neither the (effective) single-particle states are sharp nor the filling factor can be defined consistently because of the complex nature of the corresponding eigenvalues. Thus self-consistent mean-field theory cannot be employed to resolve the competition between the commuting- and anticommuting-class masses in NH systems.

\begin{figure}[t!]
\includegraphics[width=0.75\linewidth]{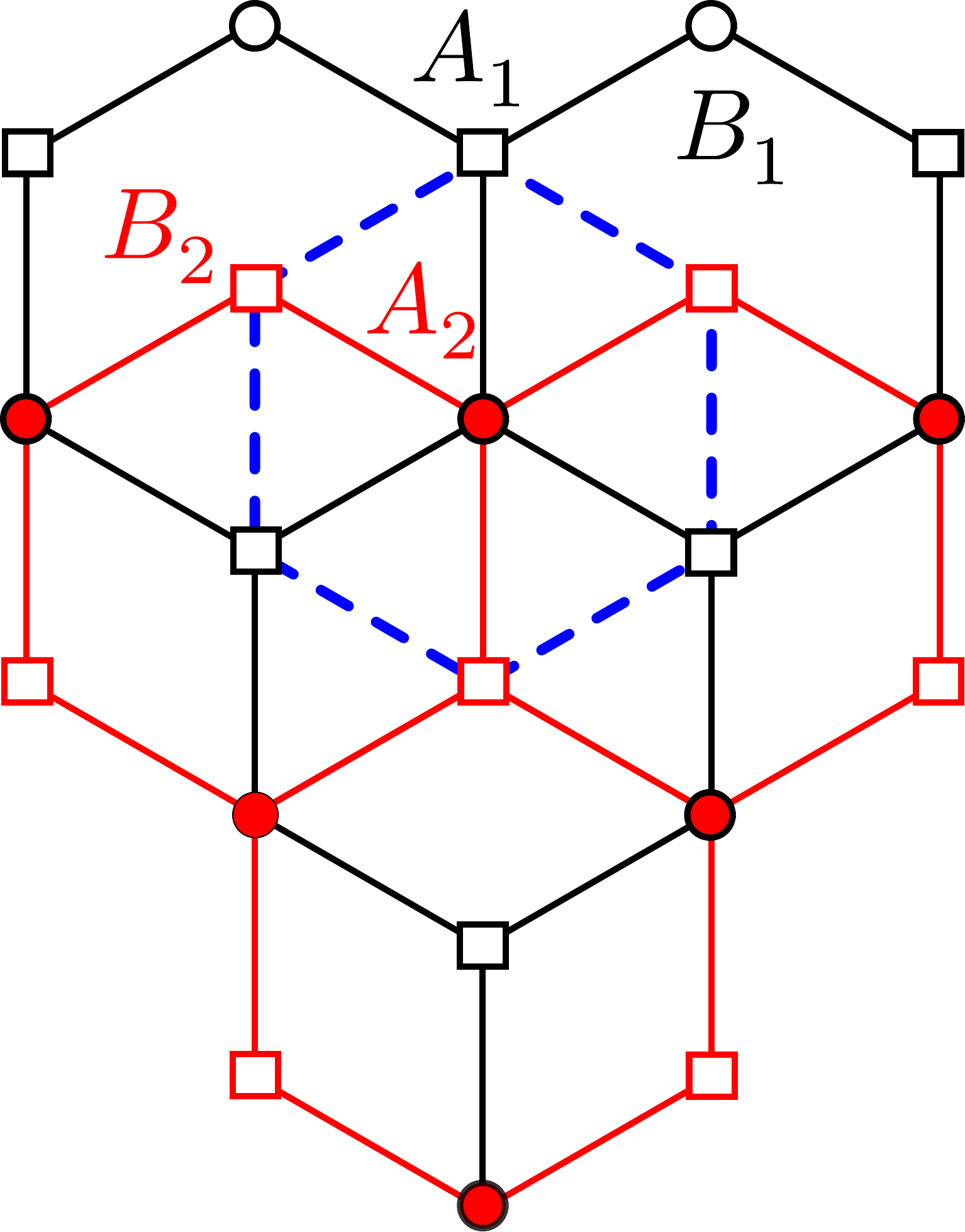}
\caption{Lattice structure of the Bernal-stacked honeycomb bilayer lattice. Here, $A$ and $B$ correspond to two sublattices of the honeycomb lattice on each layer and the subscripts 1 and 2 correspond to the layer indices. Notice that sites belonging to the $B$ sublattice from layer 1 reside at the bottom of those belonging to the $A$ sublattice from layer 2, and a direct interlayer hopping of magnitude $t_\perp$ couples them. Two out of four bands that live far from the zero energy predominantly reside on these sites, whereas the bands near the zero energy primarily live on the sites belonging to the $A_1$ and $B_2$ sublattices. The sites from these two sublattices also constitute an effective honeycomb lattice, shown by the blue dashed lines. For details, see Appendix~\ref{append:BBLG}.             
}~\label{fig:BBLGlattice}
\end{figure}

Nonetheless, in NH Dirac systems we can shed some light on such a competition quantitatively. In Dirac liquids (Euclidean or hyperbolic) there exists a nontrivial or finite critical interaction strength for any ordering ($g_c$), which we can estimate by directly computing the \emph{bare} mean-field susceptibility $\chi$ as $g_c \sim \chi^{-1}$. As detailed in Appendix~\ref{append:competition}, we compute and compare $\chi$ for both commuting- and anticommuting-class masses for graphene-based Dirac liquids. Firstly, we confirm that in agreement with our analytical prediction and findings from the lattice-based self-consistent mean-field calculations 
\begin{equation}~\label{eq:suscommute}
\frac{g^{\rm CCM}_c(\alpha)}{g^{\rm CCM}_c(0)} = \frac{\chi^{-1}_{\rm CCM}(\alpha)}{\chi^{-1}_{\rm CCM}(0)} = \left( 1-\alpha^2 \right)^{1/2}
\end{equation}
for the commuting-class mass. When extended to anticommuting-class mass, we, on the other hand, find that 
\begin{equation}~\label{eq:susanticommute}
\frac{g^{\rm ACM}_c(\alpha)}{g^{\rm ACM}_c(0)} = \frac{\chi^{-1}_{\rm ACM}(\alpha)}{\chi^{-1}_{\rm ACM}(0)} = \left( 1-\alpha^2 \right)^{3/2}.
\end{equation}
Thus a bare mean-field or susceptibility calculation suggests that the propensity toward the nucleation of anticommuting-class mass is stronger than that for commuting-class mass in NH system. In other words, the NH catalysis should favor the former class of orders. However, there is a caveat in such a rushed conclusion. From a field-theoretic renormalization group calculations for NH Euclidean Dirac liquids~\cite{NH:1}, controlled by a suitable $\epsilon$ expansion, it was shown that while $\alpha$ remains \emph{marginal} (does not change with the running scale) close to the quantum critical point controlling the transition to a commuting-class mass ordering, $\alpha$ is \emph{marginally irrelevant} around a similar critical point that however controls the transition to anticommuting-class masses, and thus $\alpha \to 0$ as the system approaches $g^{\rm ACM}_{c}$ from the disordered side. From this observation, we can conclude that nucleation of anticommuting-class mass orders should not be catalyzed by the non-Hermiticity in the system, at least in Dirac liquids. However, we believe that this competition is far from being settled and will require advanced numerical tools, such as quantum Monte Carlo simulations, for the final outcome. On the other hand, computation of bare mean-field susceptibility cannot be employed to assess this competition in Fermi liquids and flat bands, as $g_c \to 0$ therein and hence $\chi \to \infty$ in these systems. We believe that the present discussion should stimulate a surge of at least theoretical investigations geared toward unfolding the competition between anticommuting- and commuting-class masses within the proposed mechanism of NH catalysis of spontaneous symmetry breaking.

\acknowledgements 

This work was supported by NSF CAREER Grant No.\ DMR-2238679 of B.R. We thank Vladimir Juri\v ci\' c for critical reading of the manuscript. Portions of this research were conducted on Lehigh University's Research Computing infrastructure partially supported by NSF Award No.~2019035. B.R.\ is grateful to Institute for Solid State Physics, University of Tokyo where a part of this work was performed for their hospitality.

\section*{Data availability}

Numerical codes and data used and generated in this work are available in Zenodo~\cite{dataChrisNHCatNoB}.

\appendix

\section{Details of Bernal-stacked honeycomb bilayer lattice}~\label{append:BBLG}

In this appendix, we discuss the details of the NN tight-binding model and the CDW and SDW order parameters in Bernal-stacked bilayer honeycomb lattice. This system is slightly different from conventional bipartite lattices, since besides two sublattices on each layer ($A$ and $B$) we also need to keep track of the layer index $\xi=1$ and $2$. Nevertheless, all the formalism we discussed so far straightforwardly generalize to this system. The corresponding lattice structure is shown in Fig.~\ref{fig:BBLGlattice}. As mentioned in Sec.~\ref{subsec:TBnumerics}, the sites belonging to the $B$ sublattice on layer 1 reside right beneath the sites from the $A$ sublattice on layer 2, which are connected via an interlayer (dimer) hopping of amplitude $t_\perp$. The tight-binding Hamiltonian containing only spin-independent intralayer NN and interlayer dimer hopping processes, with amplitudes $t$ and $t_\perp$, respectively, takes the form~\cite{graphene:RMP}
\allowdisplaybreaks[4]
\begin{eqnarray}~\label{eq:HamilTBBBLG}
H_0 &=& - t \; \sum_{\xi=1,2} \sum_{\langle i,j \rangle} \: \sum_{\sigma=\uparrow, \downarrow} \; c^\dagger_{i_{\xi}\sigma} c_{j_{\xi} \sigma} \nonumber \\
&-& t_\perp \left( \sum_{B_1} \sum_{\sigma =\uparrow, \downarrow} c^\dagger_{B_1\sigma} c_{A_2\sigma} + {\rm H.c.} \right).
\end{eqnarray}
Here, $c^\dagger_{i_{\xi}\sigma}$ is the fermionic creation operator on the $i$th site living on the layer $\xi=1,2$ with the spin projection $\sigma=\uparrow, \downarrow$, and $c^\dagger_{B_1\sigma}$ ($c_{A_2\sigma}$) is the fermionic creation (annihilation) operator on the sites belonging to the $B$ ($A$) sublattice residing on layer $1$ ($2$).

\begin{figure}[t!]
\includegraphics[width=1.00\linewidth]{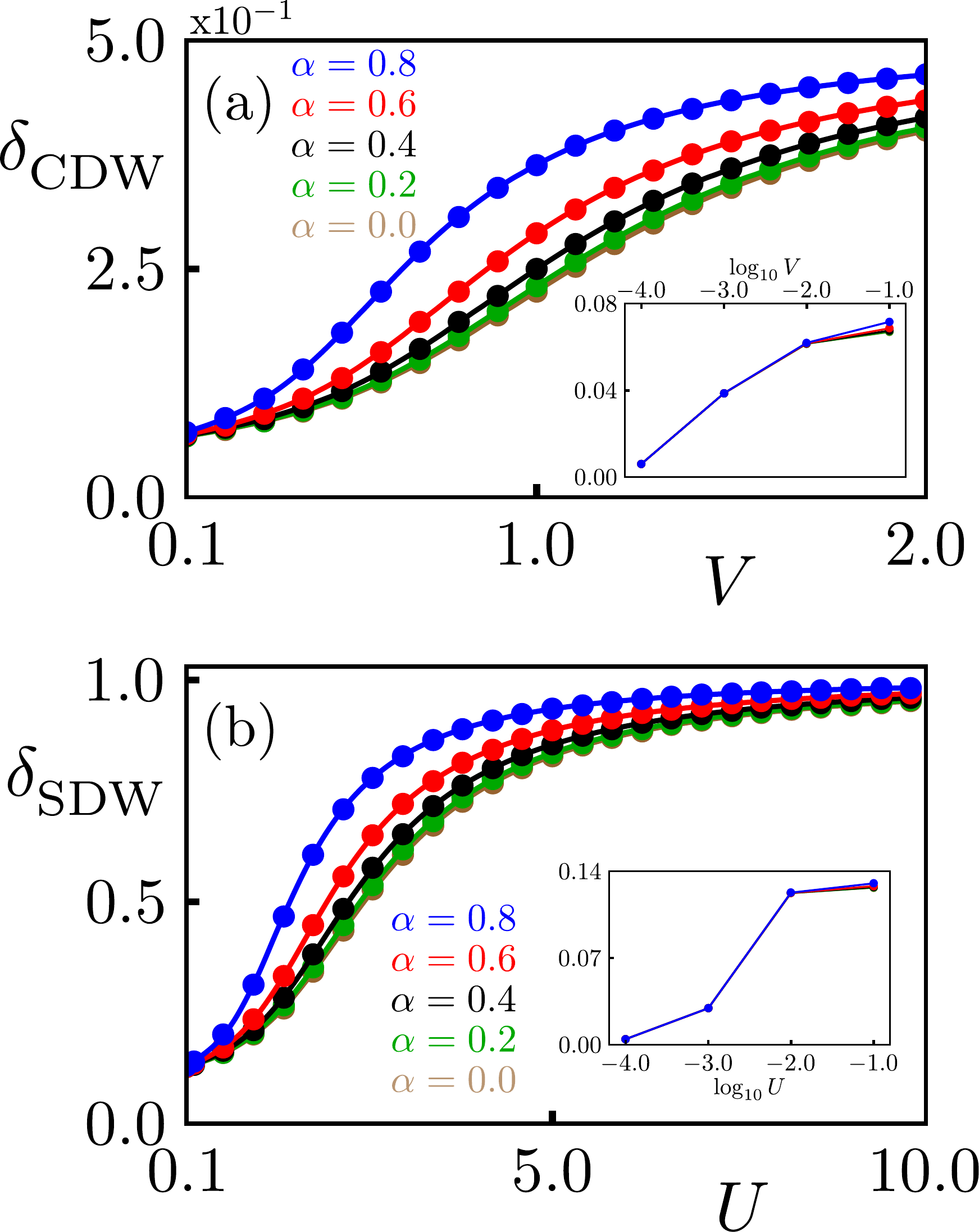}
\caption{Self-consistent solutions for the (a) charge-density-wave ($\delta_{\rm CDW}$) and (b) spin-density-wave ($\delta_{\rm SDW}$) order parameters as functions of the nearest-neighbor Coulomb ($V$) and on-site Hubbard ($U$) repulsions, respectively. These quantities are obtained upon averaging the corresponding order parameters over the entire system of the $\{ 8,4 \}$ hyperbolic lattice (see Sec.~\ref{sec:system}) for various choices of the non-Hermitian parameter ($\alpha$). Notice that $\delta_{\rm CDW}$ and $\delta_{\rm SDW}$ show putative jumps for small $V$ and $U$, respectively, in contrast to the situation when these two quantities are averaged only over the sites belonging to the zeroth generation of the lattice, as shown in Figs.~\ref{fig:CDWScalingNH}(f) and~\ref{fig:SDWScalingNH}(f) for charge-density-wave and spin-density-wave orders. Nevertheless, we find $\delta_{\rm CDW} \to 0$ ($\delta_{\rm SDW} \to 0$) smoothly as $V \to 0$ ($U \to 0$), as shown in the insets. For details, see Appendix~\ref{append:HypFB}. Here, $U$ and $V$ are measured in units of $t$ (nearest-neighbor hopping amplitude).        
}~\label{fig:HLFlatWholesystem}
\end{figure}

To describe this system, we define a $4N$-component spinor $\Psi^\top=(\Psi_\uparrow, \Psi_\downarrow)$, where the $2N$-component spinor for each spin projection $\sigma$ is now given by $\Psi^\top_\sigma=(c_{A_1 \sigma}, c_{B_1 \sigma}, c_{A_2 \sigma}, c_{B_2 \sigma})$, where $c_{A_\alpha \sigma}$ and $c_{B_\alpha \sigma}$ are $N/2$-dimensional spinors constituted by the annihilation operators belonging to the $A$ and $B$ sublattices, respectively, with spin projection $\sigma=\uparrow, \downarrow$ and residing on layer $\xi=1,2$. In this basis the Hermitian operator, associated with $H_0$ from Eq.~\eqref{eq:HamilTBBBLG} takes the form $\hat{h}_0 = \sigma_0 \otimes \hat{h}^{\rm spinless}_0$, where
\renewcommand{\arraystretch}{1.25}
\begin{equation}
\hat{h}^{\rm spinless}_0 = \left(\begin{array}{cc|cc}
{\boldsymbol 0} & {\bf t} & {\boldsymbol 0} & {\boldsymbol 0} \\
{\bf t}^\top & {\boldsymbol 0} & {\bf t}_\perp & {\boldsymbol 0} \\
\hline
{\boldsymbol 0} & {\bf t}^\top_\perp & {\boldsymbol 0} & {\bf t}^\top \\
{\boldsymbol 0} & {\boldsymbol 0} & {\bf t} & {\boldsymbol 0}
\end{array}
\right)
\end{equation} 
and ${\bf t}_\perp$ is the $N/2$-dimensional interlayer real dimer hopping matrix. In this basis, we immediately identify a Hermitian matrix that fully anticommutes with $\hat{h}^{\rm spinless}_0$, given by 
\renewcommand{\arraystretch}{1}
\begin{equation}
\hat{h}_{\rm CDW}=\left(\begin{array}{cc|cc}
{\bf I} & {\boldsymbol 0} & {\boldsymbol 0} & {\boldsymbol 0} \\
{\boldsymbol 0} & -{\bf I} & {\boldsymbol 0} & {\boldsymbol 0} \\
\hline
{\boldsymbol 0} & {\boldsymbol 0} & {\bf I} & {\boldsymbol 0} \\
{\boldsymbol 0} & {\boldsymbol 0} & {\boldsymbol 0} & -{\bf I}
\end{array}
\right).
\end{equation}
With the identification of these two mutually anticommuting Hermitian matrices, we can introduce the corresponding NH operator in Bernal-stacked bilayer honeycomb lattice following the construction from Eq.~\eqref{eq:NHHamil} with $\hat{h}_{\rm mass}=\sigma_0 \otimes \hat{h}_{\rm CDW}$ in Eq.~\eqref{eq:NHgeneral} following the general principle of constructing NH operators. Hence, all the discussions on the eigenvalue spectrum of $\hat{h}_{\rm NH}$ also apply here.

In such a NH system, we can immediately identify two mass operators belonging to the commuting-class mass family, similar to the ones shown in Eq.~\eqref{eq:massoperators} 
\allowdisplaybreaks[4]
\begin{eqnarray}~\label{eq:massoperatorsBBLG}
{\mathcal O}_{\rm CDW}=\sigma_0 \otimes \left( 
\begin{array}{cc|cc}
{\boldsymbol \Delta} & {\boldsymbol 0} & {\boldsymbol 0} & {\boldsymbol 0} \\
{\boldsymbol 0} & -{\boldsymbol \Delta} & {\boldsymbol 0} & {\boldsymbol 0} \\
\hline
{\boldsymbol 0} & {\boldsymbol 0} & {\boldsymbol \Delta} & {\boldsymbol 0} \\
{\boldsymbol 0} & {\boldsymbol 0} & {\boldsymbol 0} & -{\boldsymbol \Delta}  
\end{array}
\right) \nonumber \\
\text{and} \: {\mathcal O}_{\rm SDW}= \sigma_j \otimes \left( 
\begin{array}{cc|cc}
{\boldsymbol \Delta} & {\boldsymbol 0} & {\boldsymbol 0} & {\boldsymbol 0} \\
{\boldsymbol 0} & -{\boldsymbol \Delta} & {\boldsymbol 0} & {\boldsymbol 0} \\
\hline 
{\boldsymbol 0} & {\boldsymbol 0} & {\boldsymbol \Delta} & {\boldsymbol 0} \\
{\boldsymbol 0} & {\boldsymbol 0} & {\boldsymbol 0} & -{\boldsymbol \Delta} 
\end{array}
\right).
\end{eqnarray} 
Hence, all the discussions from Sec.~\ref{sec:NHcatalysis} promoting the concept of NH catalysis of commuting-class masses directly apply to the NH Bernal-stacked honeycomb bilayer lattice for these two mass orders. 

\begin{figure}[t!]
\includegraphics[width=1.00\linewidth]{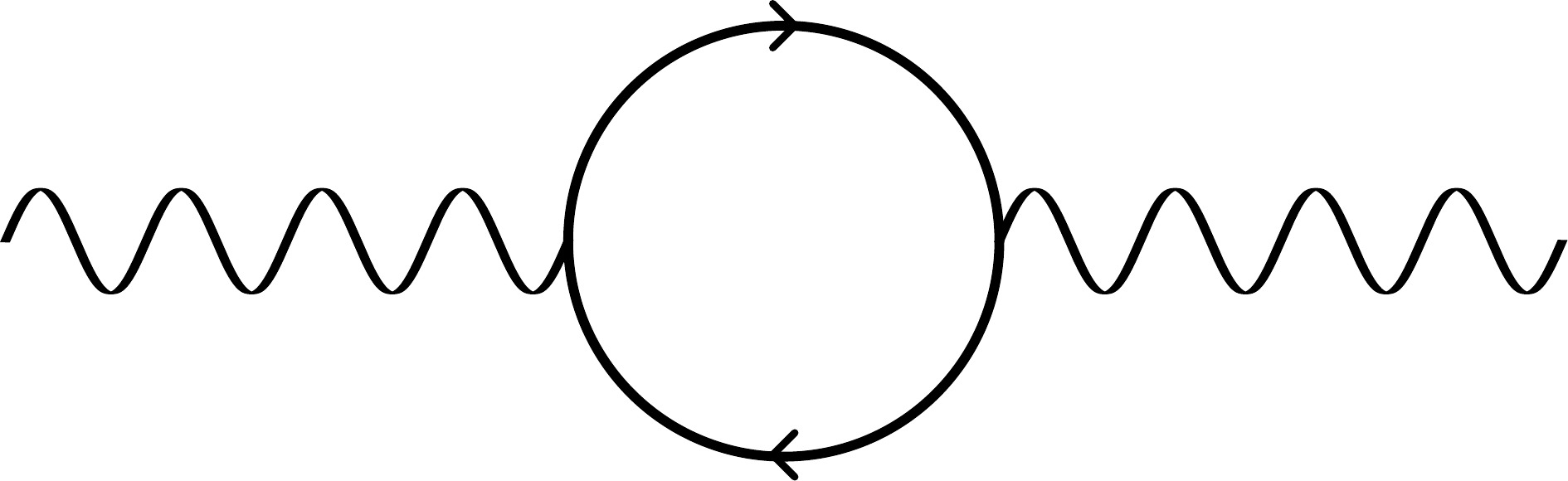}
\caption{Feynman diagram capturing the bare mean-field susceptibility; see Eq.~\eqref{eq:feyndiagram_integral}. The solid lines represent fermionic fields and the wavy lines correspond to the order parameter field. We numerically compute the contribution of this Feynman diagram at finite temperatures through a summation over Matsubara frequencies; see Eq.~\eqref{eq:matsubarasum}. The results are displayed in Fig.~\ref{fig:matsubarsum_results}. 
}~\label{fig:feyndiagram}
\end{figure} 

The Hamiltonian capturing the repulsive Coulomb interaction among the spinless fermions living on the NN sites of the honeycomb lattice on the same layer is now given by 
\begin{equation}~\label{eq:NNCoulBBLG}
H^{\rm Coul}_{\rm NN} = \frac{V}{2} \sum_{\xi=1,2} \; \sum_{\langle i,j \rangle} n^{\xi}_i n^{\xi}_j - \mu {\mathcal N},
\end{equation}
where $n^{\xi}_i$ is the fermionic density on the $i$th site of the $\xi$th layer. The Hartree decomposition of $H^{\rm Coul}_{\rm NN}$ leads to the following effective single-particle Hamiltonian 
\begin{equation}
H^{\rm Har}_{\rm NN} = V \sum_{\xi=1,2} \; \sum_{\langle i,j \rangle} \bigg( \langle n^{\xi}_{B,i} \rangle n^{\xi}_{A, j} + \langle n^{\xi}_{A,i} \rangle n^{\xi}_{B, j} \bigg) 
- \mu {\mathcal N},
\end{equation}
where $\langle n^\xi_{A,i} \rangle$ and $\langle n^\xi_{B,i} \rangle$ correspond to the site ($i$) and layer ($\xi$) dependent self-consistent average electronic density on the $A$ and $B$ sublattices, respectively, at half-filling. We measure them relative to the uniform density at half-filling according to 
\begin{eqnarray}
\langle n^{\xi}_{A,i}\rangle = \frac{1}{2} + \delta^\xi_{A,i}  
\:\:\: \text{and} \:\:\: 
\langle n^{\xi}_{B,i}\rangle = \frac{1}{2} - \delta^\xi_{B,i}.
\end{eqnarray} 
The half-filling condition is maintained by setting $\mu=V/2$ and ensuring that 
\begin{equation}
\sum_{\xi=1,2} \left( \sum_{i} \delta^\xi_{A,i} - \sum_{i} \delta^\xi_{B,i} \right) =0.
\end{equation} 
Finally, the positive definite quantities $\delta^{\xi}_A$ and $\delta^{\xi}_B$ yield the local CDW order parameter on Bernal-stacked honeycomb bilayer lattice  
\begin{equation}~\label{eq:CDWOPBBLG}
\delta^{\rm local}_{\rm CDW}=\frac{1}{2} \sum_{\xi=1,2} \; \left( \delta^\xi_{A} + \delta^\xi_{B} \right),   
\end{equation}
from which we compute $\delta_{\rm CDW}$ in the entire system, following the prescription from Sec.~\ref{sec:CDW}. Notice that in the limit $V \to \infty$, the $A_1$ and $A_2$ ($B_1$ and $B_2$) sites are completely filled (empty), for example, yielding $\delta_{\rm CDW}=0.5$.

The Hamiltonian describing on-site Hubbard repulsion in Bernal-stacked honeycomb bilayer lattice takes the form 
\begin{equation}
H^{\rm Hub}_{\rm OS}= U \sum_{\xi=1,2} \sum_{i} \left( n^\xi_{i, \uparrow} -\frac{1}{2} \right) \; \left( n^\xi_{i, \downarrow} -\frac{1}{2} \right) -\mu {\mathcal N},
\end{equation}
where $n^\xi_{i,\uparrow/\downarrow}$ is the electronic density at site $i$ on the $\xi$th layer with the spin projection $\uparrow/\downarrow$ in the $z$ direction. The Hartree decomposition of $H^{\rm Hub}_{\rm OS}$ from the above equation leads to the following effective single-particle Hamiltonian 
\begin{eqnarray}
H^{\rm Har}_{\rm OS} &=& \sum_{\xi=1,2} \sum_{x=A,B} \bigg\{ \left( \langle n^\xi_{x,\uparrow} \rangle -\frac{1}{2}\right) \left( n^\xi_{x,\downarrow} -\frac{1}{2} \right) \nonumber \\
&+& \left( \langle n^\xi_{x,\downarrow} \rangle -\frac{1}{2}\right) \left( n^\xi_{x,\uparrow} -\frac{1}{2} \right) \bigg\} -\mu {\mathcal N}.
\end{eqnarray}
For this system, we choose the following ansatz 
\begin{equation}
\langle n^\xi_{A,\sigma} \rangle= \frac{1}{2} + \sigma \; \delta^\xi_{A,\sigma}(\vec{r}), 
\:\:\:
\langle n^\xi_{B,\sigma} \rangle= \frac{1}{2} - \sigma \; \delta^\xi_{B,\sigma}(\vec{r}). 
\end{equation}
The half-filling condition is now satisfied with the choice $\mu=0$ and 
\begin{equation}
\sum_{\xi=1,2} \sum_{\sigma=\pm} \bigg(
 \sum_{\vec{r}} \; \sigma \; \delta^\xi_{A,\sigma} (\vec{r})
- \; \sum_{\vec{r}} \; \sigma \; \delta^\xi_{B, \sigma} (\vec{r}) \bigg)=0.
\end{equation}
Then, the local SDW order parameter with $\delta^{\xi}_{A/B, \uparrow/\downarrow}>0$ is given by
\begin{equation}~\label{eq:AFMOP}
\delta^{\rm local}_{\rm SDW}=\frac{1}{4} \sum_{\xi=1,2} \left( \delta^\xi_{A,\uparrow} + \delta^\xi_{A,\downarrow} + \delta^\xi_{B,\uparrow} + \delta^\xi_{B,\downarrow} \right),
\end{equation} 
from which we obtain $\delta_{\rm SDW}$ by averaging the above quantity in the entire system. Notice that in the limit $U \to \infty$, the $A_1$ and $A_2$ ($B_1$ and $B_2$) sites are solely occupied by electrons with spin projection $\sigma=\uparrow$ ($\sigma=\downarrow$), for example, yielding $\delta_{\rm SDW}=1.0$.

\section{Additional numerical results on hyperbolic flat bands}~\label{append:HypFB}

In this appendix, we discuss the scaling of the CDW and SDW order parameters in hyperbolic flat band systems realized on the $\{ 8,4 \}$ lattice in more detail. Notice that in Figs.~\ref{fig:CDWScalingNH}(f) and~\ref{fig:SDWScalingNH}(f), we display the scaling of the CDW and SDW order parameters with the NN Coulomb ($V$) and on-site Hubbard ($U$) repulsions, respectively, by averaging these quantities only over the sites belonging to the zeroth generation of the $\{ 8,4 \}$ lattice (see Sec.~\ref{sec:system}) for various choices of the NH parameter ($\alpha$). In this case these two quantities smoothly interpolate to their trivial values as $V \to 0$ and $U \to 0$, as they should. However, when we compute $\delta_{\rm CDW}$ and $\delta_{\rm SDW}$ by averaging their local values over the entire system, we find that there exist putative jumps in these two quantities for small $V$ and $U$, respectively. These outcomes are shown in Fig.~\ref{fig:HLFlatWholesystem}. Nevertheless, even in this case $\delta_{\rm CDW} \to 0$ smoothly when $V \to 0$ and $\delta_{\rm SDW} \to 0$ smoothly when $U \to 0$, as shown in the insets of Fig.~\ref{fig:HLFlatWholesystem}. Likely because of the diverging density of states, these two order parameters rise extremely rapidly when computed by averaging their self-consistent solutions over the entire system. However, such putative jumps do not alter any conclusions regarding the proposed NH catalysis of the commuting-class masses.

\begin{figure}[t!]
\includegraphics[width=1.00\linewidth]{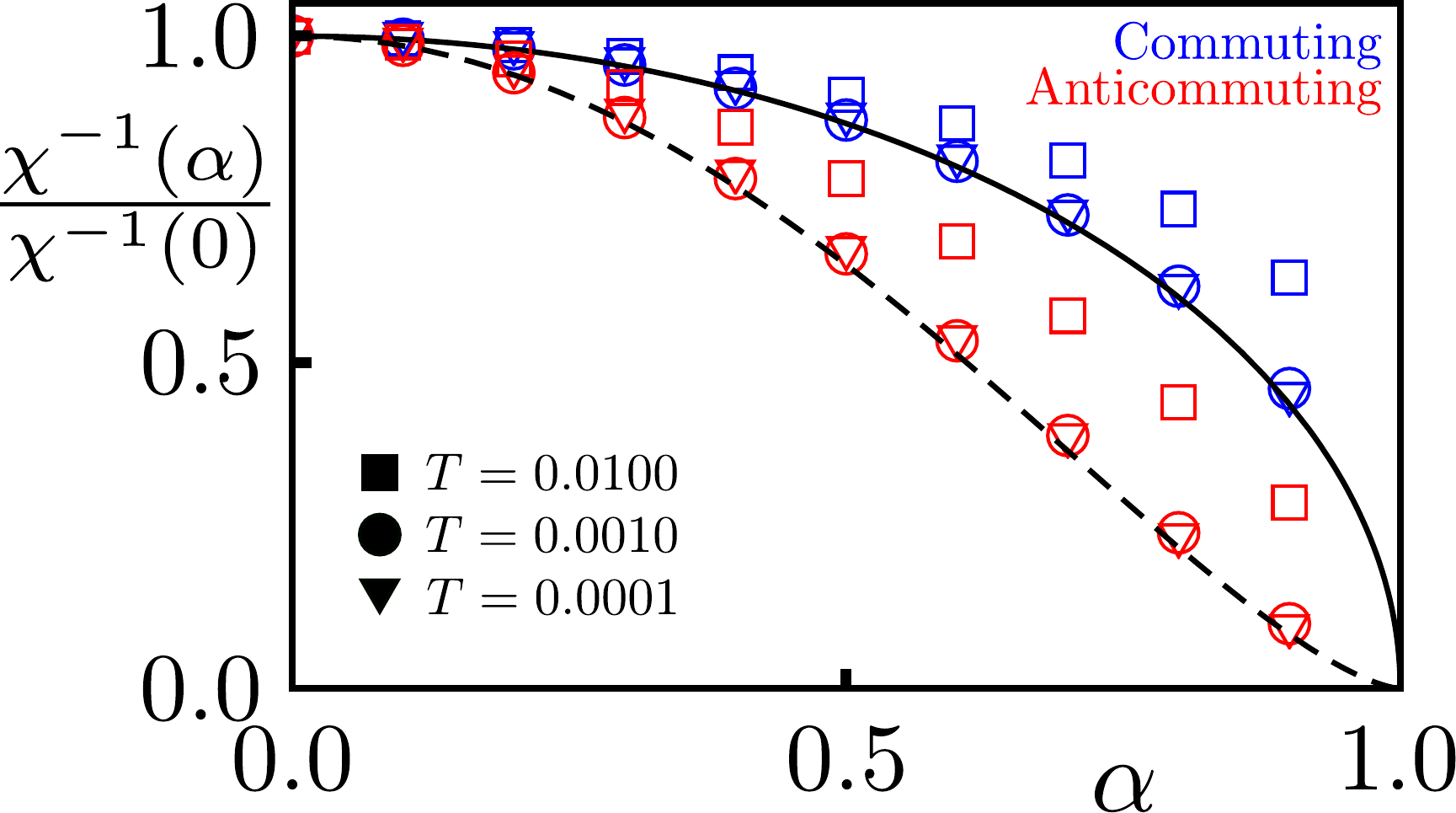}
\caption{The bare mean-field susceptibility $\chi$ obtained by computing the Feynman diagram shown in Fig.~\ref{fig:feyndiagram}, detailed in Appendix~\ref{append:competition}. Numerically extracted ratios of the inverses of the susceptibilities $\chi^{-1}(\alpha) / \chi^{-1}(0)$ as a function of the non-Hermitian parameter ($\alpha$) are shown for commuting- (blue) and anticommuting- (red) class masses at various temperatures $T$ (marked by different symbols). The computations for commuting (anticommuting)-class mass is performed on a honeycomb lattice with periodic boundary conditions for spinless (spinful) fermions containing 2400 (1536) sites. Notice that as temperature $T \to 0$, results for $\chi^{-1}(\alpha) / \chi^{-1}(0)$ converge to the analytically predicted scaling forms for commuting- and anticommuting-class masses, shown by solid and dashed black lines, respectively; quoted in Eqs.~\eqref{eq:suscommute} and~\eqref{eq:susanticommute}. Here, $T$ is measured in units of $t$ (nearest-neighbor hopping amplitude). 
}~\label{fig:matsubarsum_results}
\end{figure}

\section{Competition between commuting- and anticommuting-class masses}~\label{append:competition}

In this appendix, we present the details of the bare mean-field susceptibility calculation for commuting- and anticommuting-class masses in an Euclidean Dirac system, namely monolayer honeycomb lattice. The bare mean-field susceptibility at zero temperature associated with an ordering, described by the effective single-particle operator $\hat{{\mathcal O}}$ in the NH system, described by the NH operator $\hat{h}_{\rm NH}$, is given by 
\begin{equation}~\label{eq:feyndiagram_integral}
    \chi(\alpha) = \int\limits_{-\infty}^{\infty} \frac{d \omega}{2\pi} \; {\rm Tr} \left( \hat{{\mathcal O}} \; \frac{1}{i \omega - \hat{h}_{\rm NH}} \; \hat{{\mathcal O}} \; \frac{1}{i \omega - \hat{h}_{\rm NH}} \right),
\end{equation}
where ${\rm Tr}$ denotes trace. The corresponding Feynman diagram is shown in Fig.~\ref{fig:feyndiagram}. However, performing the integral over the Mastusubara frequency ($\omega$) is challenging to execute numerically. For this reason, we compute $\chi(\alpha)$ at a finite temperature ($T$), given by 
\begin{equation}
    \chi(\alpha, T) = T \sum \limits_{n=-N_{\rm max}}^{N_{\rm max}} {\rm Tr} \left( \hat{{\mathcal O}} \; \frac{1}{i \omega_n - \hat{h}_{\rm NH}} \; \hat{{\mathcal O}} \; \frac{1}{i \omega_n - \hat{h}_{\rm NH}} \right )
    \label{eq:matsubarasum}
\end{equation}
after setting the Boltzmann constant $k_B =1$. Notice that at any finite temperature the integral over continuous Mastubara frequency is replaced by a summation over discrete fermionic Matsubara frequencies, given by $\omega_n = (2n+1) \pi T$, where $n$ is an integer. For faster numerical convergence of $\chi(\alpha, T)$, we omit contributions from a few Matsubara frequencies around $n=0$. We also notice that for $N_{\rm max}=4000$, numerical values of $\chi(\alpha, T)$ converge and reproduce the expected values as $N_{\rm max} \to \infty$.

For the computation of $\chi(\alpha,T)$ for a commuting-class mass, we choose $\hat{{\mathcal O}}= \sigma_0 \otimes {\rm diag}.({\bf I}_{N/2}, -{\bf I}_{N/2})$ and $\hat{h}_{\rm mass} \equiv \hat{{\mathcal O}}$ in the construction of $\hat{h}_{\rm NH}$ [see Eq.~\eqref{eq:NHgeneral}]. As with such a choice of the commuting-class mass both the order parameter matrix and the NH operator are independent of the spin degrees of freedom, in this case we compute $\chi(\alpha, T)$ for spinless fermions. On the other hand, for the computation of $\chi(\alpha,T)$ for an anticommuting-class mass, we choose $\hat{{\mathcal O}}= \sigma_3 \otimes {\rm diag}.({\bf I}_{N/2}, -{\bf I}_{N/2})$ and $\hat{h}_{\rm mass} = \sigma_1 \otimes {\rm diag}.({\bf I}_{N/2}, -{\bf I}_{N/2})$. In this case, we compute $\chi(\alpha, T)$ by including the spin degrees of freedom in the system. The results are displayed in Fig.~\ref{fig:matsubarsum_results}. We note that as the temperature $T \to 0$, the scaling of the susceptibilities for commuting- and anticommuting-class masses approach their forms, predicted from analytical calculations~\cite{NH:1}, shown in Eqs.~\eqref{eq:suscommute} and~\eqref{eq:susanticommute}, respectively.


\end{document}